\documentclass[twocolumn,apj]{aastex701}
\usepackage{amsthm,amsmath,amssymb}
\usepackage{url}
\usepackage{mathrsfs}

\usepackage{mathtools}

\newcommand{\secpoint}{\mbox{$''\mskip-7.6mu.\,$}}

\newcommand {\bc}{\begin {center}}
\newcommand {\ec}{\end {center}}
\newcommand {\be}{\begin {equation}}
\newcommand {\ee}{\end {equation}}
\newcommand {\beq}{\begin {eqnarray}}
\newcommand {\eeq}{\end {eqnarray}}

\usepackage{xspace}

\def\ugc{UGC~2369S\xspace}
\def\osiris{OSIRIS\xspace}
\shorttitle{\ugc}
\shortauthors{Ding et al.}
\graphicspath{{./}{figures/}}
\newcommand{\todo}{\ifmmode \text{\color{red}\Huge{\(\bullet\)}} \else {\color{red}{\Huge$\bullet$}}\fi}
\newcommand{\tido}{\ifmmode {{\color{red}\bullet}} \else {\color{red}$\bullet$}\fi}

\newcommand{\E        }[1]{\ifmmode 10^{#1} \else $10^{#1}$\fi}
\newcommand{\til}{\ifmmode \sim \else $\sim$\fi}
\renewcommand{\~} {\ifmmode \sim \else $\sim$\fi}

\newcommand{\logNH }{\ifmmode \log (N_{\rm H}/{\rm cm}^{-2}) \else $\log (N_{\rm H}/{\rm cm}^{-2})$\fi}

\newcommand{\Mbh   }{\ifmmode M_{\rm BH} \else $M_{\rm BH}$\fi}

\newcommand{\pc}	{\ifmmode {\rm pc} \else pc\fi}
\newcommand{\ld}	{\ifmmode {\rm l.d.} \else l.d.\fi}
\newcommand{\cc}	{\ifmmode {\rm cm}^{-3}    \else cm$^{-3}$\fi}
\newcommand{\cmii}	{\ifmmode {\rm cm}^{-2}    \else cm$^{-2}$\fi}
\newcommand{\ergs}	{\ifmmode {\rm erg\,s}^{-1} \else erg s$^{-1}$\fi}
\newcommand{\ergcms}	{\ifmmode {\rm erg\,cm}^{-2}\,{\rm s}^{-1} \else erg\,cm$^{-2}$\,s$^{-1}$\fi}
\newcommand{\ergcmsA}	{\ifmmode {\rm erg\,cm}^{-2}\,{\rm s}^{-1}\,{\rm\AA}^{-1}
\else erg\,cm$^{-2}$\,s$^{-1}$\,\AA$^{-1}$\fi}
\newcommand{  \ergcmsHz  }{\ifmmode{\rm erg\,cm}^{-2}\,{\rm s}^{-1}\,{\rm Hz}^{-1}
                       \else ergs\,cm$^{-2}$\,s$^{-1}$\,Hz$^{-1}$\fi}
\newcommand{\kev}	{\ifmmode {\rm keV} \else keV\fi}

\newcommand{\mic}	{\ifmmode {\rm \mu m} \else $\mu$m\fi}
\newcommand{\vFWHM}	{\ifmmode v_{\mbox{\tiny FWHM}} \else $v_{\mbox{\tiny FWHM}}$\fi}
\newcommand{\vBLR}	{\ifmmode v_{\mbox{\tiny BLR}} \else $v_{\mbox{\tiny BLR}}$\fi}
\newcommand{\sigBLR}	{\ifmmode \sigma_{\mbox{\tiny BLR}} \else $\sigma_{\mbox{\tiny BLR}}$\fi}
\newcommand{\vNLR}	{\ifmmode v_{\mbox{\tiny NLR}} \else $v_{\mbox{\tiny NLR}}$\fi}
\newcommand{\tauBLR}	{\ifmmode \tau_{\mbox{\tiny BLR}} \else $\tau_{\mbox{\tiny BLR}}$\fi}

\newcommand{\Hubble}	{\ifmmode {\rm km\,s}^{-1}\,{\rm Mpc}^{-1} \else km\,s$^{-1}$\,Mpc$^{-1}$\fi}
\newcommand{\NDunit}	{\ifmmode {\rm Mpc}^{-3} \else Mpc$^{-3}$\fi}
\newcommand{\LFunit}	{\ifmmode {\rm Mpc}^{-3}\,{\rm mag}^{-1} \else Mpc$^{-3}$\,mag$^{-1}$\fi}
\newcommand{\MFunit}	{\ifmmode {\rm Mpc}^{-3}\,{\rm dex}^{-1} \else Mpc$^{-3}$\,dex$^{-1}$\fi}

\newcommand{\Zsun}{\ifmmode Z_{\odot} \else $Z_{\odot}$\fi}
\newcommand{\mpyr}{\ifmmode \Msun\,{\rm yr}^{-1} \else $\Msun\,{\rm yr}^{-1}$\fi}

\newcommand{\qnote}{\ifmmode q_{0} \else $q_{0}$\fi}
\newcommand{\Hnote}{\ifmmode H_{0} \else $H_{0}$\fi}
\newcommand{\hnote}{\ifmmode h_{0} \else $h_{0}$\fi}
\newcommand{\anote}{\ifmmode a_{0} \else $a_{0}$\fi}

\newcommand{\Catrip}{\ifmmode \left[{\rm Ca}\,\textsc{ii}\right\,\lambda8498, 8542, 8662 \else Ca\,\textsc{ii} $\,\lambda8498, 8542, 8662$\fi}

\newcommand{  \Halpha   }{\ifmmode {\rm H}\alpha \else H$\alpha$\fi}
\newcommand{  \ha   	}{\ifmmode {\rm H}\alpha \else H$\alpha$\fi}
\newcommand{  \Hbeta    }{\ifmmode {\rm H}\beta \else H$\beta$\fi}
\newcommand{  \hb    	}{\ifmmode {\rm H}\beta \else H$\beta$\fi}
\newcommand{  \Hgamma   }{\ifmmode {\rm H}\gamma \else H$\gamma$\fi}
\newcommand{  \Hdelta   }{\ifmmode {\rm H}\delta \else H$\delta$\fi}
\newcommand{  \Lya      }{\ifmmode {\rm Ly}\alpha \else Ly$\alpha$\fi}
\newcommand{  \Lyb      }{\ifmmode {\rm Ly}\beta \else Ly$\beta$\fi}
\newcommand{  \Pa       }{\ifmmode {\rm P}\alpha \else P$\alpha$\fi}
\newcommand{  \Pb       }{\ifmmode {\rm P}\beta \else P$\beta$\fi}
\newcommand{  \Bra      }{\ifmmode {\rm Br}\alpha \else Br$\alpha$\fi}
\newcommand{  \Brg      }{\ifmmode {\rm Br}\gamma \else Br$\gamma$\fi}
\newcommand{  \hii      }{\ifmmode {\rm H}\,\textsc{ii} \else H\,\textsc{ii}\fi}
\newcommand{  \hei      }{\ifmmode {\rm He}\,\textsc{i} \else He\,\textsc{i}\fi}
\newcommand{  \heii     }{\ifmmode {\rm He}\,\textsc{ii} \else He\,\textsc{ii}\fi}
\newcommand{  \HeIIuv   }{\ifmmode {\rm He}\,\textsc{ii}\,\lambda1640 \else He\,\textsc{ii}\,$\lambda1640$\fi}
\newcommand{  \HeIIop   }{\ifmmode {\rm He}\,\textsc{ii}\,\lambda4686 \else He\,\textsc{ii}\,$\lambda4686$\fi}
\newcommand{  \cii      }{\ifmmode {\rm C}\,\textsc{ii}  \else C\,\textsc{ii}\fi}
\newcommand{  \ciii     }{\ifmmode {\rm C}\,\textsc{iii}\right] \else C\,\textsc{iii}]\fi}
\newcommand{  \CIII     }{\ifmmode {\rm C}\,\textsc{iii}\right]\,\lambda1909 \else C\,\textsc{iii}]\,$\lambda1909$\fi}
\newcommand{  \civ      }{\ifmmode {\rm C}\,\textsc{iv}  \else C\,\textsc{iv}\fi}
\newcommand{  \CIV      }{\ifmmode {\rm C}\,\textsc{iv}\,\lambda1549 \else C\,\textsc{iv}\,$\lambda1549$\fi}
\newcommand{  \nii      }{\ifmmode [{\rm N}\,\textsc{ii}]  \else [N\,{\sc ii}]\fi}
\newcommand{  \niii     }{\ifmmode {\rm N}\,\textsc{iii} \else N\,\textsc{iii}\fi}
\newcommand{  \niv      }{\ifmmode {\rm N}\,\textsc{iv}  \else N\,\textsc{iv}\fi}
\newcommand{  \NIVuv    }{\ifmmode {\rm N}\,\textsc{iv}\,\lambda1486 \else N\,\textsc{iv}\,$\lambda1486$\fi}
\newcommand{  \nv       }{\ifmmode {\rm N}\,\textsc{v}   \else N\,\textsc{v}\fi}
\newcommand{\oi}{\ifmmode \left[{\rm O}\,\textsc{i}\right] \else [O\,{\sc i}]\fi}
\newcommand{\OI}{\ifmmode \left[{\rm O}\,\textsc{i}\right]\,\lambda6300 \else [O\,{\sc i}]$\,\lambda6300$\fi}

\newcommand{\oii}{\ifmmode \left[{\rm O}\,\textsc{ii}\right] \else [O\,{\sc ii}]\fi}
\newcommand{\OII}{\ifmmode \left[{\rm O}\,\textsc{ii}\right]\,\lambda\lambda3727,3729 \else [O\,{\sc ii}]\,$\lambda\lambda3727,3729$\fi}
\newcommand{\oiii}{\ifmmode \left[{\rm O}\,\textsc{iii}\right] \else [O\,{\sc iii}]\fi}
\newcommand{\OIII}{\ifmmode \left[{\rm O}\,\textsc{iii}\right]\,\lambda5007 \else [O\,{\sc iii}]\,$\lambda5007$\fi}

\newcommand{\NII}{\ifmmode \left[{\rm N}\,\textsc{ii}\right]\,\lambda6583 \else [N\,{\sc ii}]$\,\lambda6583$\fi}

\newcommand{\NeIII}{\ifmmode \left[{\rm Ne}\,\textsc{iii}\right]\,\lambda3968 \else [Ne\,{\sc iii}]$\,\lambda3968$\fi}

\newcommand{\NeV}{\ifmmode \left[{\rm Ne}\,\textsc{v}\right]\,\lambda3426 \else [Ne\,{\sc v}]$\,\lambda3426$\fi}

\newcommand{\HeII}{\ifmmode {\rm He}\,\textsc{ii}\,\lambda4686 \else He\,{\sc ii}$\,\lambda4686$\fi}

\newcommand{\sii}{\ifmmode \left[{\rm S}\,\textsc{ii}\right] \else [S\,{\sc ii}]\fi}

\newcommand{\SII}{\ifmmode \left[{\rm S}\,\textsc{ii}\right]\,\lambda\lambda6717,6731 \else [S\,{\sc ii}]$\,\lambda\lambda6717,6731$\fi}

\newcommand{  \OIIIuv   }{\ifmmode {\rm O}\,\textsc{iii}\,\lambda1663 \else O\,\textsc{iii}\,$\lambda1663$\fi}
\newcommand{  \oiv      }{\ifmmode {\rm O}\,\textsc{iv}  \else O\,\textsc{iv}\fi}
\newcommand{  \OIVuv    }{\ifmmode {\rm O}\,\textsc{iv}\,\lambda1402  \else O\,\textsc{iv}\,$\lambda1402$\fi}
\newcommand{  \OIVIR    }{\ifmmode {\rm O}\,\textsc{iv}\,25.9\,\mu {\rm m} \else O\,\textsc{iv}\,$25.9\,\mu$m\fi}
\newcommand{  \ovi      }{\ifmmode {\rm O}\,\textsc{vi}   \else O\,\textsc{vi}\fi}
\newcommand{  \Ovi      }{\ifmmode {\rm O}\,\textsc{vi}\,\lambda1035 \else O\,\textsc{vi}\,$\lambda1035$\fi}
\newcommand{  \nei      }{\ifmmode {\rm Ne}\,\textsc{i}   \else Ne\,\textsc{i}\fi}
\newcommand{  \neii     }{\ifmmode {\rm Ne}\,\textsc{ii}  \else Ne\,\textsc{ii}\fi}
\newcommand{  \NeiiIR   }{\ifmmode {\rm Ne}\,\textsc{ii}\,12.8\,\mu {\rm m} \else Ne\,\textsc{ii}\,$12.8\,\mu$m\fi}
\newcommand{  \neiii    }{\ifmmode {\rm Ne}\,\textsc{iii} \else Ne\,\textsc{iii}\fi}
\newcommand{  \neiv     }{\ifmmode {\rm Ne}\,\textsc{iv}  \else Ne\,\textsc{iv}\fi}
\newcommand{  \nev      }{\ifmmode {\rm Ne}\,\textsc{v}   \else Ne\,\textsc{v}\fi}
\newcommand{  \NevIR    }{\ifmmode {\rm Ne}\,\textsc{v}\,24.3\,\mu {\rm m} \else Ne\,\textsc{v}\,$24.3\,\mu$m\fi}
\newcommand{  \nevi     }{\ifmmode {\rm Ne}\,\textsc{vi}  \else Ne\,\textsc{vi}\fi}
\newcommand{  \mgi      }{\ifmmode {\rm Mg}\,\textsc{i}   \else Mg\,\textsc{i}\fi}
\newcommand{  \mgii     }{\ifmmode {\rm Mg}\,\textsc{ii}  \else Mg\,\textsc{ii}\fi}
\newcommand{  \MgII     }{\ifmmode {\rm Mg}\,\textsc{ii}\,\lambda2798 \else Mg\,\textsc{ii}\,$\lambda2798$\fi}
\newcommand{  \siii     }{\ifmmode {\rm S}\,\textsc{iii} \else S\,\textsc{iii}\fi}
\newcommand{  \siv      }{\ifmmode {\rm S}\,\textsc{iv}  \else S\,\textsc{iv}\fi}
\newcommand{  \sili     }{\ifmmode {\rm Si}\,\textsc{i}   \else Si\,\textsc{i}\fi}
\newcommand{  \silii    }{\ifmmode {\rm Si}\,\textsc{ii}  \else Si\,\textsc{ii}\fi}
\newcommand{  \Siliv    }{\ifmmode {\rm Si}\,\textsc{iv}  \else Si\,\textsc{iv}\fi}
\newcommand{  \SilIVuv  }{\ifmmode {\rm Si}\,\textsc{iv}\,\lambda1400  \else Si\,\textsc{iv}\,$\lambda1400$\fi}

\newcommand{  \caii     }{\ifmmode {\rm Ca}\,\textsc{ii}   \else Ca\,\textsc{ii}\fi}
 \newcommand{\Mgb}{\ifmmode \left{\rm Mg}\,\textsc{i}\right\,\lambda5175 \else Mg\,{\sc i}\,$\lambda5175$\fi}
\newcommand{\Cahk}{\ifmmode \left[{\rm Ca H+K}\,\textsc{ii}\right\,\lambda3935,3968 \else Ca H+K$\,\lambda3935,3968$\fi}
\newcommand{  \feii     }{\ifmmode {\rm Fe}\,\textsc{ii}  \else Fe\,\textsc{ii}\fi}
\newcommand{  \feiii    }{\ifmmode {\rm Fe}\,\textsc{iii} \else Fe\,\textsc{iii}\fi}

\newcommand{ \Lhb   }{\ifmmode L\left(\hb\right) \else $L\left(\hb\right)$\fi}
\newcommand{ \fwhb  }{\ifmmode {\rm FWHM}\left(\hb\right) \else FWHM(\hb)\fi}
\newcommand{ \Lha   }{\ifmmode L\left(\ha\right) \else $L\left(\ha\right)$\fi}
\newcommand{ \fwha  }{\ifmmode {\rm FWHM}\left(\ha\right) \else FWHM(\ha)\fi}
\newcommand{ \Lmg   }{\ifmmode L\left(\mgii\right) \else $L\left(\mgii\right)$\fi}
\newcommand{ \fwmg  }{\ifmmode {\rm FWHM}\left(\mgii\right) \else FWHM(\mgii)\fi}
\newcommand{ \Lciv  }{\ifmmode L\left(\civ\right) \else $L\left(\civ\right)$\fi}
\newcommand{ \fwciv }{\ifmmode {\rm FWHM}\left(\civ\right) \else FWHM(\civ)\fi}
\newcommand{ \fwhm  }{\ifmmode {\rm FWHM} \else FWHM\fi} 
\newcommand{ \voff  }{\ifmmode v_{\rm off} \else $v_{\rm off}$\fi} 

\newcommand{ \mumg  }{\ifmmode \mu\left(\mgii\right) \else $\mu\left(\mgii\right)$\fi}
\newcommand{ \fmg   }{\ifmmode f\left(\mgii\right) \else $f\left(\mgii\right)$\fi}
\newcommand{ \muciv }{\ifmmode \mu\left(\civ\right) \else $\mu\left(\civ\right)$\fi}
\newcommand{ \fciv  }{\ifmmode f\left(\civ\right) \else $f\left(\civ\right)$\fi}
\newcommand{  \auvo     }{\ifmmode \alpha_{\nu,{\rm UVO}} \else $\alpha_{\nu,{\rm UVO}}$\fi}
\newcommand{  \Ledd     }{\ifmmode L_{\rm Edd} \else $L_{\rm Edd}$\fi}
\newcommand{  \lamLlam  }{\ifmmode \lambda L_{\lambda} \else $\lambda L_{\lambda}$\fi}
\newcommand{  \lLl      }{\ifmmode \lambda L_{\lambda} \else $\lambda L_{\lambda}$\fi}
\newcommand{  \nuLnu    }{\ifmmode \nu L_{\nu} \else $\nu L_{\nu}$\fi}
\newcommand{  \nLn      }{\ifmmode \nu L_{\nu} \else $\nu L_{\nu}$\fi}
\newcommand{  \Luv      }{\ifmmode L_{1450} \else $L_{1450}$\fi}
\newcommand{  \Lop      }{\ifmmode L_{5100} \else $L_{5100}$\fi}
\newcommand{  \lLop     }{\ifmmode \log\left(\Lop/\ergs\right) \else $\log\left(\Lop/\ergs\right)$\fi}
\newcommand{  \Lthree   }{\ifmmode L_{3000} \else $L_{3000}$\fi}
\newcommand{  \lLthree  }{\ifmmode \log\left(\Lthree/\ergs\right) \else $\log\left(\Lthree/\ergs\right)$\fi}

\newcommand{\Fthree}{\ifmmode F_{3000} \else $F_{3000}$\fi}
\newcommand{\fuv}{\ifmmode f_{\lambda}\left(1450{\rm \AA}\right) \else $f_{\lambda}\left(1450 {\rm \AA}\right)$\fi}
\newcommand{\fthree}{\ifmmode f_{\lambda}\left(3000{\rm \AA}\right) \else $f_{\lambda}\left(3000{\rm \AA}\right)$\fi}
\newcommand{\fH}{\ifmmode f_{\lambda}\left(1.65\micron\right) \else
$f_{\lambda}\left(1.65\micron\right)$\fi}

\newcommand{\fbol}{\ifmmode f_{\rm bol} \else $f_{\rm bol}$\fi}
\newcommand{\fbolwv}{\ifmmode f_{\rm bol}\left(\lambda\right) \else $f_{\rm bol}\left(\lambda\right)$\fi}
\newcommand{\fbolopt}{\ifmmode f_{\rm bol}\left(5100{\rm \AA}\right) \else $f_{\rm bol}\left(5100{\rm \AA}\right)$\fi}
\newcommand{\fbolthree}{\ifmmode f_{\rm bol}\left(3000{\rm \AA}\right) \else $f_{\rm bol}\left(3000{\rm \AA}\right)$\fi}
\newcommand{\fboluv}{\ifmmode f_{\rm bol}\left(1450{\rm \AA}\right) \else $f_{\rm bol}\left(1450{\rm \AA}\right)$\fi}

\newcommand{  \mbh      }{\ifmmode M_{\rm BH} \else $M_{\rm BH}$\fi}
\newcommand{  \lmbh     }{\ifmmode \log\left(\mbh/\Msun\right) \else $\log\left(\mbh/\Msun\right)$\fi} 
\newcommand{  \lledd    }{\ifmmode L/L_{\rm Edd} \else $L/L_{\rm Edd}$\fi}
\newcommand{  \Lbol     }{\ifmmode L_{\rm bol} \else $L_{\rm bol}$\fi}
\newcommand{  \lbol     }{\ifmmode L_{\rm bol} \else $L_{\rm bol}$\fi}
\newcommand{  \lLbol    }{\ifmmode \log\left(\Lbol/\ergs\right) \else $\log\left(\Lbol/\ergs\right)$\fi} 
\newcommand{  \Lagn     }{\ifmmode L_{\rm AGN} \else $L_{\rm AGN}$\fi}
\newcommand{  \lagn     }{\ifmmode L_{\rm AGN} \else $L_{\rm AGN}$\fi}

\newcommand{  \tgrow     }{\ifmmode t_{\rm growth} \else $t_{\rm growth}$\fi}
\newcommand{  \tUni      }{\ifmmode t_{\rm Universe} \else $t_{\rm Universe}$\fi}

\newcommand{  \Mindot	}{\ifmmode \dot{M}_{\rm infall} \else $\dot{M}_{\rm infall}$\fi}
\newcommand{  \Mbhdot	}{\ifmmode \dot{M}_{\rm BH} \else $\dot{M}_{\rm BH}$\fi}
\newcommand{  \Maddot	}{\ifmmode \dot{M}_{\rm AD} \else $\dot{M}_{\rm AD}$\fi}

\newcommand{  \as	}{\ifmmode a_{\rm *} 		\else $a_{\rm *}$\fi}
\newcommand{  \avec	}{\ifmmode \vec{a}_{\rm *} 	\else $\vec{a}_{\rm *}$\fi}
\newcommand{  \re	}{\ifmmode \eta      	\else $\eta$\fi}

\newcommand{  \mseed    }{\ifmmode M_{\rm seed} \else $M_{\rm seed}$\fi}
\newcommand{  \mbul     }{\ifmmode M_{\rm Bulge} \else $M_{\rm Bulge}$\fi} 
\newcommand{  \mstar    }{\ifmmode M_{*} \else $M_{*}$\fi} 
\newcommand{  \mgal     }{\ifmmode M_{*} \else $M_{*}$\fi} 
\newcommand{  \mhost    }{\ifmmode M_{\rm Host} \else $M_{\rm Host}$\fi}
\newcommand{  \mm       }{\ifmmode M_{*}/M_{\rm BH} \else $M_{*}/M_{\rm BH}$\fi}
\newcommand{  \mmsmall  }{\ifmmode M_{\rm BH}/M_{*} \else $M_{\rm BH}/M_{*}$\fi}
\newcommand{  \mmlarge  }{\ifmmode M_{*}/M_{\rm BH} \else $M_{*}/M_{\rm BH}$\fi}
\newcommand{  \mmwp     }{\ifmmode \left(M_{*}/M_{\rm BH}\right) \else $\left(M_{*}/M_{\rm BH}\right)$\fi}
\newcommand{  \ml       }{\ifmmode M_{*}/L_{*} \else $M_{*}/L_{*}$\fi}
\newcommand{  \mlwp     }{\ifmmode \left(M_{*}/L\right) \else $\left(M_{*}/L\right)$\fi}
\newcommand{  \mlk      }{\ifmmode \left(M_{*}/L_{K}\right) \else $\left(M_{*}/L_{K}\right)$\fi}
\newcommand{  \sigs     }{\ifmmode \sigma_{*} \else $\sigma_{*}$\fi}
\newcommand{  \Reff     }{\ifmmode R_{\rm e} \else $R_{\rm e}$\fi}

\def \chandra {{\em Chandra\ }}

\def \chandra {{\em Chandra\ }}

\def\kmps{\hbox{$\km\s^{-1}\,$}}

\newcommand{\bj}{\ifmmode b_{\rm J} \else $b_{\rm J}$\fi}

\newcommand{\iab}{\ifmmode i_{\rm AB} \else $i_{\rm AB}$\fi}

\newcommand{\jab}{\ifmmode J_{\rm AB} \else $J_{\rm AB}$\fi}
\newcommand{\hab}{\ifmmode H_{\rm AB} \else $H_{\rm AB}$\fi}
\newcommand{\kab}{\ifmmode K_{\rm AB} \else $K_{\rm AB}$\fi}

\newcommand{\jveg}{\ifmmode J_{\rm Vega} \else $J_{\rm Vega}$\fi}
\newcommand{\hveg}{\ifmmode H_{\rm Vega} \else $H_{\rm Vega}$\fi}
\newcommand{\kveg}{\ifmmode K_{\rm Vega} \else $K_{\rm Vega}$\fi}

\def\arcsec{\hbox{$^{\prime\prime}$}}

\newcommand{  \Chisq    }{\ifmmode \chi^{2} \else $\chi^{2}$}
\newcommand{  \nelec    }{\ifmmode n_{e} \else $n_{e}$\fi}     % electron density
\newcommand{  \nh       }{\ifmmode N_{H} \else $N_{\rm H}$\fi}     % hydrogen density
\newcommand{  \Ncol     }{\ifmmode N_{col} \else $N_{col}$\fi} % column density
\newcommand{  \NH       }{\ifmmode N_{H} \else $N_{\rm H}$\fi}     % column density

\def\deg{\hbox{$^\circ$}}
\def\sun{\hbox{$\odot$}}

\def\arcsec{\hbox{$^{\prime\prime}$}}

\def\ion#1#2{#1$\;${\small\rm\@Roman{#2}}\relax}

\newcommand{\OIIIa}{\ifmmode \left[{\rm O}\,\textsc{iii}\right]\,\lambda4959 \else [O\,{\sc iii}]\,$\lambda4959$\fi}
\newcommand{\NIIa}{\ifmmode \left[{\rm N}\,\textsc{ii}\right]\,\lambda6548 \else [N\,{\sc ii}]\,$\lambda6548$\fi}
\newcommand{\SIIa}{\ifmmode \left[{\rm S}\,\textsc{ii}\right]\,\lambda6716 \else [S\,{\sc ii}]\,$\lambda6716$\fi}
\newcommand{\SIIb}{\ifmmode \left[{\rm S}\,\textsc{ii}\right]\,\lambda6732 \else [S\,{\sc ii}]\,$\lambda6731$\fi}
\newcommand{\NeVa}{\ifmmode \left[{\rm Ne}\,\textsc{v}\right]\,\lambda3346 \else [Ne\,{\sc v}]\,$\lambda3346$\fi}
\newcommand{\NeVb}{\ifmmode \left[{\rm Ne}\,\textsc{v}\right]\,\lambda3426 \else [Ne\,{\sc v}]\,$\lambda3426$\fi}
\newcommand{\NeIIIa}{\ifmmode \left[{\rm Ne}\,\textsc{iii}\right]\,\lambda3869 \else [Ne\,{\sc iii}]\,$\lambda3869$\fi}
\newcommand{\NeIIIb}{\ifmmode \left[{\rm Ne}\,\textsc{iii}\right]\,\lambda3968 \else [Ne\,{\sc iii}]\,$\lambda3968$\fi}

\newcommand{\mgb}{\ifmmode \left{\rm Mg}\,\textsc{i}\right \else Mg\,{\sc i}\fi}

\def\degree{{\mbox{$^{\circ}$}}}
\def\arcsec{{\mbox{$^{\prime \prime}$}}}

\def\g{{\rm\thinspace g}}

\def\K{{\rm\thinspace K}}

\def\km{{\rm\thinspace km}}

\def\m{{\rm\thinspace m}}

\def\pc{{\rm\thinspace pc}}

\def\s{{\rm\thinspace s}}

\newcommand{\NaIb}{\ifmmode {\rm Na}\,\textsc{i}\,\lambda5896 \else Na\,{\sc i}$\,\lambda5896$\fi}
\newcommand{\NaID}{\ifmmode {\rm Na}\,\textsc{I}\,\lambda5890,5896 \else Na\,{\sc I}$\,\lambda5890,5896$\fi}

\newcommand{\HeIIir}{\ifmmode {\rm He}\,\textsc{ii}\,\lambda8237 \else He\,{\sc ii}$\,\lambda8237$\fi}
\newcommand{\HeIir}{\ifmmode {\rm He}\,\textsc{i}\,\lambda10830 \else He\,{\sc i}$\,\lambda10830$\fi}

\newcommand{\SIII}{\ifmmode \left[{\rm S}\,\textsc{iii}\right]\,\lambda9531 \else [S\,\textsc{ii}]\,$\lambda9531$\fi}

\newcommand {\Lsoftint} {\ifmmode L^{\rm in}_{\mathrm{2-10\ keV}} \else $L^{\rm in}_{\mathrm{2-10\ keV}}$\fi}
\newcommand {\ergpersec} {\ifmmode {\rm erg~s}^{-1} \else erg~s$^{-1}$ \fi}

\def\micron{{\mbox{$\mu{\rm m}$}}}
\def\arcsec{{\mbox{$^{\prime \prime}$}}}

\def\degree{{\mbox{$^{\circ}$}}}

\def\arcsec{{\mbox{$^{\prime \prime}$}}}

\def\g{{\rm\thinspace g}}

\def\K{{\rm\thinspace K}}

\def\km{{\rm\thinspace km}}

\def\m{{\rm\thinspace m}}

\def\pc{{\rm\thinspace pc}}

\def\s{{\rm\thinspace s}}

\def\kmps{\hbox{$\km\s^{-1}\,$}}

\def\apjl{\hbox{ApJL}}
\def\apjs{\hbox{ApJS}}

\def\aap{\hbox{AAP}}

\def\micron{{\mbox{$\mu{\rm m}$}}}
\def\arcsec{{\mbox{$^{\prime \prime}$}}}

\def\degree{{\mbox{$^{\circ}$}}}

\newcommand{\nuvr}{\ifmmode {\rm NUV}-r \else NUV-$r$\fi}
\newcommand{\mh}{\ifmmode M_{\rm H_2} \else $M_{\rm H_2}$\fi}
\newcommand{\mhi}{\ifmmode M_{\rm HI} \else $M_{\rm HI}$\fi}

\newcommand{\must}{\ifmmode \mu_{\ast} \else $\mu_{\ast}$\fi}
\newcommand{\hmol}{\ifmmode H_2 \else $H_2$\fi}
\newcommand{\rmol}{\ifmmode R_{\rm mol} \else $R_{\rm mol}$\fi}
\newcommand{\tdep}{\ifmmode t_{\rm dep}({\rm H_2}) \else $t_{\rm dep}({\rm H_2})$\fi}
\newcommand{\tdepHI}{\ifmmode t_{\rm dep}({\rm HI}) \else $t_{\rm dep}({\rm HI})$\fi}
\newcommand{\fgas}{\ifmmode f_{\rm H_2} \else $f_{\rm H_2}$\fi}
\newcommand{\fhi}{\ifmmode f_{\rm HI} \else $f_{\rm HI}$\fi}
\newcommand{\xco}{\ifmmode \alpha_{\rm CO} \else $\alpha_{\rm CO}$\fi}

\newcommand{\SiX}{\ifmmode \left[{\rm Si}\,\textsc{x}\right]\,\lambda14300 \else [Si\,{\sc x}]\,$\lambda14300$\fi}
\newcommand{\SiVI}{\ifmmode \left[{\rm Si}\,\textsc{vi}\right]\,\lambda19640 \else [Si\,{\sc vi}]\,$\lambda19640$\fi}
\newcommand{\SXI}{\ifmmode \left[{\rm S}\,\textsc{xi}\right]\,\lambda19196 \else [S\,{\sc xi}]\,$\lambda19196$\fi}
\newcommand{\SVIII}{\ifmmode \left[{\rm S}\,\textsc{viii}\right]\,\lambda9915 \else [S\,{\sc viii}]\,$\lambda9915$\fi}
\newcommand{\SIX}{\ifmmode \left[{\rm S}\,\textsc{ix}\right]\,\lambda12520 \else [S\,{\sc ix}]\,$\lambda12520$\fi}
\newcommand{\FeXIII}{\ifmmode \left[{\rm Fe}\,\textsc{xiii}\right]\,\lambda10747 \else [Fe\,{\sc xiii}]\,$\lambda10747$\fi}
\newcommand{\SiXI}{\ifmmode \left[{\rm Si}\,\textsc{xi}\right]\,\lambda19320 \else [Si\,{\sc xi}]\,$\lambda19320$\fi}

\providecommand{\e}[1]{\ensuremath{\times 10^{#1}}}

\def\km{{\rm\thinspace km}}
\def\kmps{\hbox{$\km\s^{-1}\,$}}
\def\s{{\rm\thinspace s}}

\def\arcsec{\hbox{$^{\prime\prime}$}}

\def \chandra {{\em Chandra\ }}

\newcommand{\alphaox}{\ifmmode \alpha_{\rm ox} \else $\alpha_{\rm ox}$\fi}

\begin{document}

\correspondingauthor{Yuanze Ding}

\title{UGC~2369S: a Kpc Scale Triple Merger Candidate Identified in a Nearby Luminous Infrared Galaxy}

\author[0000-0002-5770-2666]{Yuanze Ding}
\affiliation{Cahill Center for Astronomy and Astrophysics, California Institute of Technology, 1216 E California Blvd, Pasadena, CA 91125, USA}
\email[show]{yding@caltech.edu}

\author[0000-0002-7998-9581]{Michael J. Koss}
\affiliation{Eureka Scientific, 2452 Delmer Street, Suite 100, Oakland, CA 94602-3017, USA}
\email{mike.koss@eurekasci.com}  

\author[0000-0002-4226-8959]{Fiona A. Harrison}
\affiliation{Cahill Center for Astronomy and Astrophysics, California Institute of Technology, 1216 E California Blvd, Pasadena, CA 91125, USA}
\email{fiona@srl.caltech.edu}

\author[0000-0002-4834-7260]{Charles C. Steidel}
\affiliation{Cahill Center for Astronomy and Astrophysics, California Institute of Technology, 1216 E California Blvd, Pasadena, CA 91125, USA}
\email{ccs@astro.caltech.edu}

\author[0000-0002-5504-8752]{Connor Auge}
\affiliation{Eureka Scientific, 2452 Delmer Street, Suite 100, Oakland, CA 94602-3017, USA}
\email{connor.auge@gmail.com}  

\author{Jared Gillette}
\affiliation{Eureka Scientific, 2452 Delmer Street, Suite 100, Oakland, CA 94602-3017, USA}
\email{gillettejarred@gmail.com}

\author[0000-0002-5698-8703]{Erica Hammerstein}
\affiliation{Department of Astronomy, University of California, Berkeley, CA 94720-3411, USA}
\email{ekhammer@berkeley.edu}

\author{Ruancun Li}
\affiliation{Max-Planck-Institut f\"ur extraterrestrische Physik, Giessenbachstra{\ss}e, 85748 Garching, Germany}
\affiliation{Kavli Institute for Astronomy and Astrophysics, Peking University, Beijing 100871, People's Republic of China}
\email{liruancun@pku.edu.cn}

\author[0000-0002-1292-1451]{Macon Magno}
\affiliation{George P. and Cynthia Woods Mitchell Institute for Fundamental Physics and Astronomy, Texas A\&M University, College Station, TX,
77845, USA}
\email{macon.a.magno@tamu.edu}

\author[0000-0001-8931-1152]{Ignacio del Moral-Castro}
\affiliation{Instituto de Astrofísica, Facultad de Física, Pontificia Universidad Católica de Chile, Av. Vicu\v{n}a  Mackenna 4860, 782-0436 Macul,Santiago, Chile}
\email{ignaciodelmoralcastro.astro@gmail.com}

\author[0000-0003-2196-3298]{Alessandro Peca}
\affil{Eureka Scientific, 2452 Delmer Street, Suite 100, Oakland, CA 94602-3017, USA}
\affiliation{Department of Physics, Yale University, P.O. Box 208121, New Haven, CT 06520, USA}
\email{peca.alessandro@gmail.com}

\author[0000-0001-5231-2645]{Claudio Ricci}
\affiliation{Department of Astronomy, University of Geneva, ch. d'Ecogia 16, 1290, Versoix, Switzerland}
\affiliation{N\'ucleo de Astronom\'ia de la Facultad de Ingenier\'ia, Universidad Diego Portales, Av. Ej\'ercito Libertador 441, Santiago 22, Chile}
\affiliation{Kavli Institute for Astronomy and Astrophysics, Peking University, Beijing 100871, People's Republic of China}
\email{Claudio.Ricci@unige.ch}

\author[0000-0002-3139-3041]{Yiqing Song}
\affiliation{European Southern Observatory, Alonso de Córdova, 3107, Vitacura, Santiago, 763-0355, Chile}
\affiliation{Joint ALMA Observatory, Alonso de Córdova, 3107, Vitacura, Santiago, 763-0355, Chile}
\email{ysongastro@gmail.com}

\author[0000-0001-7568-6412]{Ezequiel Treister}
\affiliation{Instituto de Alta Investigaci{\'{o}}n, Universidad de Tarapac{\'{a}}, Casilla 7D, Arica, Chile}
\email{etreister@academicos.uta.cl}

\author{Zhuyun Zhuang}
\affiliation{Cahill Center for Astronomy and Astrophysics, California Institute of Technology, 1216 E California Blvd, Pasadena, CA 91125, USA}
\affiliation{Department of Physics and Astronomy, Northwestern University, 2145 Sheridan Road, Evanston, IL 60208, USA}
\email{zzhuang@astro.caltech.edu}

\begin{abstract}
We present high spatial resolution ($\lesssim$1\secpoint0), multi-wavelength observations of UGC 2369S, a nearby luminous infrared galaxy showing three distinct cores separated on kpc scales in near-infrared (NIR) imaging with significant X-ray emission. Utilizing optical/NIR adaptive optics (AO), radio, \chandra X-ray, as well as archival HST imaging, we perform a comprehensive study of AGN activity, obscuration, and host properties. As one of the clearest cases of a triple-nucleus merger at $\simeq$3 kpc separations, UGC 2369S is the first to be studied with high-resolution observations at multiple wavelength. We find that the northern core, having possibly the most massive black hole in the system ($\rm M_{BH}\simeq10^{8}\,M_{\odot}$) is consistent with a heavily obscured AGN. However, its high dust extinction ($\rm A_v>5$), hydrogen column density ($N_\mathrm{H}\gtrsim 10^{25}\,\rm cm^{-2}$) and non-detection of optical coronal lines and coronal X-ray emission leave the identification inconclusive. The other two cores show no evidence for black-hole activity and instead exhibit signatures of tidal disruption. From stellar mass surface density and stellar velocity dispersion maps, we infer that the strongly varying gravitational potential in this three-body system may have cannibalized the stellar bulge of the southwestern core, leaving a metal enriched remnant. An ongoing survey focusing on similar triple systems could help us understand how they evolve and help benchmark numerical simulations, providing insight into gravitational wave predictions and the formation of the most massive black holes.
\end{abstract}

\section{Introduction}
\label{sec:Intro}

Under the currently favored cold dark matter ($\Lambda$CDM) cosmology paradigm \citep{BFPR1984,Garnavich1998ApJ,Hopkins:2008:356}, galaxy mergers are one of the major routes for growing baryonic structures in the universe.  Mergers provide an efficient mechanism for removing angular momentum, for fueling the nuclear region, and for triggering starbursts and Active Galactic Nuclei (AGN). Mergers may be one of the processes that connect supermassive black holes (SMBHs) with their host galaxies \citep{Hopkins:2006:1,Ferrarese:2000:L9a,Magorrian:1998:2285,Gebhardt2000ApJ}.

%On the other hand, triplet nuclei are relevant for future gravitational wave observations. The strong N-body interactions could greatly decrease the evolution time scale in the system, boosting coalescence rate; more importantly, the Kozai-Lidov resonance introduced by the outer perturber could result in high-eccentricity orbits, producing gravitational radiation ‘spikes’ and high order harmonics during close encounters  \citep{Hoffman:2007:957}. These are important for the prediction of the potential signal for the future Laser Interferometer Space Antenna \citep[LISA;][]{Verbiest:2016:1267}, because the occurrence of high-eccentricity coalescences brings progressively higher frequency harmonics, and could extend LISA’s sensitivity into the mass range of $10^{8-9} M_{\odot}$, or lengthen the duration of its sensitivity to $10^{6-7}M_{\odot}$ events \citep{Hoffman:2007:957}. %While observational evidence for merger-AGN connection is still conflicted \cite{Comerford:2015:219,Kocevski:2012:148,Treister:2012:L39,Villforth:2017:812}.
% the existence of UMBH in at least some bright cluster galaxies have been established \citep{McConnell:2011:215,Hlavacek-Larrondo2012}.

Triplet nuclei in merging galaxies produce a range of phenomena with signatures rather different from those with two or fewer SMBHs, primarily owing to the chaotic dynamics \citep[e.g.,][]{Perna2020A&A,Grajales-Medina2023,Kollatschny:2020:A79}. Recent cosmological hydrodynamic simulations have suggested the critical role of triplet mergers in producing the most massive BHs \citep{Ni2022ApJ}; in their simulations, virtually all of the ultramassive black holes (UMBH, $M_\mathrm{BH}\simeq10^{10} M_{\odot}$) and half of the $10^{9} M_{\odot}$ SMBHs have undergone a triple quasar phase by $z=2$ \citep{Hoffman2023}. In simulation, the most massive SMBHs in triples experience a phase of rapid growth, during which their mass increases by more than an order of magnitude following the triple formation, both by efficient gas accretion and BH mergers \citep{Hoffman2023}.  

The link between triplet mergers and UMBH may be important because virial BH mass measurements have suggested the existence of dozens of UMBH at the centers of z$=2-3$ hyper-luminous QSOs \citep{Trainor2012ApJ, Prochaska2014ApJ}. These UMBHs exhibit very high far-infrared (FIR) luminosity, consistent with submillimeter galaxies \citep{Hill2019}. However, due to the limitations of current instrumentation, resolved studies of close ($\lesssim\,$a few kpc) triplet mergers are only possible at low redshifts. Studying low redshift triplet merger systems is thus relevant to understanding whether many-body mergers can feed BHs more efficiently, and how such rapid gas accretion events shape the CGM and influence the galaxy star formation history \citep[SFH,][]{Tumlinson2017ARA&A,2022MNRAS.512.3703B}.

The detection of triple SMBHs is observationally challenging. However, recent studies have reported triple SMBHs found through the detection of AGN features. Utilizing the large AGN sample provided by the BAT and SDSS surveys, \citet{Koss2012ApJ} found a triple AGN candidate with companions at 15 and 87\,kpc (NGC 835). More recently, \citet{Pfeifle_2019} reported one of the first compelling cases of a triple AGN in a late-stage merger (SDSS J0849+1114). The authors used multi-wavelength data to confirm three actively accreting SMBHs with $<\,10\,\rm kpc$ separations. \citet{Foord2021ApJ} analyzed archival \chandra\ observations of seven triple galaxy mergers, finding four new dual-AGN candidates, while confirming the presence of a triple AGN in SDSSJ0859+1114. \citet{Yadav2021} reported a candidate triple AGN in NGC 7733-7734, and noted that such systems may be common in gas-rich galaxy groups and at high redshift.
 
The all-sky survey of the Infrared Astronomical Satellite \citep[IRAS,][]{Neugebauer1984} revealed the existence of a population of galaxies with extraordinarily high IR luminosity -- the so-called Ultraluminous Infrared Galaxies (ULIRGs) and Luminous IR Galaxies \citep[LIRGs,][]{Soifer1987ApJ,Sanders:1988:74,Sanders:1988:L35}. ULIRGs have $8-1000\micron$ IR luminosity $L(\rm IR)$ above $10^{12} L_{\odot}$, while LIRGs typically have $L(\rm IR)$ between $10^{11} L_{\odot}$ and $10^{12} L_{\odot}$. These galaxies are of interest because the classical galaxy evolution picture requires (U)LIRGs to appear during the final coalescence of the galaxies, when massive gas inflows feed the BH as well as trigger intense starbursts \citep{Sanders:1988:74,DiMatteo:2005:604,Hopkins:2008:356}. Indeed, $91\%$ of LIRGs and 93\% of ULIRGs are associated with mergers \citep{Haan:2011:100}, making them the ideal test ground for galactic feedback models. Searching for triples in (U)LIRGs tends to select these objects during an advanced merger stage, allowing a systematic study of the accompanying baryonic processes.
%a systematic study of the chaotic behaviour in complex systems, where gas can be expelled to the intergalactic medium (IGM) and other nearby galaxies \citep{Gao:2000}. 

\ugc is the southern irregular galaxy in the UGC 2369 interacting galaxy pair (Figure~\ref{fig:overview}). Unlike its faint edge-on companion to the north, \ugc is formally classified as a LIRG \cite[$\log L(\mathrm{IR})/L_{\odot}=11.57 $,][]{Vega2008A&A}. \ugc has been argued to host at least one radio AGN, based on its high radio brightness temperature \citep[T$_{b}\gtrsim 10^{6}\,$K,][]{Smith1998ApJ}, and unusually steep continuum slope \citep{Vardoulaki2015A}. However, the identification of an AGN is still ambiguous because the IR emission can be fully explained by a starburst \citep{Vega2008A&A}, though a weak 24.3 \micron\ [Ne \textsc{V}] was found in its MIR spectrum \citep{Roper2010AAS}. Early HST/ACS observations unveiled the existence of three individual cores with broadband imaging, suggesting hierarchical close mergers in the system \citep{Roper2010AAS}. The separation between each stellar core is $\simeq1\rm \, kpc$, representing one of the most compact triple candidates ever studied. The primary focus of this study is to search for multi-wavelength AGN signatures in this chaotic three-body system and to compare its properties with those of the general LIRG population.

This paper presents a multi-wavelength (X-ray, optical, NIR, radio) high spatial resolution study of \ugc (Table~\ref{tab:obs_data}). We refer to the southeastern, southwestern, and northern cores in the system as object~1, 2, and 3 (Figure~\ref{fig:optical image}), respectively. We describe observations and data reductions in Section~\ref{sec:obs_and_data_reduction}; Section~\ref{sec:analysis} discusses data analysis and fitting methods; Section~\ref{sec:discuss} discusses the results and their physical interpretations. We assume the Planck 2018 Cosmology \citep{Planck2020A&A}. The assumed redshift of the northern core $\rm z=0.0318$ corresponds to a luminosity distance of 145.2\,Mpc. 1\arcsec\ subtends 0.66\,physical kpc at $\rm z=0.0318$.

\begin{deluxetable*}{cccccccc}
\tablenum{1}
\caption{{\bf Summary of observations analyzed in this study}}
\label{tab:obs_data}
\setcounter{table}{1}
\tablehead{
\colhead{Observatory}&
\colhead{Instrument} &
\colhead{Date} &
\colhead{Filter/Mode} &
\colhead{Proposal ID} &
\colhead{Band} &
\colhead{Res (\arcsec)} &
\colhead{Exp (ks)}
}
\decimalcolnumbers
\startdata
\chandra&ACIS-S&2024-11-25&\nodata &29090&$\rm 0.5-8\,keV$&0.5&10.08\\
\chandra&ACIS-S&2023-12-03&\nodata &29089&$\rm 0.5-8\,keV$&0.5&10.08\\
\chandra&ACIS-S&2023-12-03&\nodata &28148&$\rm 0.5-8\,keV$&0.5&18.08\\
\chandra&ACIS-S&2002-12-14&\nodata &04058&$\rm 0.5-8\,keV$&0.5&10.18\\
HST&NICMOS&2008-09-09&F160W&11235&\nodata&0.08&2.49\\
HST&ACS&2006-07-22&F415W&10592&\nodata&0.08&1.26\\
HST&ACS&2006-07-22&F814W&10592&\nodata&0.08&0.72\\
VLA&\nodata &2024-11-01&A-array&SC250046&K&0.1& 0.18\\
VLA&\nodata &2023-09-27&A-array&23A-324&K&0.1& 0.18\\
VLA&\nodata &2014-11-01&C-array&14A-471&Ka&0.6&\nodata\\
VLA&\nodata &2016-10-07&A-array&16A-204&Ka&0.06&\nodata\\
VLT&MUSE&2021-11-10&NFM-AO&108.22C1.001&$4600-9000$\,\AA&0.05&2.4\\
Keck&KCWI&2024-12-04&BL/RL+small slicer&2024B/C355&$3500-9000$\,\AA&$0.3-0.6$&2.4\\
Keck&OSIRIS&2023-10-04&Kbb&2023B/Y281&$1.97-2.38$\,\micron&0.1&3.6\\
%Spitzer&IRAC&2008&Channel 1-4&   &2-8\micron&2&18\\
%Spitzer&MIPS&2008&Channel 1-3& &24-100\micron&5&18\\
\enddata
\end{deluxetable*}

\section{Observations and Data Reduction}
\label{sec:obs_and_data_reduction}
\begin{figure}
    \centering
    \includegraphics[width=0.5\textwidth]{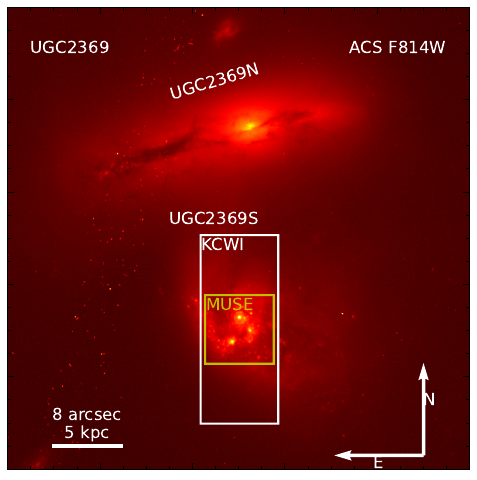}
    \caption{HST/ACS F814W imaging of UGC2369. The target highlighted in this work is the southern luminous IR galaxy \ugc. The FOVs of our KCWI and MUSE IFU observations are highlighted with boxes.}
    \label{fig:overview}
\end{figure}

We observed \ugc with multiple instruments from radio to X-ray as part of our effort to systematically study local triple mergers. Figure~\ref{fig:overview} shows the archival HST image of the entire UGC 2369 system. \footnote{All wavelength calibrated to vacuum. All HST data used in this paper can be found in MAST: 10.17909/280v-ry85} Three cores are detected in both continuum images and emission line maps, and can be easily identified with essentially all instruments that we utilized (see Figure~\ref{fig:optical image}).

\subsection{MUSE Data}

%Optical integral field unit (IFU) observations with adaptive optics were conducted by Multi Unit Spectroscopic Explorer \citep[MUSE,][]{Bacon:2010:773508a} in the Narrow-Field Mode (NFM), covering an FOV of $7\secpoint5$ by $7\secpoint5$ on the sky, sampled by 0\secpoint025 spaxels. Calibration and data reduction were done using the ESO VLT/MUSE pipeline in the ESO \textsc{Reflex} environment \citep{Freudling:2013:A96}. Individual sky-subtracted frames were coadded and manually aligned using the \oiii\ emission before stacking with \textsc{QFitsView}. The PSF FWHM was 0\secpoint09 at \oiii\ for the night, with a nominal spectral resolution $R\simeq1740-3450$ in the $4800-9300$\AA\ bandwidth. The line spread function estimation and vacuum wavelength correction are done through the MUSE Python Data Analysis Framework (MPDAF).

%%%

We conducted optical integral field unit (IFU) observations with adaptive optics with the Multi Unit Spectroscopic Explorer \citep[MUSE,][]{Bacon:2010:773508a} in the Narrow-Field Mode (NFM), covering a FOV of $7\secpoint5$ by $7\secpoint5$ on the sky, sampled by 0\secpoint025 spaxels with a nominal spectral resolution R $\simeq$ 1740 - 3450 in the 4800-9300\AA{} bandwidth. The calibration and data reduction were performed using the ESO MUSE pipeline \citep[][; version 2.8.9]{Weilbacher:2020:641A}, including bias correction, flat fielding, wavelength calibration, sky subtraction, flux calibration, geometric reconstruction of the data cube, and exposure combination using white-light images under the ESO Reflex 
\citep[Recipe flexible execution workbench;][]{Freudling:2013:A96} software, that gives a graphical and automated way to execute the Common Pipeline Library \citep[CPL;][ ESO CPL Development Team 2014]{Banse:2004:314}  reduction recipes with EsoRex \citep[ESO Recipe Execution Tool;][]{ESO_CPL:2015:04003E}, within the Kepler workflow engine \citep{Kepler:2004}. Dedicated offset observations of the sky background, $\sim$35'' away from the target, were obtained and used for the sky subtraction. %The PSF FWHM is 0\secpoint09 at \oiii\ .
%%%

\subsection{OSIRIS Data}
We observed \ugc using near infrared (NIR) IFU instrument OH-Suppressing Infrared Integral Field Spectrograph
\citep[OSIRIS,][]{Larkin:2006:441} with adaptive optics. We used the 0\secpoint05 scale with a field of view (FOV) of $0\secpoint8\times3\secpoint2$ in the classical object-sky-object dithering pattern, with an exposure of 600\,s and a sky offset of 20\arcsec\ in Kbb filter covering $1.96-2.3\,$\micron. Due to the rectangular nature of the FOV, the galaxy was observed in the N–S direction (i.e., PA = 0$\degree$), approximately corresponding to the alignment of the northern and southeastern cores (object~1 and object~3, Figure~\ref{fig:optical image}). The data were reduced using the OSIRIS data reduction pipeline (version 4.2) to perform dark-frame subtraction, crosstalk removal, sky subtraction, rectification, data cube assembly, and the wavelength solution, which is manually refined based on OH lines. The spectra were telluric-corrected and flux-calibrated using the software xtellcor \citep{Cushing:2004:362}. The PSF FWHM is $0\secpoint1$ in our Kbb filter.

\subsection{KCWI Data}

We performed seeing-limited IFU observations with the Keck Cosmic Web Imager \citep[KCWI,][]{KCWI} on UT 4 December 2024. The instrument was aligned north-south (position angle PA = 0$\degree$). The total effective exposure time was 2400\,s for the blue grating and 1800\,s for the red grating. Each 1200s blue exposure is paired with $3\times300$\,s red channel exposures to mitigate the high cosmic ray (CR) contamination on the red CCD. Observing conditions were excellent, with seeing ranging from 0\secpoint3 to 0\secpoint6 as measured by the Canada-France-Hawaii Telescope (CFHT) Differential Image Motion Monitor (DIMM). KCWI was configured with the small slicer and the low resolution BL/RL gratings, centered at wavelengths of 4500\,\AA\ and 7250 \AA, respectively, covering spectral ranges of $3500$--$5600\,$\AA\ and $5600$--$9030\,$\AA\ with the 5600\,\AA\ dichroic. 

In this setup, the slices were 0\secpoint35 wide in the east-west direction, with a pixel scale of 0\secpoint147/pixel in the north-south direction and an 8\secpoint4$\times20\secpoint4$ field of view. The average spectral resolution was $\rm R\simeq3600$ for both gratings. Twilight flat fielding was conducted before taking any science images. Flux calibration was performed using the standard star Feige 110, which was observed immediately before the science target. We combined three standard star exposures after refitting the post-DRP inverse sensitivity curve. The telluric corrections are performed separately with a \textsc{pypeit} \citep{pypeit:joss_pub,pypeit:zenodo} based GUI data reduction tool KSkywizard\footnote{An open source tool available from here:  https://github.com/zhuyunz/KSkyWizard}. Data reduction for the blue channel was primarily completed with the standard KCWI DRP. We created an additional sky mask from the intermediate 2D images to remove astronomical sources from the sky model. The red channel reduction was conducted by a customized pipeline\footnote{The customized pipeline can be downloaded here: https://github.com/yuguangchen1/KCWI\_DRP/tree/KCWIKit}, which reads in CR masks generated by \textsc{kcwikit} \citep{KcwiKit,C21} from three consecutive red channel exposures. The DAR-corrected cubes were fed into KSkywizard, which performed bad wavelength cropping, flux calibration, telluric corrections, and further sky subtraction improvement with the Zurich Atmospheric Purge package \citep[\textsc{zap},][]{2016MNRAS.458.3210S}. Exposures in each channel were cross-correlated, aligned with white-light images, and drizzled onto a common spatial grid with a 0\secpoint146 pixel scale in both directions.

\subsection{Chandra Data}

\chandra observed \ugc on-axis, in a program designed to study candidate triple merger systems (PI: M. Koss). We reduced \chandra data with the standard \chandra\ Interactive Analysis of Observations \citep[\textsc{ciao},][]{CIAO2006SPIE} software version 4.16 and CALDB 4.11.0. We utilize the \textsc{chandra\_repro} function built into \textsc{ciao-4.16} to reprocess the data with the latest calibrations. One point source as well as some diffuse emissions are detected at the expected positions in the broadband X-ray image (Figure~\ref{fig:X-ray_image}). The astrometry accuracy has been checked by comparing the $0.5-7\,$keV image source position with those reported in NED. The point source is consistent with object~3 and no significant offsets (beyond PSF scale) were found. With \chandra\ Ray Tracer (\textsc{ChaRT}), we use the X-ray spectrum extracted from a large 4\arcsec\ radius aperture around the point source to estimate the PSF, while estimating local background from a circular-annulus source-free region with inner radius 4$^{''}$ and outer radius 40$^{''}$.

\subsection{VLA Data}

We observed \ugc with the VLA in A-array, K-band (22 GHz) receiver. The project codes involved with these observations are VLA/23A-324 (PI: Geferson Lucatelli) and VLA/SC250046 (PI: Michael Koss). VLA/23A-324 used 3C48 as a flux calibrator, while VLA/SC250046 used 3C147. The beam sizes for both VLA programs are very similar to one another; VLA/SC250046 has a beam size of $0\secpoint105 \times 0\secpoint083$ with a position angle of $-55.63\degree$, while VLA/23A-324 has a beam size of $0\secpoint108 \times 0\secpoint083$ with a position angle of $62.77\degree$. Additionally, the gain calibrator for VLA/23A-324 is J0238+1636, while VLA/SC250046 used J0256+1334. These observations had an on-source time between $170-180$ seconds, and with their respective gain calibrators observed for an equal amount of time. Both observations achieved an RMS of $\sim15\,\mu$Jy. In both observations, we first observed the flux calibrators mentioned above, then moved on to the X- and K-band attenuation scans. We reduced the raw data and processed it through the standard VLA reduction pipeline \textsc{casa}. Since both observations had the source near the phase center, we expect the bandwidth-smearing effects to be minimal in the vicinity of the source. These observations were cleaned using the \verb|tclean| task in \textsc{casa} to a threshold of 0.05 mJy ($\sim3\sigma$ above the noise) with Briggs weighting (\verb|robust|=0.5), and the primary beam was corrected by setting \verb|pblimit| to $-0.001$. After making the images, we inspected for RFI effects within the image and saw none. The flux densities are estimated using \textsc{PyBDSF} from the images.

Additional Ka-band (33\,GHz) archival data are part of the GOALS-ES \citep{Song2022ApJ,Linden2019ApJ} campaign. GOALS \citep{Armus:2009:559} is dedicated to studying a sample of over 200 local (U)LIRGs, extracted from the IRAS Revised Bright Galaxy Sample, while GOALS-ES is its companion radio survey. GOALS-ES is a multifrequency, multiresolution snapshot VLA survey that maps the brightest radio continuum sources from GOALS that have a declination smaller than 20\deg, allowing follow-up studies with ground-based facilities from both hemispheres. The details of these data are in \citet{Song2022ApJ} and \citet{Linden2019ApJ}.

\section{Results and Analysis}
\label{sec:analysis}

\subsection{Optical Spectral Fitting}
\label{sec:MUSE_obs}
%stellar kinematics and BH mass
Using \textsc{sourcextractor} \citep{Bertin:1996:393}, we detect three cores in the optical pseudo-broadband images covering major nebular lines in the IFU data. We extract 1-D spectra from each core with a fixed aperture of 1\secpoint5 diameter. The nebular line profile is later fitted with a customized version of \textsc{pyqsofit} \citep{pyqsofit,Ren2022ApJ} on the binned cube and the 1-D spectra of each core. To obtain the best-fit stellar continuum, for each 1-D spectrum, we first conduct a Penalized PiXel-Fitting \citep[\textsc{pPXF}][]{Cappellari:2017:798,Cappellari2023} fit with the X-shooter Spectral Library (XSL) and Salpeter initial mass function \citep[IMF][]{Verro2022A&A}. Gaussians are fitted to the emission lines after subtracting the stellar continuum. A single narrow component typically cannot adequately describe the line profile due to the prevalence of outflow. We introduce an extra outflow component that shifts from the rest-frame wavelength within a velocity range ($\lesssim 700\rm\, km~s^{-1}$). Given the possible origins (from outflow or nuclear structure), we tie the line widths and velocity shifts of the various narrow line components, while keeping the line 1-sigma widths larger than 1/4 of the instrumental resolution. Accurate redshift measurement is crucial in reducing model degeneracy, especially given that each core is undergoing proper motion as large as a few hundred \kmps relative to the system's mass center. Based on KCWI data, we characterize the redshift of each core with the stellar velocity shift measured from Ca II absorption, masking any nearby contaminating emission lines (Figure~\ref{fig:stellar_abs_spec}). We found a two-component, double-Gaussian model almost always provides the best fitting result for our data. The various parameters are later marginalized with MCMC sampler \cite[emcee,][]{emcee}.

We map SFH and stellar kinematics across the FOV primarily using KCWI, due to KCWI's coverage of the various stellar absorption lines in $\sim3600-4200$\AA, and the significantly higher S/N around the Ca \textsc{ii} triplets in red channel. Ca \textsc{ii} triplet is especially useful because the other stellar absorption lines in the blue channel overlap with emission lines from the starburst region, making it almost impossible to infer stellar kinematics from blue channel data alone. We use the package \textsc{pPXF} and the X-Shooter stellar population models \citep{Verro2022A&A}, instantiated through the GIST framework \citep{Bittner2019A&A}. The model setup is similar to \citet{Koss:2022:1}, with the standard four moments (MOMENTS = 4) and a multiplicative (MDEGREE = 4) polynomial instead of a specific extinction curve to develop the best composite stellar population to fit the galaxy continuum and absorption lines and makes the fit insensitive to a specific reddening curve. We employ the regularization scheme and set the regularization error to 0.01, which is comparable to the weights of templates \citep{Cappellari:2017:798}. To improve the S/N, Voronoi binning scheme has been adopted on the IFU cube, with a target S/N of 20 for MUSE and 35 for KCWI in restframe $7800-8800\,$\AA\, after masking the spaxels with S/N$\,<1$. The stellar velocity dispersion of each core is derived from an aperture of diameter 1\secpoint5, similar to \citet{Koss:2022:1}. We calculate the inferred BH mass from the $M_{\rm BH}-\sigma$ relation \citep{Kormendy:2013:511}. These measurements are summarized in Table~\ref{tab:core_params_clean}. We did not jointly fit the red and blue channels with \textsc{pPXF} when mapping the SFH, because KCWI CCD undersampled the seeing PSF ($\simeq0.4''$), resulting in prominent discontinuities around 5600\AA\ where the blue and red data connect. Instead, we only use the blue data for SFH mapping, which is a suitable alternative method since there are no significant stellar features, except for the Ca II triplet, in such a young population, and no sub-pixel-level ($\simeq0\secpoint15$) variation of the kinematics is expected in the seeing-limited data. We present the spatially resolved best-fit stellar kinematics in Figure~\ref{fig:stellar_kinematics_map} and the light-weighted age in Figure~\ref{fig:age_KCWI}.

Optical nebular lines provide essential diagnostics for the ionizing source of the extended gas. We conduct emission line fitting on each Voronoi bin using a technique similar to that discussed above to fit the nuclear spectra. We begin with test fits using the preferred two-component model. If this initial trial results in an \oiii\ line (typically the faintest BPT line in our data) with S/N$ < $1, we abandon the fit and switch to a simpler model with only one component for each line, keeping each line's velocity and FWHM tied to each other. We calculate a flux-weighted FWHM and velocity shift map for all detected emission lines. For our investigation, the outflow component with the higher FWHM is of greater interest, so we created a velocity shift map that presents only the velocity shift of the outflow component. We present these results in Figure~\ref{fig:emission_line_map}.

\subsubsection{Extinction, Excitation and SFR in Optical}
\begin{figure*}
    \centering
    \includegraphics[width=0.48\textwidth]{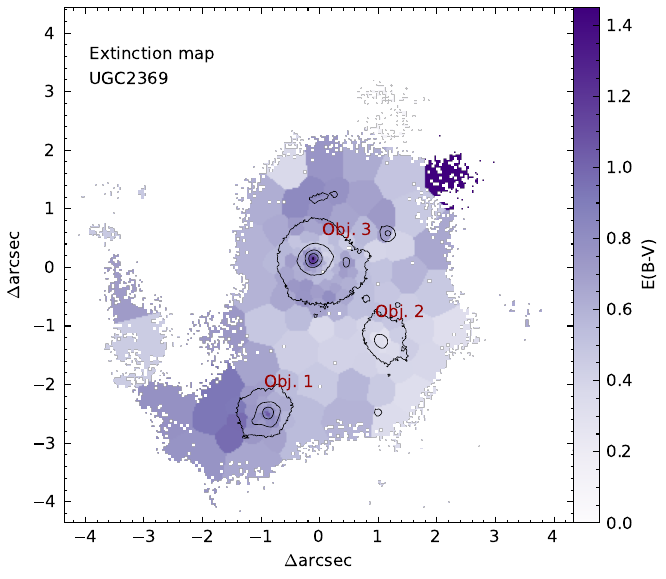}
    \includegraphics[width=0.48\textwidth]{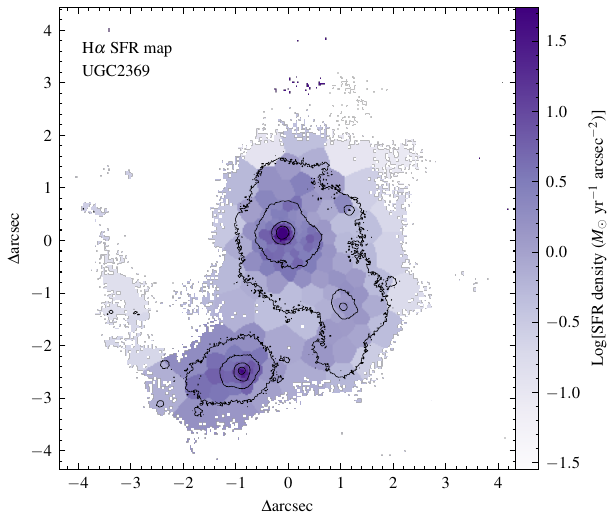}
    \caption{\textbf{Left}: Extinction map calculated from $\rm H\alpha/H\beta$ line ratio. Increased extinction is observed at the centers of objects 1 and 3, indicating the presence of abundant dust at their cores. On the other hand, object 2 has no apparent peak in extinction. \textbf{Right}: Extinction corrected H$\alpha$ SFR density.}
    \label{fig:extinction_sfr_map}
\end{figure*}

\begin{figure*}
    \centering
    
    \includegraphics[height=0.44\textwidth]{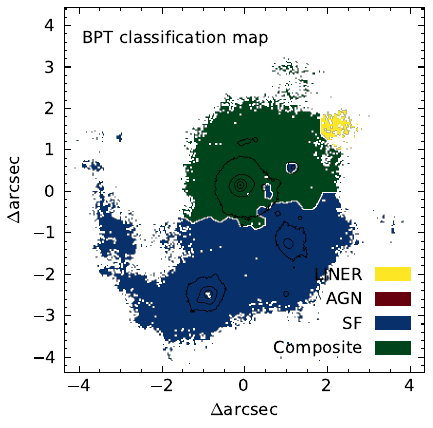}
    \includegraphics[height=0.44\textwidth]{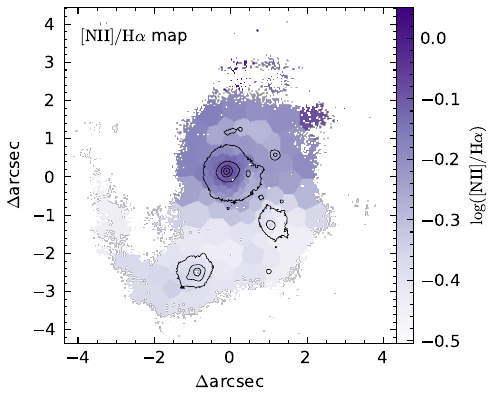}
    \caption{\textbf{Left}: Spaxel-wise BPT classification using the \oiii/\hb\ -- \nii/\ha\ BPT diagram \citep{Baldwin:1981:5,Kewley:2006:961}. \textbf{Right}: \nii/\ha\ line ratio map. The primary driving factor of the composite classification is \nii/\ha\ ratio, which increases rapidly towards object~3.}
    \label{fig:BPT_map}
\end{figure*}

\begin{figure}
    \centering
    \includegraphics[width=0.49\textwidth]{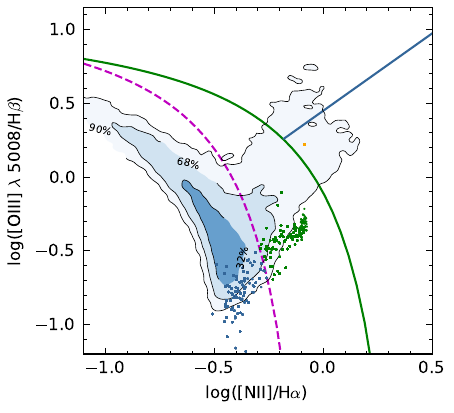}
    \caption{Extinction corrected line ratios compared between SDSS galaxies (contours) and \ugc\ on \nii\ -- \oiii\ BPT plane. All Voronoi bins are plotted and they correspond to regions in Figure~\ref{fig:BPT_map}. The lines are \citet{Kewley:2001:37} maximum starburst (green), \citet{Kauffmann:2003:1055} pure star formation (purple) and \citet{Schawinski:2007:512} LINER -- Seyfert separation (blue).}
    \label{fig:BPT_points}
\end{figure}

The FOV of MUSE in narrow field mode (NFM) is large enough to cover all three cores, enabling accurate mapping of the gas properties. We derive a 2D nebular extinction map using the hydrogen Balmer line ratio. With $\rm H\alpha/H\beta$ ratio, we have \citep[e.g.,][]{Momcheva2013AJ}:

\begin{align}
    E(B-V)%&\frac{E(\rm H\alpha-H\beta)}{k(\mathrm{H}\beta)-k(\mathrm{H}\alpha)}\\
    =&\frac{2.5}{k(\mathrm{H}\beta)-k(\mathrm{H}\alpha)}\log\left[\frac{(\rm H\alpha/H\beta)_{obs}}{\rm (H\alpha/H\beta)_{int}}\right]
    \label{eqn:NIR_extinction}
\end{align}
 \noindent
 %we have $\rm (Br\gamma/Br\delta)_{int}=1.51$ \citep{Osterbrock2006agna.book}. Similarly,
Assuming Case B recombination with T=$10^{4}$, we have $\rm (H\alpha/H\beta)_{int}=2.86$ \citep{Osterbrock2006agna.book}. We adopt the \citet{Cardelli:1989:245} galactic extinction curve to calculate the total-to-selective extinction $k(\lambda)$ \footnote{There is a misunderstanding of the \citet{Calzetti:2000:682} extinction curve in \citet{Momcheva2013AJ}. \citet{Calzetti:2000:682} curve is intended for continuum attenuation, as by design it is meant for averages over a mixture of attenuated and unattenuated sightlines, and is inappropriate for nebular extinction. }. We obtain the color excess E(B-V) map for each nucleus (left panel of Figure~\ref{fig:extinction_sfr_map}) and also the extinction-corrected line map. We utilize the line map to estimate the star formation rate (SFR). We use the standard H$\alpha$ SFR calibration \citep{Murphy2011ApJ,Kennicutt:2012:531} to derive the extinction-corrected, spatially resolved star formation density in Figure~\ref{fig:extinction_sfr_map}. We separately report the E(B-V) values estimated from 1D spectra (extracted from a 1\secpoint5 diameter circular aperture centered on each core) in Table~\ref{tab:core_params_clean}. We report nebular line flux measured from the 1D spectra in Table~\ref{tab:line_fluxes_nist}.
%In principle, the H$\alpha$ SFR is only valid in the sense of spatial average.
%, thanks primarily to the metallicity sensitivity of [N II]/H$\alpha$
With an appropriate extinction curve, we obtain line intensity maps over the MUSE FOV, enabling traditional nebular line diagnostics \citep{Baldwin:1981:5}. Of the three BPT diagrams usually used as AGN indicators \citep{Kewley:2006:961}, \nii/\ha\ is more sensitive to the presence of low-level AGN than \sii/\ha or \oi/\ha, thanks to the metallicity sensitivity of \nii/\ha\ -- this line ratio scales almost linearly with nebular metallicity until it saturates at \nii/\ha$\simeq-0.5$; AGN contributions shift \nii/\ha\ beyond the saturation point, enabling the identification of galaxies with very weak AGN \citep{Kauffmann:2003:1055,Kewley:2006:961}. We find that two cores are consistent with a star-forming H II region, while object 3 is classified as a composite object due to its high \nii/\ha\ ratio ($\gtrsim-0.2$, see Figure~\ref{fig:BPT_map} and~\ref{fig:BPT_points}). %All three of them lie within the star-forming sequence in the \sii/\ha\ and [O \textsc{i}]/\ha\ diagrams.% It may contain a metal-rich stellar population plus an AGN./

\begin{figure}
    \centering
    \includegraphics[width=0.4\textwidth]{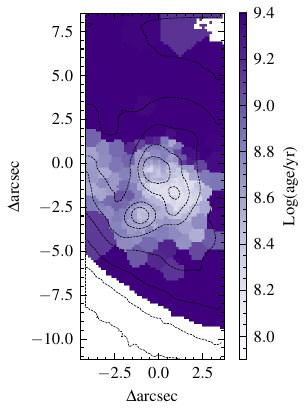}
    \caption{Light-weighted age distribution (with $7800-8800$\AA\ flux level contour) derived from the \textsc{pPXF} fit suggests that the central starburst region is surrounded by a relatively old stellar population with age $\gtrsim$ a few Gyrs. The fitting is based on KCWI blue channel data covering restframe $3650-5400$\AA. Object~3 and 1 have similarly old stellar populations, while object~2 assembled most of its mass in a recent starburst.}
    \label{fig:age_KCWI}
\end{figure}

The extinction-corrected line map permits an estimation of the SFR from the nebular lines. Following \citet{Kennicutt:2012:531}, we estimate the short-time-scale SFR from the extinction-corrected H$\alpha$ flux. We find that the distribution of the SFR surface density is generally consistent with the continuum hot spots visible in broadband imaging (Figure~\ref{fig:extinction_sfr_map}). A caveat is that the H$\alpha$ SFR was calibrated on the $\simeq 0.8\,\rm kpc$ scales. At the very center of object~1 and 3, the calibration may break down as the assumption of spatial averaging is no longer valid. Nevertheless, combined with the stellar age distribution derived from KCWI data, it is likely that the SFR density hot spots contain young stellar populations with a light-weighted age of a few hundred Myr (Figure~\ref{fig:age_KCWI}).

\begin{figure*}
    \centering
    \includegraphics[width=0.43\textwidth]{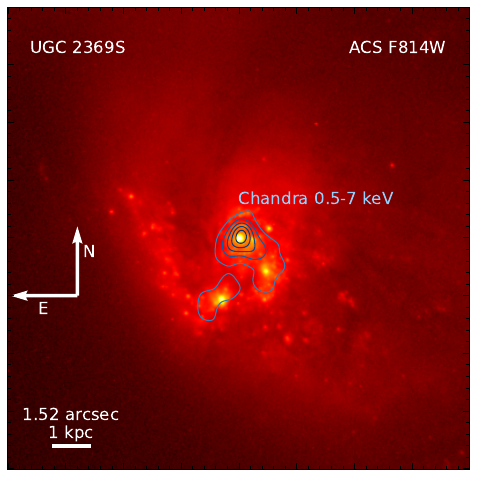}
    \includegraphics[width=0.46\textwidth]{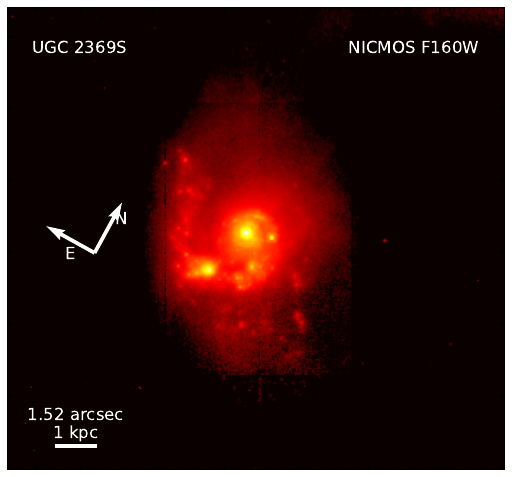}
    \includegraphics[width=0.31\textwidth]{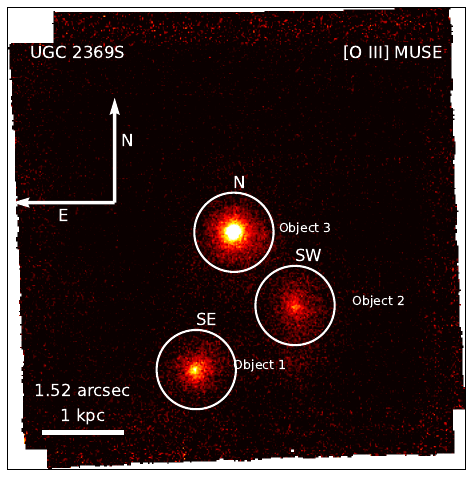}
    \includegraphics[width=0.31\textwidth]{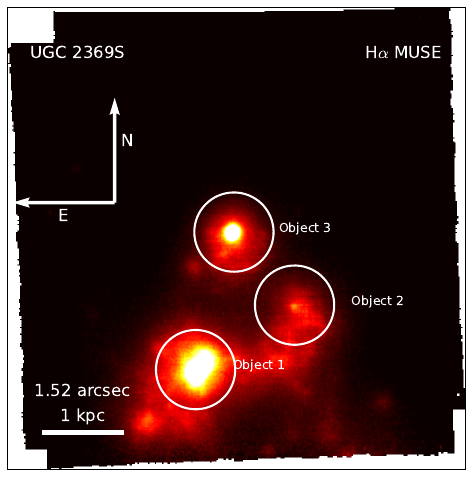}
    \includegraphics[width=0.31\textwidth]{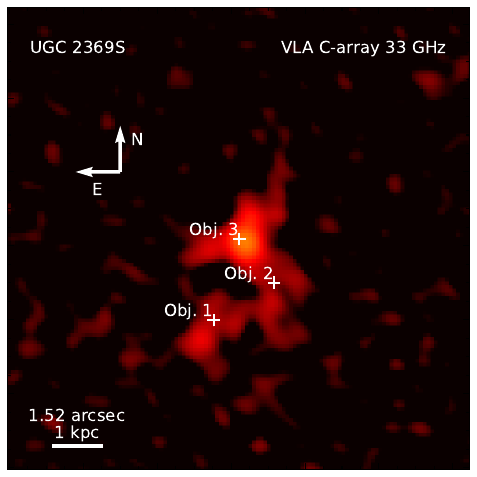}

    \caption{Multi-wavelength observations of \ugc. \textbf{Top}: \textit{HST} ACS and NICMOS images. The contours overplotted on the ACS image indicate the \chandra\ $0.5-7\,$keV X-ray emission. The \chandra\ counts have been convolved with a Gaussian of $\sigma=0.25^{''}$; \textbf{Bottom}: MUSE and VLA 33 GHz observations. Three cores are detected above 5-sigma. Circular apertures with 1.5 arcsec diameter are indicated in MUSE pseudo-narrow band images, from which we extract spectra and estimate the stellar kinematics.}
    \label{fig:optical image}
\end{figure*}

\begin{figure*}
    \centering
    \includegraphics[width=0.95\textwidth]{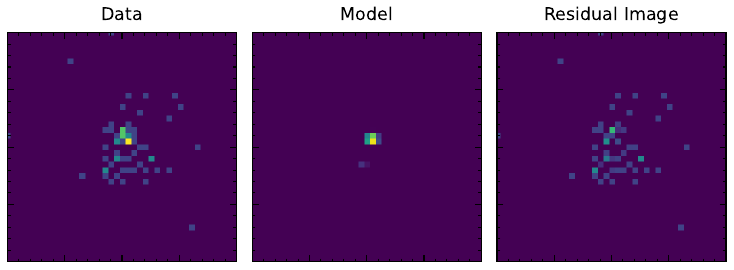}
    \caption{\chandra\ broad band (ObsID: 28148, $0.5-7\,$keV) image of \ugc. The cores are fitted with three PSFs generated by \textsc{ChaRT}. We fix the relative displacement between these three cores to the value found in the MUSE observation. Only the northern core has an unresolved point-like structure. We note that even for object~3, a significant amount of flux is generated in a diffuse star-forming region.}
    \label{fig:X-ray_image}
\end{figure*}

\begin{figure*}
    \centering
    \includegraphics[width=0.49\textwidth]{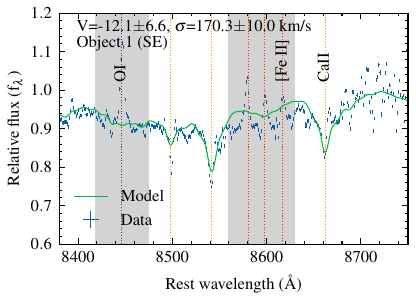}
    \includegraphics[width=0.49\textwidth]{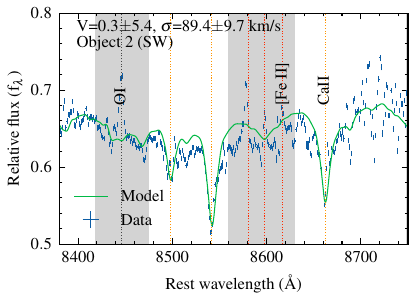}
    \includegraphics[width=0.49\textwidth]{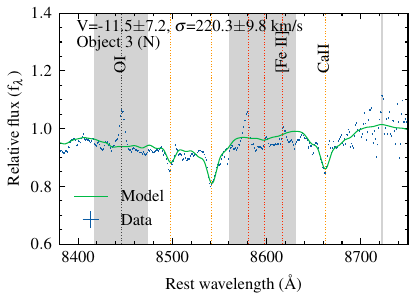}
    \caption{Spectra extracted from KCWI data with circular apertures as presented in Figure~\ref{fig:optical image}. The spectra are fitted by \textsc{pPXF} to extract stellar kinematics. The grey shaded areas are excluded from the fit. The velocity shift (V) is measured relative to the best redshift determined from prior run with stellar absorption, so it is expected to have very small value. The rest frame positions of the Ca II absorption are marked with orange dotted lines. The stellar absorptions are all redshifted relative to the nebular lines and [Fe II] emission, implying that the bulk of the line-emitting gas is outflowing.}
    \label{fig:stellar_abs_spec}
\end{figure*}

%[SII] 6716/6731\,\AA\ 
The \SII\ line map is useful for constraining outflows because its doublet line ratio provides valuable information of electron density ($n_{\rm e}$) \citep[e.g.,][]{Arribas2014}. We map the electron density over the MUSE FOV, utilizing the results in \citet{Osterbrock2006agna.book}, assuming an electron temperature \footnote{The collisionally excited [O\,\textsc{ii}]$\lambda\lambda$7320,7332 and [O\,\textsc{ii}]$\lambda\lambda$3727,3729 are well detected ($>3\sigma$) in the integrated 1D KCWI spectrum around object~3. To verify the self-consistency of our $\rm T_{e}$ assumption, we measure the electron temperature with the line ratio [O\,\textsc{ii}]$\lambda\lambda$7320,7332/3727,3729 using KCWI data. We use the averaged (over 1\secpoint5 diameter) electron density of 921 $\rm cm^{-3}$ around object~3. The resulting electron temperature is T$_{\rm OII}\simeq9228\,$K. } of $10^{4}$\,K. The line decomposition in the \sii\ spectral region is not stable due to the weak line intensity, and we thus use the total flux combining narrow and broad component to estimate the electron density. The value that we obtain is around 300\,$\rm cm^{-3}$ at 0.5$^{''}$ rising slightly above 1000\,$\rm cm^{-3}$ towards the core of object~3 (results shown in Figure~\ref{fig:edensity_map}). These values are slightly higher than the averages from LIRG ($\simeq296-459\,\rm cm^{-3}$), and are closer to typical ULIRG values \citep[$\simeq244-600\,\rm cm^{-3}$,][]{Arribas2014}.

\subsubsection{Neutral Gas Outflow}

The \NaID\ doublet absorption lines are good indicators of neutral gas outflows in AGN and starburst galaxies \citep{Baron2022MNRAS.509.4457B}. We performed an additional multi-component Gaussian fit with a customized PyQSOFit \citep{Ren2022ApJ} routine based on the earlier emission line fitting result. During fitting, we simultaneously modeled the contribution of the \hei\ emission and the Na \textsc{id} complex system, the latter of which traces the neutral gas component of the ISM. The Na \textsc{ID} H/K absorption lines are added in the routine as multiplicative absorption components of the form \citep{Rupke2005ApJ,Rupke2005ApJS,Shih2010ApJ}:
\begin{align}
I_\mathrm{abs} = 1 -C_{f}+ C_f \exp\Bigg[ & -\tau_\mathrm{H} \exp\frac{(\ln\lambda - \ln\lambda_\mathrm{H})^{2}}{2(\sigma_\mathrm{H}/c)^{2}} \notag \\
& -\tau_\mathrm{K} \exp\frac{(\ln\lambda - \ln\lambda_\mathrm{K})^{2}}{2(\sigma_\mathrm{K}/c)^{2}} \Bigg]
\end{align}
where $C_{f}$ is the covering fraction; $\tau$ is the optical depth at line center; $\sigma$ is the 1-sigma velocity dispersion; $\lambda_\mathrm{H}$, $\lambda_\mathrm{K}$ are the vacuum wavelengths of the H, K absorption lines, which vary with a maximum velocity shift of 1000$\,\kmps$. Using at most two such components with different velocity shifts is sufficient for our data. We present further discussion of the results in Section~\ref{sec:outflow}.

In addition to this absorption component, redshifted \NaIb\ emission is a typical feature of neutral outflow \citep{Baron2020MNRAS}. Due to the resonant scattering occurring in the gas shell, the emission from the receding side of the neutral gas is more likely to escape from the absorbing sodium atoms in the approaching shell. However, dust could destroy the scattered photons, suppressing the redshifted NaID features \citep{Prochaska2011ApJ}. This may be happening in \ugc as we see only very weak Gaussian-like 'bumps' around 5896\AA\ across the FOV. We add this component in addition to other nebular lines, though in most cases, it has an S/N$\,\lesssim0.5$, thus being consistent with noise.

\begin{figure*}
    \centering
    \includegraphics[width=0.48\textwidth]{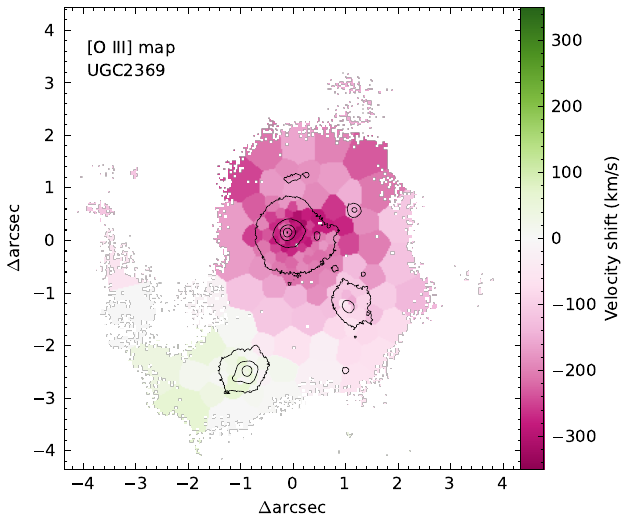}
    \includegraphics[width=0.48\textwidth]{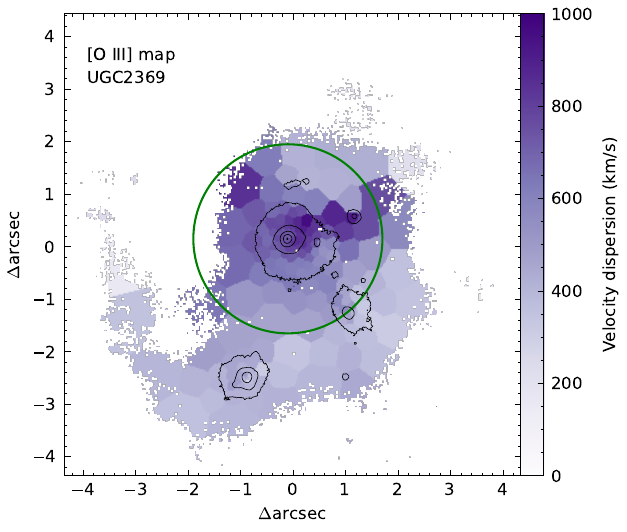}
    \caption{MUSE \oiii line map. The black contour indicates the isophotal level in the rest frame $7800$--$8800\,\mathrm{\AA}$. \textbf{Left}: velocity shift of the component with the highest FWHM, using the systemic redshift of object 3 as rest frame reference. The velocity shift pattern indicates the existence of a large-scale disk-like structure in \ugc with a maximum line of sight velocity on the order of $100$--$200\,\mathrm{km\,s^{-1}}$; \textbf{Right}: velocity dispersion measured from the \oiii 5008 line. If there are multiple components with S/N$>1$ we show the FWHM of the broad component. The green circle indicates the assumed radius of the outflowing gas shell (Section~\ref{sec:outflow}). As seen from the velocity dispersion map, gas kinematics get increasingly turbulent around object 3.}
    \label{fig:emission_line_map}
\end{figure*}

\subsection{NIR Emission Line Diagnostics}
\label{sec:OSIRIS_obs}
Our OSIRIS K-band observation captured object~3 and object~1 (see Figure~\ref{fig:optical image} and Table~\ref{tab:core_params_clean}). This IFU observation detects various H$_{2}$ rovibrational transitions and narrow recombination lines (Figure~\ref{fig:obj3_spectra}). We analyze OSIRIS data by fitting Gaussian profiles on a spaxel-by-spaxel basis, and we model the continuum with a polynomial. 

H$_{2}$ lines provide valuable information on the gas excitation and molecular gas in a manner largely independent of the assumed extinction curve. 

Similar to Section~\ref{sec:MUSE_obs}, we generate an extinction map using the $\rm Br\gamma/Br\delta$ line ratio, while assuming an intrinsic $\rm (Br\gamma/Br\delta)_{int}$ ratio of 1.51 \citep{Osterbrock2006agna.book}. Compared to the map derived from optical Balmer lines, this results in a similar E(B-V), when taking spectrum from a 1\secpoint5 diameter.

\begin{figure}
    \centering
    \includegraphics[width=0.49\textwidth]{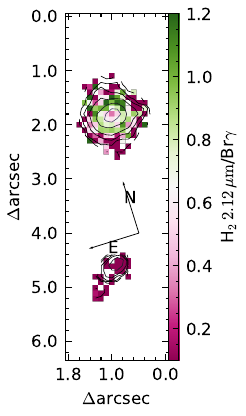}
    \caption{Line ratio map made from \textit{OSIRIS-AO} observation, covering object 1 and 3. Object 3 is surrounded by spaxels that can be formally classified as AGN ($\rm H_2/Br\gamma>0.6$, indicated with green color). The contour shows the flux level of H$_{2}\,1-0\, S(1)$ emission.}
    \label{fig:NIR_diagnostic}
\end{figure}

Similar to the optical BPT diagram, in the NIR, AGNs are characterized by H$_{2}$ 2.121 \micron/Br$\gamma$ flux ratios between 0.6 and 2. Starburst/H II galaxies display line ratios $<0.6$ while LINERs are characterized by values larger than 2 in either ratio \citep{Rodriguez-Ardila2005MNRAS}. Object 3 is surrounded by spaxels with spectrum suggests AGN excitation based on the NIR diagnostic (Figure~\ref{fig:NIR_diagnostic}).

\begin{figure}
    \centering
    \includegraphics[width=0.4\textwidth]{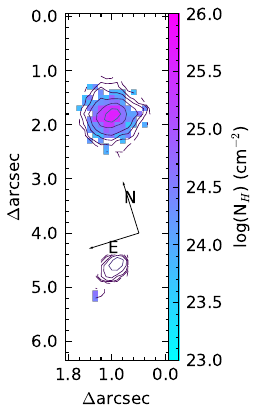}
    \caption{Hydrogen column density ($N_\mathrm{H}$) estimated from \osiris data cube. Contours indicate the isophots estimated from the rest-frame $\rm H_{2}\,1-0\,S(1)$ wavelength. No estimate of $N_\mathrm{H}$ is available towards the core of object 1 (SE) because no hydrogen molecular lines are detected. Only objects 1 and 3 are covered due to the limited FOV. We obtained $N_\mathrm{H}$ on the order of $10^{25.5}$ towards the core of object 3, possibly explaining the X-ray weakness.}
    \label{fig:NH_map}
\end{figure}

The abundant hydrogen molecular lines enable estimation of the hydrogen column density ($N_\mathrm{H}$) independent from X-ray observations, as the latter method could be biased when the source is heavily obscured ($N_\mathrm{H}\gtrsim10^{24}\,\rm cm^{-2}$). We estimate the total line emitting (hot) molecular gas mass using the extinction-corrected H$_{2}\,1-0$S(1) emission line \citep{Scoville1982,Wolniewicz1998}:
\begin{equation}
    M_\mathrm{H_{2}}=5.1\times10^{13}\left(\frac{D_\mathrm{L}}{\rm Mpc}\right)\left(\frac{F_\mathrm{H_{2}1-0S(1)}}{\rm erg~s^{-1}~cm^{-2}}\right)M_{\odot}
\end{equation}
where $D_\mathrm{L}=145\,\rm Mpc$ is the luminosity distance and we assume that the gas is in local thermal equilibrium, assuming a typical\footnote{ We measure the vibrational temperature from the integrated 1D spectrum around object 3, using line ratio 1-0S(1)/2-1S(1), which results in an averaged temperature of 2903 K, while the rotational temperature is estimated from 1-0S(2)/1-0S(0) to be 1393 K. These two T measurements are consistent with a T = 2000 K thermal, collisional excitation component blended with $\simeq$20\% nonthermal UV fluorescent excitation \citep{Black1987ApJ, Mazzalay2013}} excitation temperature of T$_\mathrm{exc}=2000\,$K \citep{Mazzalay2013}. Applying extinction correction (assuming E(B-V)$=0.88$ as obtained from the Paschen decrement), we obtain a total hot H$_{2}$ mass of 1318 $M_{\odot}$ over the OSIRIS FOV. This value is dominated by the hot molecular gas around object~3, representing only the hot surface of molecular gas clouds. To estimate the corresponding cold molecular gas mass we employ the conversion factor: $M_\mathrm{H_{2}}(\rm {\rm cold})/M_\mathrm{H_{2}}({\rm hot}) = (0.3-1.6) \times 10^{6} M_{\odot}$ from \citet{Mazzalay2013}. We obtain a total cold molecular gas mass estimation of $(0.39-2.07)\times10^{9}\, M_{\odot}$, broadly consistent with the value obtained from \textit{Spitzer-IRS} SED modeling \citep{Vega2008A&A}. We rescaled the molecular gas mass to obtain the total gas mass at each spaxel according to the best IR SED model given by \citet{Vega2008A&A}. This enables a rough, order-of-magnitude estimation of $N_\mathrm{H}$. We conservatively show the lower limit of $N_\mathrm{H}$ in Figure~\ref{fig:NH_map}. In summary, our OSIRIS observation indicates an $N_\mathrm{H}$ on the order of $10^{25.5}\rm\,cm^{-2}$ towards the bright, unresolved core of object 3, consistent with a heavily obscured nuclear region.

\begin{figure}
    \centering
    \includegraphics[width=0.52\textwidth]{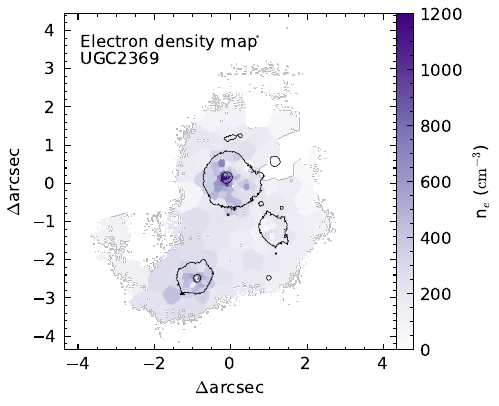}
    \caption{Electron density map derived from \SII\ doublet ratio, assuming electron temperature of $10^{4}\rm \,K$. The values we obtained are slightly higher than the average value from LIRG, and reaches $>10^{3}\,\rm cm^{-3}$ in the core of object~3. This is a significant fraction of \oii's critical electron density ($3\times10^{3}\,\rm cm^{-3}$) and might indicate the existence of a buried narrow line region (more discussion in Section~\ref{sec:gas_excitation}).}
    \label{fig:edensity_map}
\end{figure}

\begin{figure}
    \centering
    \includegraphics[width=0.5\textwidth]{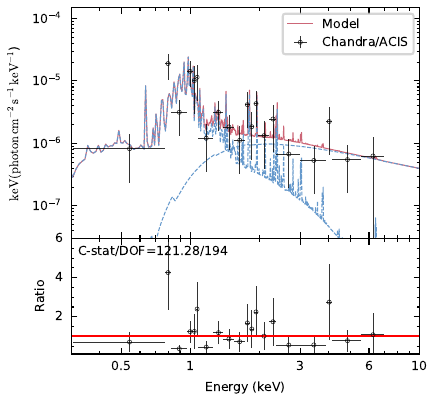}
    \caption{\chandra\ combined spectrum extracted from an 1\secpoint5 diameter aperture centered on object~3. The fitted model has a hard component with a photon index of $\Gamma<2.48$. Data in this figure have been visually rebinned to S/N$>1$.}
    \label{fig:chandra_spec}
\end{figure}

%\subsection{Spitzer Data} 

%Photometric measurements are performed with a point spread function (PSF) fitting technique. We measured PSFs for each individual instrument and filter configuration by fitting field stars (cross-matched with \textit{Gaia} DR3) with a non-parametric PSF model \citep{Li2024,Bertin2011}. The resulting PSF images are then used to derive the total source flux.

\subsection{X-ray Spectral Modeling}
\label{sec:xray}
We fit the X-ray photometric data, using the \chandra\ PSF model to estimate the centroid position and possible extended emission. Figure~\ref{fig:X-ray_image} shows that the X-ray emission from \ugc is predominantly due to diffuse sources embedded in the star-forming region, though there is an unresolved point source visible at the center of object 3.

We extract spectra from a circular aperture with 0.75$^{''}$ radius around object~3, estimating the local background from a circular annulus region with inner radius 4$^{''}$ and outer radius 40$^{''}$ free of any significant contaminating sources. All available spectra are combined with the \textsc{ciao} built-in function \textsc{combine\_spectra}. The combined spectrum is binned to have at least one count per bin, avoiding empty channels (using \textsc{ftgrouppha}). During the fitting, the modified Cash statistic, built into \textsc{Xspec} \citep{Arnaud:1996:17}, was applied. The final spectrum has 199 bins in the $0.3-7\,\rm keV$ band.

We fit the \chandra\ spectrum from object~3 using \textsc{Xspec} version 12.13.0. We adopt a \textsc{apec} component to fit the hot gas in the star-forming region, attenuated by galactic HI gases \citep[$N_\mathrm{H}=7.7\times10^{20}\,\rm cm^{-2}$][]{Kalberla:2005:775}. We add power-law coronal emission and a neutral absorber at the rest frame of $\rm z=0.0318$ to describe the potential hard component from an AGN and the nuclear obscurer\footnote{Our final model, in \textsc{Xspec}'s terminology, is: TBabs*zTBabs(APEC+zPowerlaw); The $\Delta\rm AICc$ is computed by comparing a simpler model without the \textsc{zPowerlaw} component.}. The addition of a power-law component improves the fit by $\Delta \rm C-stat/DOF=13.1/2$, which is $\gtrsim3\sigma$ in a Gaussian approximation. With the corrected Akaike Information Criterion \citep[AICc,][essentially AIC plus an additional penalty term that increases with the number of parameters and decreases with the sample size]{Akaike1974,CAVANAUGH1997201}, we find the model with the power-law component is preferred ($\Delta\rm AICc=8.91$). We note that the Bayesian Information Criterion \citep[BIC,][]{Kass1995} also prefers the more complicated model, though not with high evidence ($\Delta\rm BIC=2.5$). We add an extra neutral absorber at the rest frame to the power-law component (the resulting absorption level is $N_{\rm H}=1.5^{+5.4}_{-1.1}\times10^{22}\rm\,cm^{-2}$).% because we have very limited counts above 2\,keV where power-law component dominate (see Figure~\ref{fig:\chandra\_spec}).

From a purely statistical perspective, we cannot exclude the presence of a hard power-law component in object 3 with $\Gamma<2.48$. Given the very high level of nuclear obscuration unveiled by NIR and radio observations ($N_\mathrm{H}\simeq 10^{25}$, Section~\ref{sec:OSIRIS_obs},~\ref{sec:VLA}), this power-law component may be the emission from large number of high-mass X-ray binaries (HMXBs) in the nuclear starburst region outside the thick dust screen. We note that the level of X-ray flux from \ugc can be entirely explained by the SFR estimated in Section~\ref{sec:SFR_Mstar} \citep[using Table~3 of][]{Lehmer2016ApJ}.

%We caution that the original model library for $M/L$ calibration was not specifically designed for heavily dust obscured system so our $M_{\star}$ values are at best order-of-magnitude estimates.

\subsection{Radio Continuum Analysis}
\label{sec:VLA}
At radio wavelengths one can usually overcome the effects of dust obscuration and study the centers of obscured objects in great detail.

\ugc\ was observed with the VLA at 33 GHz (Ka band) using both A and C arrays \citep[see][for details]{Song2022ApJ,Linden2019ApJ}. All three cores are detected with the C-array (0\secpoint63 resolution), but only object~3 is detected with the A-array (0\secpoint06 resolution). With 33 GHz C-array data, assuming 10\% calibration uncertainty, the measured flux densities for object~3, 2 and 1 within a 1\secpoint5 diameter aperture are $1.46\pm0.15\,$mJy, $0.26\pm0.05\,$mJy, $0.35\pm0.05\,$mJy, corresponding to SFR of $14.7\pm1.53$, $2.63\pm0.49$, $3.55\pm0.54\,M_{\odot}$/yr using the standard calibration in \citet{Murphy2012ApJ}, and assuming a non-thermal spectral index $\alpha=-0.85$ \citep[this was the average non-
thermal spectral index found among the 10 star-forming regions by][, $S_{\nu}\propto \nu^{\alpha}$)]{Murphy2011ApJ}.

Object~3 remains unresolved in the 0\secpoint06 resolution 33\,GHz A-array data, with a flux density of 0.68$\pm0.01\,$mJy at 33\,GHz, assuming an effective radius of 60\,pc \citep{Song2022ApJ}. We also did a 2D Gaussian fit using the Common Astronomy Software Applications \citep[\textsc{casa},][]{McMullin:2007:127}\footnote{\url{https://casa.nrao.edu/}} \textsc{imfit} and the fitted flux is 0.82\,mJy with a deconvolved major and minor axis of 50 and 60\,pc (in Gaussian FWHM). For object~1 and 2, taking the image RMS noise of 11.7\,$\mu$Jy/beam (from \citealt{Song2022ApJ} Table~A1), the 3-$\sigma$ limit would be 0.035\,mJy assuming they are both unresolved.

More recent A-array, K-band (22\,GHz) VLA observations (program 23A-324 and SC250046) confirmed that only the unresolved northern core of \ugc is detected in A-array. Program 23A-324 found that \ugc\ has a flux density of $780 \pm 61$ $\mu$Jy. From program VLA/SC250046, we report a flux density of $630 \pm 32\,\mu$Jy. The 33 and 22 GHz flux densities of 0.82 mJy and 0.78 mJy, respectively, correspond to a spectral index of $\alpha = 0.12$ (where $S_\nu \propto \nu^\alpha$). Such a flat spectrum is consistent with emission from a compact, self-absorbed AGN corona \citep{Raginski2016MNRAS}, but inconsistent with a purely optically thin synchrotron emission from star formation \citep[$\alpha \approx -0.7$,][]{2016era..book.....C, 2019A&A...630A..83Z, 2024MNRAS.528.5346A}. However, a substantial contribution from thermal free--free emission in star-forming regions ($\alpha \approx -0.1$) cannot be ruled out \citep{2016era..book.....C, 2025MNRAS.539..808M}. Assuming that the \citet{1993ApJ...405L..63G} relation for stellar coronae ($\rm L_\text{R}/L_\text{X} \sim10^{-5}$) holds for SMBH \citep{Laor:2008:847}, and the unresolved radio emission is produced by an accretion disk corona \citep{Raginski2016MNRAS}, we predict the X-ray flux to be around $\sim 5 \times 10^{-12}\rm\,erg~s^{-1}~cm^{-2}$. This is roughly two orders of magnitude brighter than the observed value, indicating heavy obscuration. We summarize the 22\,GHz measurements in Table~\ref{tab:radio-predicted-xray}.

%Table~\ref{tab:radio-predicted-xray} displays the fluxes and luminosities of UGC 2369S.
\begin{deluxetable*}{ccccc}
\tablecaption{22\,GHz Radio and Predicted X-ray Properties of UGC 2369S}
\tablehead{
 \colhead{Year} & \colhead{Radio Flux Density} & \colhead{Predicted X-Ray Flux} & \colhead{22\,GHz Radio Luminosity} & \colhead{Predicted X-ray Luminosity} \\  & mJy &  $10^{-12}$ erg s$^{-1}$ cm$^{-2}$ & erg s$^{-1}$ & erg s$^{-1}$}
\colnumbers
\startdata
 2023 & 780 $\pm$ 61 & 6.24 $\pm$ 0.49 & 38.19 $\pm$ 0.03 & 43.19 $\pm$ 0.03 \\
 2024 & 630 $\pm$ 32 & 5.04 $\pm$ 0.26 & 38.10 $\pm$ 0.02 & 43.10 $\pm$ 0.02 \\
\enddata
\label{tab:radio-predicted-xray}
\tablecomments{VLA A-array radio measurements of UGC 2369S over two years, along with predicted X-ray flux and luminosity assuming that the radio emission is from the accretion disk corona. Columns are:  (1) Observation year, (2) Radio flux density, (3) Predicted X-ray flux, (4) Radio luminosity, and (5) Predicted X-ray luminosity.}
\end{deluxetable*}

\subsection{Star-formation Rate and Stellar Mass}
\label{sec:SFR_Mstar}
%we estimate and map the stellar mass over the MUSE FOV with a mass-to-light ratio $(M/L)$ corrected HST \textit{H}-band (WFC3/F160W) magnitude rescaling method \citep{Zibetti2009MNRAS}. 
As a LIRG, \ugc\ has a very large V-band attenuation, approaching 10 mag at the core, making any optical SED modeling highly sensitive to the assumed attenuation curve. Consequently, rather than using SED fits, we derive the star-formation rate (SFR) by rescaling the VLA 33\,GHz flux \citep[see further details in Section~\ref{sec:VLA},][]{Murphy2012ApJ} and the attenuation corrected H$\alpha$ flux \citep{Kennicutt:2012:531}. For similar reasons, we calculate HST $g-i$ color (F435W$-$F814W), from which we obtain $(M/L)_{H-band}$ according to the power-law fitting parameters in \citet{Zibetti2009MNRAS}, with an estimated calibration uncertainty of 0.3\,dex. The results are shown in Figure~\ref{fig:Mstar_map}. The total stellar mass within 1\secpoint5 aperture for each cores is summarized in Table~\ref{tab:core_params_clean}. We did not use the more accurate two-color method because the corresponding $M/L$ look-up table covers only a very limited parameter space. We caution that this method does not work in the very dusty regime due to the lack of dusty galaxies in the original sample, which results in the prominent dipping in our stellar mass map within the central $\simeq0\secpoint4$ (Figure~\ref{fig:value_radial}).

\section{Discussion}
\label{sec:discuss}
Building on the multi-band analysis above—which constrains dust extinction, SFR, the stellar and molecular-gas masses—we now link these measurements to the system’s physical conditions. We first assess the excitation state of the line-emitting gas, then analyze the origin and dynamics of the candidate triple nucleus, and finally evaluate the energetics and kinematics of the outflows.

\begin{deluxetable}{cccccc}
\caption{Summary of the Various Line Diagnostic Results}
\tablehead{
\colhead{Obj. \#}&
\colhead{Position}&
\colhead{Optical} &
\colhead{NIR} &
%\colhead{L$_\mathrm{0.5-2\,keV, 1^{"}}$} &
\colhead{X-ray} &
\colhead{Radio}
}
\decimalcolnumbers
\startdata
1 & SE &SF & SF  & SF &SF \\
2 & SW&SF & \nodata  & SF &SF \\
3 & N &Composite & AGN  & Unclear &AGN \\
\enddata
\tablecomments{
Col. (1): Object number;
Col. (2): Position of the cores on Figure~\ref{fig:optical image} relative to the pointing center (SE: southwest, SW: southeast, N: north);
Col. (3): Classification based on the \nii\ BPT diagram (SF: star-forming);
Col. (4): Classification based on H$_{2}/\rm Br\gamma$ ratio (Figure~\ref{fig:NIR_diagnostic});
Col. (5): Classification based on Chandra X-ray flux and morphology. Only object~3 has an unresolved point source, but the high $N_{\rm H}$ implied by NIR observations means no unambiguous conclusion can be reached with Chandra data;
Col. (6): Both 33\,GHz and 22\,GHz A-array observations indicate an unresolved core at object~3. Earlier observations classified UGC 2369S as radio AGN based on the spectral index and high brightness temperature \citep{Smith1998ApJ,Vardoulaki2015A}.
}
\label{tab:Line_diagnostic}
\end{deluxetable}

\subsection{Gas Excitation}
\label{sec:gas_excitation}
The northern core of \ugc\ appears to be the only plausible site for an accreting SMBH in this system, with multiple lines of evidence pointing to a heavily obscured AGN. However, the thick dust screen together with starburst make an unambiguous AGN identification particularly challenging, because AGN signatures are both attenuated and blended with strong stellar emission. We focus on object~3 in the following discussion to understand the authenticity of the AGN signatures.

We assessed the presence of an AGN using optical and NIR diagnostics \citep{Kewley:2006:961, Baldwin:1981:5, Rodriguez-Ardila2005MNRAS}. For the northern core (object~3), we find AGN-like line ratio (as presented in Sections~\ref{sec:MUSE_obs} and \ref{sec:OSIRIS_obs}). We also examine the WISE color, which is red ($W1-W2=0.703\pm0.029$) and satisfies the 90\% reliability-optimized AGN selection criterion, providing independent support for an AGN \citep{Assef:2018:23}. These results are summarized in Table~\ref{tab:Line_diagnostic}.

\begin{figure}
    \centering
    \includegraphics[width=1.0\linewidth]{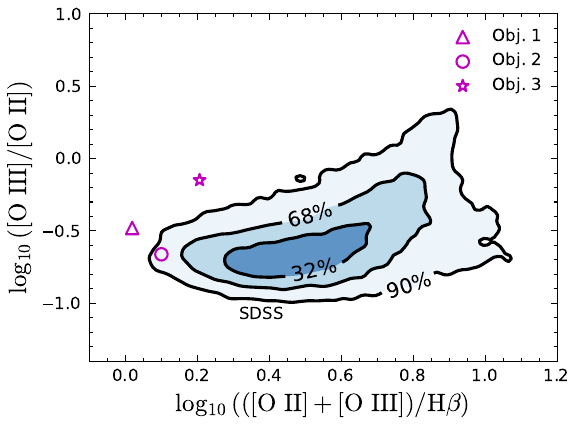}
    \caption{O32 -- R23 compared between SDSS galaxies and the three cores in this study, using extracted KCWI spectra. Object~3 has higher O32 when compared with SDSS galaxies at similar R23 (similar metallicity).}
    \label{fig:R23vO32_abund_map}
\end{figure}

A potential false positive for AGN-like excitation can arise from low metallicity \citep{Steidel2014,Steidel2016,Strom2017}. We explore this further by comparing the ionization-sensitive parameter O32:
\begin{equation}
    \mathrm{O32}=\log\left(\frac{[\mathrm{O\,III}]\,\lambda\lambda 5008,4960}{[\mathrm{O\,II}]\,\lambda\lambda 3729,3727}\right),
\end{equation}
with the metallicity-sensitive parameter R23:
\begin{equation}
\mathrm{R23}=\log\left(\frac{[\mathrm{O\,III}]\,\lambda\lambda 5008,4960+[\mathrm{O\,II}]\,\lambda\lambda 3729,3727}{\mathrm{H}\beta}\right).
\end{equation}
using the Balmer extinction (Table~\ref{tab:core_params_clean}) and spectra extracted with a $1\secpoint5$-diameter aperture. We report the measurements in Figure~\ref{fig:R23vO32_abund_map}. The $\mathrm{R23}$ value for object~3 is low, indicating high metallicity, whereas its $\mathrm{O32}$ is higher than $>90\%$ of SDSS star-forming galaxies (SFGs) at the same R23. We further measure the strong line empirical gas phase metallicity with O3N2 \citep[using the definition and calibration in][]{Pettini2004}, obtaining $12+\log(\rm O/H)=8.78\pm0.3$. Together, these results suggest that object~3 (northern core) of \ugc\ likely has mildly super-solar metallicity and exhibits relatively high excitation (compared with SDSS galaxies at the same R23, Figure~\ref{fig:R23vO32_abund_map}).
%% high 

An anomalous abundance pattern—specifically, a high N/O ratio—may also explain the elevated \nii/\ha\ ratio that drives the composite classification of \ugc without requiring an AGN contribution. Super-solar N/O can be generated in starburst galaxies by strong winds from asymptotic giant branch stars, as previously stated in an X-ray study of M82 \citep{Fukushima2024A&A}. Since nitrogen and oxygen share similar ionization potentials, the total N/O ratio can be directly estimated from their ionic abundance ratio \citep{Draine2011piim}. We determined the electron temperature (T$_{\rm e}$) from the \oii$\lambda\lambda$7320,7332/3727,3729 line ratio using \textsc{pyneb}, applying a flux-weighted average electron density of 921\,cm$^{-3}$ (measured over a 1\secpoint5 diameter around object~3; see Table~\ref{tab:line_fluxes_nist} for observed line fluxes). We find $T_{\rm OII}\simeq9228\,$K. From these parameters, we calculate the $\rm N^{+}/O^{+}$  ionic abundance ratio, yielding $\rm N/O\simeq N^{+}/O^{+}\simeq0.46$. This result is more than three times the solar value \citep[assuming $\rm (N/O)_{\odot}=0.138$,][]{Draine2011piim} and is unusually high even when compared to some of the most extreme metal rich starburst galaxies \citep{Coziol1999A&A,Mouhcine2002A&A}. Rather than an intrinsic chemical anomaly, this extreme apparent ratio could be an artifact of collisional de-excitation in dense gas (typical for AGN narrow line region). The $\rm^2D$ level of $\rm O^{+}$ has a critical density of only $3300\,\rm cm^{-3}$, while the $\rm ^{1}D$ upper level of $\rm N^{+}$ has critical density of $\simeq7.68\times10^{4}$ \citep{Draine2011piim}. Although our volume-averaged density is lower, local gas clumps within the core of object~3 can approach or exceed the $\rm O^{+}$ critical electron density (Figure~\ref{fig:edensity_map}), significantly quenching \OII\ emission. These results therefore favor an underlying AGN that possesses high electron density narrow line region and/or commonly displays nitrogen enhancement \citep{Osterbrock2006agna.book}, as an explanation based purely on star forming HII region would demand highly unusual ISM properties. %, plausibly due to an accreting SMBH located behind a dust screen.

Radio observations can penetrate the dust screen. \ugc\ is classified as a radio AGN based on its steep radio spectral index between 1.4 and 8.4\,GHz \citep{Vardoulaki2015A}, flat and even inverted spectral index between 22 and 33\, GHz, consistent with stratified, self-absorbed BH corona (Section~\ref{sec:VLA}). The X-ray flux can be compared with the radio flux using the radio--X-ray flux ratio calibrated by local quasars. Assuming a spectral index of -0.85 (as in Section~\ref{sec:VLA}), we find $L_{2\text{--}10\,\mathrm{keV}}/L_{100\,\mathrm{GHz}}=1.7$, consistent with $\log(N_\mathrm{H})\gtrsim25$ inferred from the OSIRIS data \citep{Ricci_2023ApJL}, noting that their model provides only an upper limit for $L_{2\text{--}10\,\mathrm{keV}}/L_{100\,\mathrm{GHz}}$ above $\log(N_\mathrm{H})\gtrsim24.5$. Such a high $N_\mathrm{H}$ naturally explains the X-ray weakness and the non-detection of high-ionization (ionization potential $>100$\,eV) coronal lines \citep[CLs;][]{Lamperti:2017:540}.

In summary, the radio, NIR, and optical data are all consistent with an AGN under heavy obscuration in the northern core of \ugc. However, the \chandra\ X-ray data do not require high $N_\mathrm{H}$ and are compatible with emission from HMXBs located outside a thick dust screen. We note that a recent \textit{JWST}/NIRSpec IFS study found that Arp~220 exhibits similar characteristics \citep{Perna2024arXiv}. High levels of obscuration are expected in mergers and, in particular, in (U)LIRGs.
%, implying that this kind of candidate with a heavily obscured AGN may be a significant sub-group in the LIRG family.
%In the scenario that an actively accreting BH being blanketed by thick molecular gases and dust grains, its X-ray and optical emissions can be significantly suppressed, while the star forming regions embedded in the molecular cloud dominate the flux in those wavelength. 
\subsection{Origin of the Cores}
Triplet mergers may be an important phase of baryonic structure assembly and a potential channel for producing strong gravitational wave sources. However, in the case of \ugc, the nature of objects~1 and 2 remains ambiguous: they could be tidally stripped, SMBH-hosting satellite nuclei, or instead newly formed (massive) star clusters. Here we reassess the observational evidence for tidal stripping and evaluate the most likely origin of objects~1 and~2.

\begin{figure}
    \centering
    \includegraphics[width=1.0\linewidth]{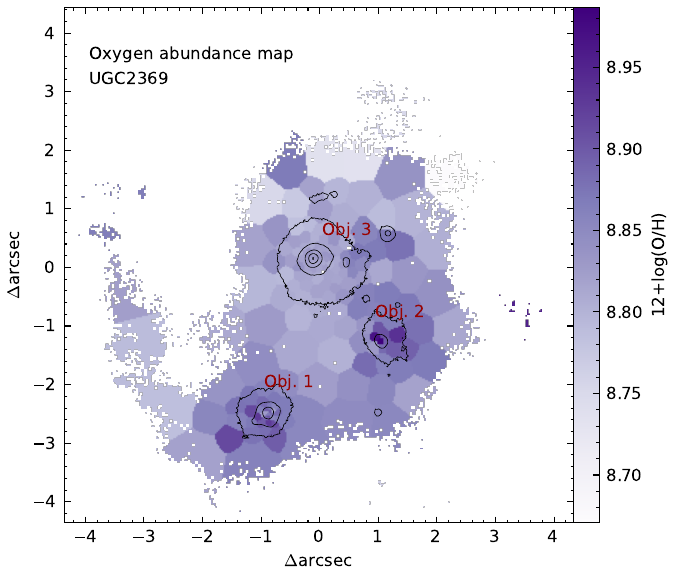}
    \caption{Extinction corrected oxygen abundance map using the strong line (O3N2) method described in \citet{Pettini2004}. Both object~1 and 2 show slightly super-solar ($\gtrsim$0.2\,dex) abundance at the core. The lack of abundance peak at object~3 is likely due to heavy dust obscuration and the dilution by the starburst region outside the thick dust screen.}
    \label{fig:O_abund_map}
\end{figure}

Chemical enrichment provides a sensitive diagnostic of evolutionary history. We map the gas-phase oxygen abundance using extinction-corrected strong-line ratios (O3N2; Figure~\ref{fig:O_abund_map}) following \citet{Pettini2004}. Both objects~1 and~2 exhibit central O/H values that are $\gtrsim 0.2\,\mathrm{dex}$ higher than the surrounding off-nuclear ISM, indicating locally enriched gas in the cores. Such a contrast is naturally expected if the cores are the chemically evolved nuclei of stripped satellites, whereas it is difficult to produce (and maintain) this level of offset across nearby star-forming complexes over the short merger timescale (a few $\times10$~Myr).

\begin{figure}
    \centering
    \includegraphics[width=0.5\textwidth]{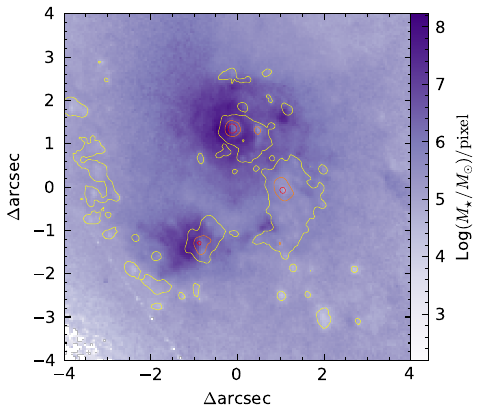}
    \caption{Stellar mass map estimated from $g-i$ color and H-band magnitude, with HST $g$-band (F435W) contour being overplotted. The western core shows a deficiency in stars, which is consistent with the relatively low dust extinction and SFR, indicating a history of tidal disruption. Note that this method tends to underestimate the stellar mass in dusty region, due to the power-law scaling between $g-i$ color and mass to light ratio.}
    \label{fig:Mstar_map}
\end{figure}

\begin{figure*}
    \centering
    \includegraphics[width=1.0\linewidth]{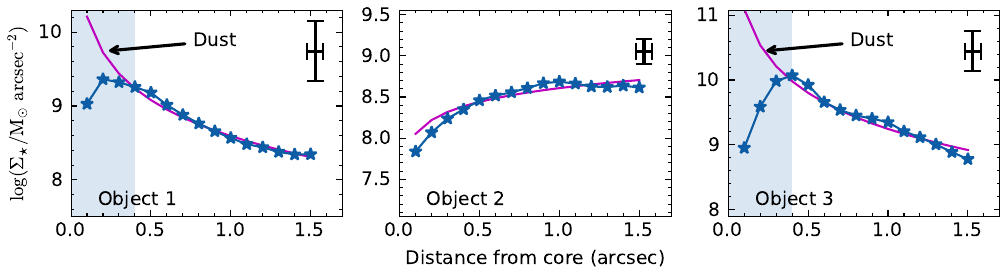}
    \includegraphics[width=1.0\linewidth]{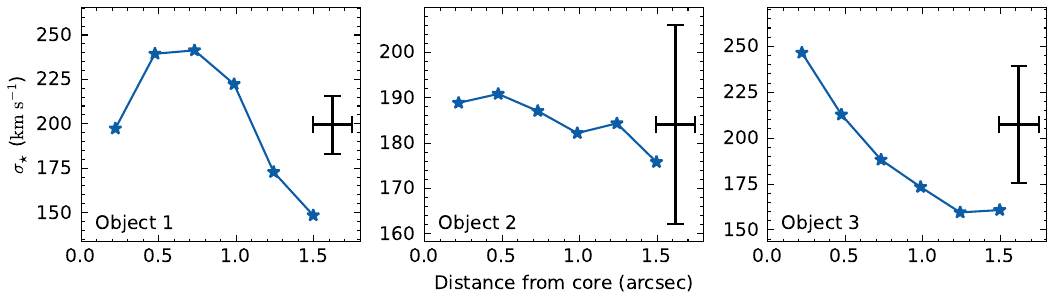}
    \caption{{\bf Top}: Stellar mass surface density profile for each core, fitted by power-laws (magenta). The dips at small radii for object~1 and 3 are due to heavy dust obscuration. Spatial resolution is $\simeq0\secpoint05$. {\bf Bottom}: Stellar velocity dispersion radial distribution based on seeing limited KCWI observation. We indicate the radial bin size and median standard deviation within each bin with reference errorbars on the right of each subfigure. 
    These measurements are consistent with object~1 and 2 being tidally stripped satellites. While the values for Object 2 in Figure~\ref{fig:stellar_abs_spec} may appear inconsistent with those presented here, the discrepancy arises from the differing weighting methods employed: Figure~\ref{fig:stellar_abs_spec} displays a flux-weighted spectrum, whereas the values in this section represent a direct spatial average of individual spaxels within each annulus.}
    \label{fig:value_radial}
\end{figure*}

We also examine how key physical quantities vary with radius. Figure~\ref{fig:value_radial} presents 1D radial profiles derived from Figures~\ref{fig:Mstar_map} and~\ref{fig:stellar_kinematics_map}. For each object we define concentric annuli and compute azimuthally averaged stellar mass surface density ($\Sigma_{\star}$) and stellar velocity dispersion ($\sigma_{\star}$) in radial bins.

Object~2 shows a monotonic rise in $\Sigma_{\star}$ toward the outskirt out to $\sim1\arcsec.0$, as expected for a disrupted satellite. The apparent central $\Sigma_{\star}$ dips in objects~1 and~3 are likely driven by strong dust attenuation: heavy extinction suppresses the \textit{g} band flux and can bias the inferred mass-to-light ratio, leading to underestimated stellar masses in the most obscured regions. The $\sigma_{\star}$ profiles of objects~1 and~2 are flat toward the center, with a pronounced central dip in object~1. 

These trends are qualitatively consistent with tidal heating and dynamical disturbance during stripping: tidal shocks inject kinetic energy, redistribute stars to larger radii, and can reduce the central concentration of the bound remnant, thereby accelerating mass loss from the outer layers \citep{mo_galaxy_2012}. Taken together, the elevated central metallicities and the radial structural/kinematic trends, as well as their large stellar mass favor an interpretation of objects~1 and~2 as stripped satellite nuclei rather than purely in-situ young star clusters.

\subsection{Dynamics of A Chaotic Merger System}
\label{sec:core_dynamics}
%The velocity map in \oiii\ indicates the system is rotating on a large scale ($\simeq3\,\rm kpc$), with no apparent disk structures around each individual core being visible (Figure~\ref{fig:emission_line_map}). The flux-weighted FWHM increases around object~3, consistent with a starburst outflow associated with the increased SFR.

In this section, we discuss the dynamics of the SMBHs in the system, assuming that it is consist of three SMBH at each stellar core. An inspiraling satellite BH could affect the evolution of an SMBHB long before it sinks into the center by perturbing the large-scale gravitational field and scattering more stars into the loss cone. This process may or may not be efficient enough to prevent close three-body interactions between them \citep{Hoffman:2007:957}. If the outer intruder (object~1) comes close enough before it causes a sufficient hardening of the (inner) binary, then a strong three-body encounter occurs.

We have projected the inner binary semi-major axis $a_\mathrm{in}=1\secpoint79$ and outer binary separation $ R_\mathrm{out}=2\secpoint7$, which correspond to 1.18 and 1.78\,proper-kpc at the rest frame. If $R_\mathrm{out}/a_\mathrm{in}$ is large enough, we expect this hierarchical triple to exhibit very regular behavior. The onset of an unstable interaction can be estimated by the following criteria \citep{Mardling2001MNRAS}:

\begin{equation}
    R_\mathrm{out}<2.8 a_\mathrm{in}\left[\frac{(1+q_\mathrm{out})(1+e_\mathrm{out})}{\sqrt{1-e_\mathrm{out}}}\right]^{2/5}
\end{equation}
where $e_\mathrm{out}$ is the eccentricity of the outer binary system, $q_\mathrm{out}$ is the mass ratio between the intruder and the inner binary system. Assuming a face-on geometry, we find \ugc is in an unstable three-body interaction stage unless the orbit is very circular ($e_\mathrm{out}\lesssim 0.12$). We note that $e_\mathrm{out}>0.12$ is a likely situation as three-body interactions tend to produce eccentric orbits \citep{Hoffman:2007:957}.

Object~2 has the smallest stellar velocity dispersion in the system. The stellar mass ratio (Table~\ref{tab:core_params_clean}) between object~3 and~2 is $\simeq$4.83 and represents a minor merger if it was a merging satellite in this hierarchical merger sequence \footnote{One possibility is that object~2 is a SF complex formed in situ due to the fragmentation and local collapse of gas-rich, turbulent disk \citep{Zanella2019MNRAS}. However, it would be difficult to explain how such a structure could grow more than tens of Myr without being destroyed by tidal forces, since we have shown that unstable three-body processes control the cores.}. 

A small satellite galaxy like object~2 is subjected to tidal forces, which would lead to a complete disruption in a few pericenter passage \citep{Koehn2023A&A}, so it is not surprising to see a core with a low stellar mass density if it was a satellite galaxy not long ago (Figure~\ref{fig:Mstar_map}). We speculate that tidal heating effects may be responsible for the removal of gas and stars from object~2, because triplet systems typically feature an extreme gradient and time variation of the gravitational field in simulations (on host galaxy scale). 

One indicative feature in the stellar dynamics (Figure~\ref{fig:stellar_kinematics_map}) is that stars at large radii in the smaller cores have higher velocity dispersion (Figure~\ref{fig:value_radial}). This is because the halo star and gas have orbital periods comparable to the encounter time scale. Therefore, they see a gravitational potential that changes strongly but not periodically during an orbit, and the stars suffer strongly from violent relaxation \citep{Lynden-Bell:1969:690}.

The tidal radius is a good indicator of the strength of the tidal disturbance on a stellar cluster \citep{mo_galaxy_2012}. An estimation of the tidal radius can be obtained from the mass distribution \citep{mo_galaxy_2012,King1962}:
\begin{equation}
r_\mathrm{t}=\left[\frac{m(r_{t})/M(R_{0})}{2+\Omega^{2}R^{3}_{0}/GM(R_{0})-d\ln M/d\ln R}\right]
\end{equation}
where $m(r_\mathrm{t})$ is the satellite mass within the tidal radius; $M(R_{0})$ is the host mass within a separation $R_{0}$; $\Omega$ is the angular speed and is equal to $G(M+m)/R^{3}_{0}$ in the limit where $m$ and $M$ are point masses traveling in a circular orbit. We conclude that object~1 and~2 have $r_\mathrm{t}\simeq 139\rm\,pc$ and $47\rm \,pc$ respectively, corresponding to 0.216 and 0.07\,arcsec (pixel scale is 0.05 arcsec in Figure~\ref{fig:Mstar_map}). Note that the angular speed is estimated with the redshift of individual cores. This includes only the line-of-sight velocity component, which means the tidal radius could be lower. This result supports a scenario in which both cores are experiencing tidal stripping to some extent. Due to strong tidal stripping, close (sub-kpc) multi-AGNs are likely to be rare \citep{Hoffman2023,Mannerkoski2021ApJ}. It was shown numerically that the probability for all three SMBHs to be active increase monotonically with their maximum separation, and $>$ 30\% of the triple SMBH at separation $<$ 80 kpc may only be observed as duals.

%Our result confirms a theoretical expectation that close (sub-kpc) multi-AGN is likely to be rare \citep{Hoffman2023,Mannerkoski2021ApJ}. Tidal force inject energy into the satellites' ISM. As they evolve towards a new equilibrium, their mass distribution expands, causing stars with higher velocities to migrate to larger radii. This is indeed happening in both object~1 and 2 as we can see both the emission line width and the stellar velocity dispersion are decreasing towards the cores. As a consequence, the satellite’s profile changes to be less concentrated, thereby facilitating the particles in the outer layers to become unbound. 

It is rather difficult to predict the merger timescale for this complex system, especially since the hosts have mostly merged, leaving only a stellar bulge (which is also missing for object~2). A direct application of Chandrasekhr's dynamical friction timescale \citep{chandrasekhar1943}, assuming a singular isothermal sphere with the property of object~3's stellar bulge, produces a merger time $\lesssim 5\,$Myr which may be an upper limit as the chaotic three-body dynamics should expedite the process \citep{Liu:2019:21,Binney2008gady.book}.

\subsection{Outflow Dynamics}
\label{sec:outflow}
Estimating the total energy and mass outflow rate from the stellar cores is important in understanding whether the outflows can efficiently remove gas, as expected by the standard AGN feedback model \citep{Hopkins:2008:356}.

Total mass contained in ionized gas outflow can be estimated by assuming pure case B recombination, after estimating electron density with the \sii\ doublet line ratio method \citep{Osterbrock2006agna.book}:
\begin{equation}
    M_{\rm gas}=2.82\times10^{9} L_{\rm H\beta,43}\,n_{\rm e,100}^{-1}\,M_{\odot}
\end{equation}
Where $L_{\rm H\beta,43}$ is the extinction corrected H$\beta$ luminosity in units of $10^{43}\,\rm erg~s^{-1}$, including only pixels with $W_{80}$ $>500\,\rm km~s^{-1}$ \citep[velocity width that encloses 80\% of total flux,][]{Liu2013MNRASb}; $n_{\rm e,100}$ is the electron density in units of 100\,cm$^{-3}$. 

We assume an outflow shell radius of $R_{\rm out}=1.8^{''}$, based on Figure~\ref{fig:emission_line_map}. Summing over this $R_{\rm out}=1.8^{''}=1.19\,\rm kpc$ circular region around object~3, we obtain a total ionized gas mass in outflow of $4\times10^{7}M_{\odot}$. The simplest method to estimate total kinetic energy in the outflow is:
\begin{equation}
    E_{\rm k}=\frac{1}{2}M_{\rm gas}v_{\rm out}^{2}
\end{equation}
where $v_{\rm out}$ is the outflow velocity \footnote{We set $v_{\rm out}$ to the velocity shift of the broad, blue-shifted outflow component for simplicity.}. We find that the total kinetic energy is $E_{\rm k}=2\times10^{55}\,\rm erg$. The outflow time scale is around $t_{\rm out}\simeq 0.5\,{\rm kpc}/v_{\rm gas}\simeq1\,\rm Myr$, resulting in an estimated energy outflow rate of $\dot{E}_{\rm k}=6.5\times10^{41}\,\rm erg~s^{-1}$, similar to some of the most powerful outflows in Type-2 AGN \citep{Harrison2014}.

We could obtain a more accurate but model dependent estimate by assuming an outflow geometry. A spherically symmetric outflowing gas shell has mass outflow rate \citep{Rodriguez2013MNRAS}: 
\begin{equation}
    \dot{M}_{\rm out}=\frac{3M_{\rm gas}v_{\rm out}}{R_{\rm out}}
\end{equation} 
and kinetic energy rate:
\begin{equation}
    \dot{E}_{\rm out}=\frac{\dot{M}_{\rm out}}{2}(v^{2}_{\rm out}+3\sigma^{2})
\end{equation}
where $\sigma$ and $v_{\rm out}$ are the velocity dispersion and velocity shift of the broad outflow component. The calculations are conducted in each spaxel separately and then summed together. We find $\dot{M}_{\rm out}=23\,M_{\odot}/\rm yr$ and $\dot{E}_{\rm out}=1.9\times10^{42}\,\rm erg~s^{-1}$. These values are consistent with our earlier estimate, indicating mild outflows. These two methods typically underestimate the outflowing mass and energy rate, as we only consider the clump of the line-emitting gas shell with the largest velocity shift. Hence, the total mass involved in the outflow could be higher \citep{Greene2011ApJ}.

Another plausible scenario involves a constant-velocity spherical outflow with a power-law radial luminosity density distribution \citep{Liu2013MNRASa,Liu2013MNRASb}. In this model, the entirety of the observed line-emitting gas is associated with the outflow, while the redshifted receding component is obscured by dust. Under these assumptions, the velocity width containing 80\% of the total flux \citep[$W_{80}\simeq1.08\rm\, FWHM$ for Gaussian profile,][]{Liu2013MNRASb} is constant across the imaging plane, and is a tracer of the physical outflow velocity ($W_{80}/1.3\simeq v_{0}$ where $v_{0}$ is outflow velocity), as it is largely free from inclination effects. We then have outflow kinetic rate \citep{Liu2013MNRASb}:
\begin{equation}
\begin{split}
    \frac{\dot{E}}{4.1\times10^{44}\,\rm erg~s^{-1}}&=\left(\frac{\rm SB_{\rm H\beta}}{5\times10^{-15}\,\rm erg~s^{-1}~cm^{-2}~arcsec^{-2}}\right) \\
    &\times\left(\frac{D}{7\,\rm kpc}\right)\left(\frac{W_{80}}{988\,\rm km~s^{-1}}\right)^{3}\left(\frac{n_{\rm e}}{10\,\rm cm^{-3}}\right)^{-1}
\end{split}
\end{equation}
where $\rm SB_{H\beta}(D)$ is the surface brightness of \hb\ at $D$, and $D$ is the size of the outflow shell where most of the \hb\ emitting atoms are situated; we take $D=1.19\,\rm kpc$ as the radius enclosing the kinematically disturbed area. The electron density $n_{\rm e}$ is measured from \sii. The mass outflow rate in the observed ionized gases is $\dot{M}=2\dot{E}/v_{0}^{2}$:
\begin{equation}
\begin{split}
    \frac{\dot{M}}{2240\, M_{\odot}\rm~yr^{-1}}&=\left(\frac{\rm SB_{\rm H\beta}(D)}{5\times10^{-15}\,\rm erg~s^{-1}~cm^{-2}~arcsec^{-2}}\right) \\
    &\times\left(\frac{D}{7\,\rm kpc}\right)\left(\frac{W_{80}}{988\,\rm km~s^{-1}}\right)\left(\frac{n_{\rm e}}{10\,\rm cm^{-3}}\right)^{-1}
\end{split}
\end{equation}
Ruling out spaxels that have $W_{80}/1.3<236\,\rm km~s^{-1}$ (thus are unlikely to be part of the outflow), and taking average in 1\secpoint1-1\secpoint8, we get $\dot{M}=20\, M_{\odot}\rm~yr^{-1}$ and $\dot{E}=1.3\times10^{42}\,\rm erg~s^{-1}$, similar to the values obtained with the shell model. 
%Considering the outflow as an energy conserving bubble expanding in a uniform medium could provide an upper limit for the outflow energy and mass rate \citep{Greene2012, Harrison2014}. The resulting energy rate is \citep{Nesvadba2006, Heckman1990, Veilleux2005ARA&A,Harrison2014}: 
%\begin{equation}
%\dot{E}_{\rm out}\simeq 1.5\times10^{46}R_{\rm out,10}^2v_{1000}^{3}n_{0.5}\,\rm erg~s^{-1}
%\end{equation}
%where $n_{0.5}= \frac{n}{0.5\,\rm cm^{-3}}\simeq 2$ is the hydrogen number density of ambient medium; $v_{1000}$ is outflow velocity in unit of 1000\,$\rm km/s$; $R_{\rm out,10}$ is outflow shell radius in unit of 10\,kpc. We obtain a kinetic energy rate of $1.42\times10{45}\,\rm erg~s^{-1}$ and a mass rate of $\dot{M}_{\rm out}\simeq8.8\times10^{5} M_{\odot}\rm ~yr^{-1}$. These are very large values, with the outflow being able to completely drain the gas reservoir in only tens of kyr.

The properties of the neutral gas outflow can be calculated using the ansatz discussed in \citet{Shih2010ApJ} and \citet{Rupke2005ApJS}. The neutral outflow is assumed to be spherically symmetric, and distributed on a thin shell being roughly $R_{\rm out}=1.8^{''}$ away from the center. We deproject the observed velocity onto the radial direction and convert the apparent spaxel area (multiplied by the covering factor) to the solid angle subtended to the center (object~3). The radially averaged mass outflow rate is then:
\begin{equation}
    dM/dt=\mu m_{p} R_{\rm out}\int N_{\rm H}(\theta,\phi)v(\theta,\phi)d\Omega,
\end{equation}
where $\mu$ is the mean molecular weight, $m_{p}$ is the mass of a proton, and $v(\theta,\phi)$ is the radial velocity deprojected from the observed value. The hydrogen column density associated with the neutral outflow at each spaxel $N_{\rm H}(\theta,\phi)$ can be obtained from:
\begin{equation}
    N_\mathrm{H}=N(\rm NaI)(1-y)^{-1}10^{-(a+b)}
\end{equation}
where $y=\rm(1-NaI/Na)=0.9$, $a=\log[N_\mathrm{Na}/N_\mathrm{H}]=-5.69$ and $b=\log[N_\mathrm{Na}/N_\mathrm{H,total}]-\log[N_\mathrm{Na}/N_\mathrm{H,gas}]=-0.95$ are fixed at solar values. The Na column density can be inferred from the optical depth of the measured absorption lines \citep{Spitzer1978ppim.book.....S,Rupke2002ApJ}:
\begin{equation}
    N(\rm NaI)=\frac{\tau_{c}b}{1.497\times10^{-15}\lambda_{1}f_{1}}
\end{equation}
where $f_{1}$, $\lambda_{1}$, and $\tau_{c}$ are the oscillator strength, vacuum wavelength, and central optical depth of the NaI$\lambda5896$ line. $b=\sqrt{2} \sigma$ is the Doppler parameter in $\rm km~s^{-1}$. We use the vacuum wavelength of the transition $\lambda_{1}=5897.55\,$\AA\ and $f_{1}=0.3180$, while adopting the strongest NaID velocity component with $C_{f}>0.1$. Summing over all spaxels within $R_{\rm out}$, we get a total mass outflow rate of $7.8 \,M_{\odot}\rm/yr$ and an energy rate of $1.0\times 10^{42}\rm\, erg~s^{-1}$. This is likely to be a lower limit because absorption lines only probe the gas along our line of sight. Given that at most $\simeq 50\%$ of the neutral gas shell is taken into account in this calculation and the neutral gas shell may have a larger radius, the neutral outflow rate is likely to be comparable to the SFR.

Combining the lower limits of both neutral and ionized outflows, we obtain a mass-loading factor $\mu\gtrsim1.9$, which is consistent with those obtained in larger LIRG surveys \citep{Cazzoli2016A&A,Rupke2005ApJS}. However, significant uncertainties exist as the spatial extent and the gas kinematics inside the dust-obscured central starburst region remain unclear.
%very large even compared with luminous AGN \citep[e.g.,][]{Martin1999,Heckman2000,Newman2012,Cicone2014,Harrison2014}. The lower limit of combined outflow energy is $\simeq 2.3\times10^{43}\,\rm erg~s^{-1}$. 

Comparing the outflow energy rate we got with what could be provided by starformation may illuminate the driving process of the observed outflow. For every solar mass of stars formed, the stellar winds and supernovae can yield at most $7\times10^{41}\rm\,erg~s^{-1}$ worth of kinetic energy of the outflow \citep{Leitherer1995,Veilleux2005ARA&A}, placing an upper limit on the energy rate of $\dot{E}_{\rm SF,out}=1.029\times10^{43}\,\rm erg~s^{-1}$. This is a factor of three larger than the lower limit on the combined energy rate we obtained, suggesting that an AGN outflow is not required to explain the observed gas kinematics.
%by a factor of two lower than \textbf{the lower limit} of the observed energy outflowing rate, indicating the existence of AGN-powered outflow.

%Drag by a dense halo or tidal debris may be especially important in high-luminosity starburst \citep{Veilleux:2002:315}.
A large fraction of this outflowing gas is likely to join the recycling flow and fuel a late-time starburst. However, a fraction may escape and enrich the IGM. A typical way to estimate the escape fraction is to compare the outflow velocity with the local escape velocity derived from a gravitational model of the host galaxy. A simple, truncated isothermal sphere model provides a rough estimate of escape velocity at radius \textit{r} \citep{Veilleux2005ARA&A}:
\begin{equation}
    v_{\rm esc}(r)=\sqrt{2}v_{\rm rot}[1+\ln(r_{\rm max}/r)]^{1/2}
\end{equation}
where $v_{\rm rot}$ is rotation speed and $r_{\rm max}$ is the truncation radius. If halo drag is negligible, gas that exceeds $v_{\rm esc}$ may vent into the IGM. This escape velocity is not sensitive to the exact value of $r_{\rm max}/r$ and is able to provide a good estimate of the escape fraction of outflowing material.

The rotation speed $v_{\rm rot}$ could be inferred from stellar velocity shift. KCWI has the constraining power for pixel-wise stellar kinematics (results shown in Figure~\ref{fig:stellar_kinematics_map}). Due to its relatively low, seeing limited spatial resolution ($\simeq0\secpoint5-0\secpoint9$), we estimate the rotation velocity from the inclination corrected shear velocity at the continuum half light radius $R_{\rm eff}$ in a 0\secpoint3 bin, avoiding regions dominated by starburst outflows. $R_{\rm eff}=3.55''$ is determined from the HST/ACS F814W image. The velocity shear is defined as $v_{\rm shear}=\frac{1}{2}(v_{\rm max}-v_{\rm min})$, where $v_{\rm max}$ and $v_{\rm min}$ are 2.5 percentile at each end of the velocity distribution so that outliers and artifacts are excluded \citep{2010ApJ...724.1373G}. The morphology of \ugc is complex, such that no reliable inclination value can be inferred. We use a generic value $i=52\pm26^{\circ}$ (equivalent to $\sin i=0.79\pm0.35$) derived from ULIRG surveys \citep{Bellocchi2013A&A}. The resulting escape velocity is $v_{\rm esc}=235\pm104\,\rm km~s^{-1}$. If we take the velocity shift of the broad component as the outflow velocity, we found that $<73\%$ of the gas may escape into the IGM. The escape fraction would be 43\% if $i=52^{\circ}$ was assumed. One feature worth noting is that almost all ionized gas that has $v_{\rm shift}<-235\,\rm km~s^{-1}$ is in a thin, turbulent ($v_{\rm turb}\gtrsim700\,\rm km~s^{-1}$) filament to the west of Object~3 (Figure~\ref{fig:emission_line_map}). While this structure is not traced by SFR density hotspots and high dust extinction, it appears to be perpendicular to the stellar kinematic major axis, which a starburst outflow could potentially explain. 

In summary, the kinematics and dynamics of the outflowing gas in \ugc do not require the existence of an AGN and could be explained almost entirely by central starburst.

\section{conclusion}

Mergers are believed to be important in transporting gas toward galactic nuclei and enabling rapid BH accretion \citep{Hopkins:2008:356}. A recent cosmological simulation has suggested that triplet merger systems are responsible for most of the UMBH ($M_\mathrm{BH}\simeq10^{10}M_{\odot}$) \citep{Ni2022ApJ}. However, only a handful of such systems have been studied in the past, leaving the demographics and properties of triplet mergers largely unknown. 

In this paper, we analyzed multi-wavelength data of an interacting galaxy \ugc, which is one of the few cases with clear evidence of ongoing violent three-body interaction. MUSE-AO data permit high-resolution mapping of the extinction through a typical nebular line ratio. We find that two of the three cores have significant nuclear dust, with A$_{\rm v}\gtrsim3$ and $\sim 10$ in the nuclear region of the northern core (object~3, see Figure~\ref{fig:optical image}). The inferred SFR surface density map from extinction corrected \ha\ flux shows a similar structure to the extinction map, unveiling SF hot spots within the galaxy and increased SFR in the nuclear region of objects~1 and~3 (Figure~\ref{fig:extinction_sfr_map}). The star formation history inferred from full-spectrum fitting indicates the whole system is of a young stellar population, overlapping with a high SFR region as revealed by the strong diffuse X-ray and radio flux. Object~1 and 3 have similarly old light-weighted stellar populations, while object~2 appears to be dominated by a recent starburst ($\tau\lesssim100\,\rm Myr$). Our result supports the general picture that violent mergers can efficiently torque pristine gas toward the nuclei, resulting in boosted SFR and a radial stellar age gradients \citep{mo_galaxy_2012}.

Our optical, NIR and radio band analysis all imply that object~3 hosts an AGN buried by heavy obscuration, with a $\sim+0.2\,$dex super-solar gas phase metallicity in each of the core. With the same dataset, we rule out AGN associated with objects~1 and 2. While starburst and the accompanying X-ray binary population alone are sufficient to produce the X-rays as observed by \chandra in \ugc, it is certain that the ISM properties of object~3 is unusual and deviated from $\simeq90\%$ of the SDSS galaxies (Figure~\ref{fig:R23vO32_abund_map} and \ref{fig:BPT_points}). It remains unclear whether the high \nii/\ha\ is due to anomalous abundance pattern in this system \citep[e.g., M82 has $\sim2$ times solar N/O,][]{Fukushima2024A&A}, but high electron density ($\gtrsim10^{3}\,\rm cm^{-3}$) and/or nitrogen enhancement typical in AGN narrow line region are natural explanations for the high \nii/\oii\ line ratio. The extremely high $N_\mathrm{H}$ ($\gtrsim10^{25}\,\rm cm^{-2}$) inferred from OSIRIS data means that it is difficult to understand the nature of the nuclear region, as the material is optically thick even for X-ray photons of energy $\simeq 7\,$keV. Future observations with \textit{NuSTAR} targeting the Compton hump ($\gtrsim 15\,$keV) may be the final piece of the puzzle to confirm (or rule out) the possible existence of an AGN \citep{Boorman2024}. If \ugc does have a Seyfert-2 nuclei under the dust screen with an intrinsic X-ray flux on the order of $10^{-12}\,\rm erg~s^{-1}~cm^{-2}~keV^{-1}$ (estimated by assuming point-like radio emission from object~3 is coronal), there should be a detectable amount of hard X-ray photons above 10\,keV for \textit{NuSTAR}. In addition, ALMA high-resolution continuum observation at $\sim$100\,GHz may be used to search for self-absorbed synchrotron from a hybrid AGN corona \citep[e.g.,][]{Ricci_2023ApJL}.

We reconstructed the line-of-sight stellar kinematic profile from MUSE and KCWI data. This kinematic information suggests that the smaller cores may have undergone tidal disruption, exhibiting increased velocity dispersion in their outskirts. Moreover, object~2 has a stellar mass radial distribution increasing monotonically from the center. Given its high metallicity, diffuse core, and kinematic profile, we propose that object~2 has been tidally stripped during recent close encounters with object~3. Our measurements suggest that both object~1 and 2 are likely stripped satellite nuclei. This result is consistent with the theoretical prediction that very close ($\lesssim 1\,\rm kpc$) BH triples are less likely to be observed as multi-AGN, due to the strong tidal effect that may destroy the stellar bulge and unbind the gas \citep{Mannerkoski2021ApJ,Hoffman2023}. Without typical AGN features, it is challenging to confirm the existence of multiple SMBHs in a merger system.

After subtracting best-fit stellar templates, we measured the kinematics of all the prominent nebular lines and Na ID absorption lines in the system with MUSE-AO data. We find that the gas around object~3 shows evidence of disturbance consistent with a spherical outflow. The combined mass and energy rates from ionized and neutral outflows do not require AGN contributions and are generally consistent with typical LIRG \citep{Cazzoli2016A&A}. Nevertheless, the escape fraction of these outflows could be significant ($\simeq50\%$). While the gas kinematics inside the dust-obscured region remain unknown, there is no evidence of AGN-driven outflow at kpc scales for \ugc. Our ongoing IFU/\chandra survey, including $\sim$8 more triplet systems like this, will provide a further census of this elusive population. Finally, the information on cold molecular gas is currently missing; this is important for providing a better estimate of the total outflowing energy and the dynamics of the cold gas. Thus, future ALMA observations are important.

\begin{acknowledgments}
This work is supported by NASA grants under contract No. NNG08FD60C. 
M.K. acknowledges support from NASA through ADAP award 80NSSC22K1126. This work was performed in part at the Aspen Center for Physics, which is supported by National Science Foundation grant PHY-2210452. 
ET and CR acknowledge support from the ANID CATA-BASAL program FB210003. 
ET acknowledges support from FONDECYT Regular - 1190818, 1200495, 1241005, and  1250821. FEB acknowledges support from the ANID Millennium Science Initiative Program ICN12\_009. 
IMC acknowledges support from the ANID program FONDECYT Postdoctorado 3230653 and ANID, BASAL, FB210003
ZZ acknowledges the financial support of the NASA FINESST Program under contract No. 80NSSC22K1755.

Google Gemini (version 3.1 Pro) was utilized during the drafting of this manuscript solely for the purpose of refining sentence structure and correcting grammatical errors. The authors reviewed all AI-suggested edits to ensure scientific accuracy and retain full responsibility for the published work.

\software{
\textsc{pyneb} \citep{pyneb2015A&A...573A..42L,pyneb2020Atoms...8...66M,pyneb2023Atoms..11...63M},
\textsc{astropy} \citep{Astropy2013A&A...558A..33A,Astropy2018AJ....156..123A},
\textsc{casa} \citep{McMullin:2007:127},
\textsc{xspec} \citep{Arnaud:1996:17},
\textsc{ciao} \citep{CIAO2006SPIE},
\textsc{ds9} \citep{ds9_2000ascl.soft03002S},
\textsc{matplotlib} \citep{matplotlib_2007CSE.....9...90H},
\textsc{numpy} \citep{numpy_2020Natur.585..357H},
\textsc{photutils} \citep{photutils_larry_bradley_2025_14889440},
\textsc{zap} \citep{2016MNRAS.458.3210S},
\textsc{pypeit} \citep{pypeit:joss_pub,pypeit:zenodo}
}

%support from the ANID BASAL project FB210003. This work was

\end{acknowledgments}

\appendix
\section{Spatially Resolved Stellar Kinematics}
\begin{figure*}
    \centering
    \includegraphics[width=0.48\textwidth]{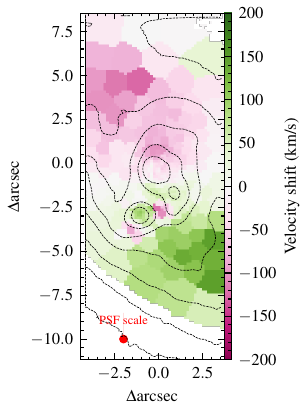}
    \includegraphics[width=0.48\textwidth]{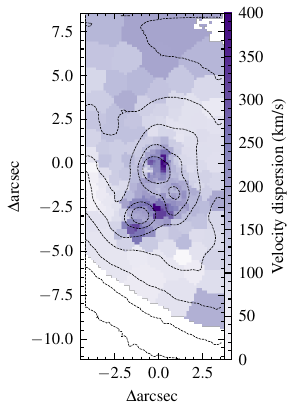}
    \caption{Stellar kinematics map as derived by \textsc{pPXF} fittings. These values are measured through the Ca II triplet absorptions at a rest-frame wavelength of $\sim8500\,\mathrm{\AA}$. The black contour indicates the isophotal level in the rest frame $7800$--$8800\,\mathrm{\AA}$. North is up; east is to the left. \textbf{Left}: velocity shift of the Ca II absorption features relative to the systemic redshift of object~3 (north core).; \textbf{Right}: stellar velocity dispersion.}
    \label{fig:stellar_kinematics_map}
\end{figure*}

In Figure~\ref{fig:stellar_kinematics_map}, we show stellar kinematics measured by \textsc{pPXF} from each Voronoi bin. The velocity shift uses the redshift of object~3 as the zero point.
%But these results should be treated with cautious because in the case that no significant Ca II features are present, the \textsc{pPXF} code could mistakenly consider background features as stellar absorptions. This could result in very small $\sigma_{\star}$ values and is not easy to be identified until the 1D spectra are inspected individually.
\restartappendixnumbering

\begin{figure*}
    \centering
    \includegraphics[width=0.49\textwidth]{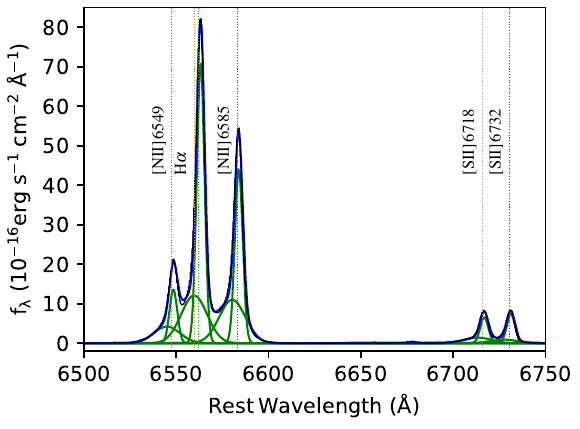}
    \includegraphics[width=0.49\textwidth]{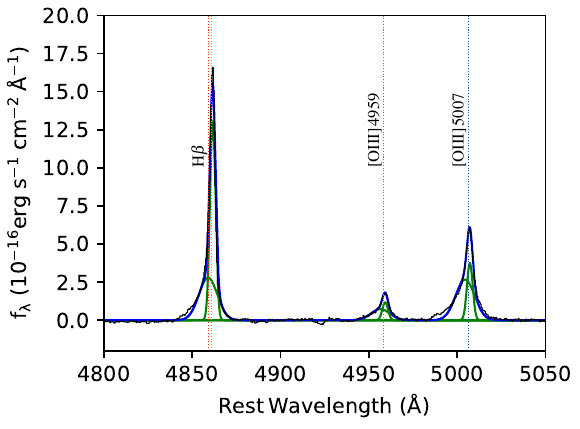}
    \includegraphics[width=0.49\textwidth]{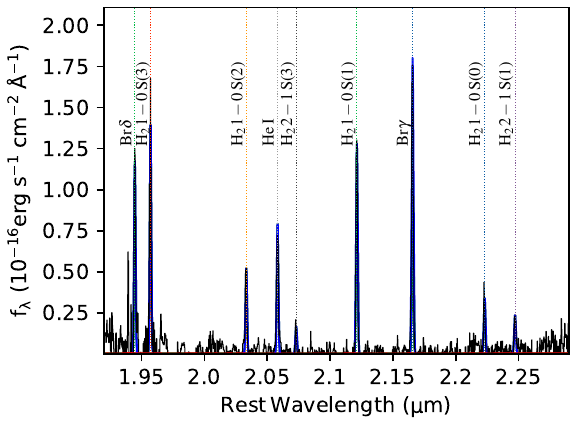}
    \caption{Continuum subtracted KCWI and OSIRIS spectra extracted from a 1\secpoint5 diameter aperture centering on the northern core (object~3) of \ugc.  We detect a blue-shifted broad component in \ha\ and \hb\ but not in other hydrogen recombination lines as seen in K-band OSIRIS exposure, arguing against its BLR origin. %(\textbf{Left}): MUSE \oiii band. White circles indicate the extraction radius for nuclear 1-D spectra; \textbf{Right}: 
    }
    \label{fig:obj3_spectra}
\end{figure*}

\begin{figure*}
    \centering
    \includegraphics[width=0.49\textwidth]{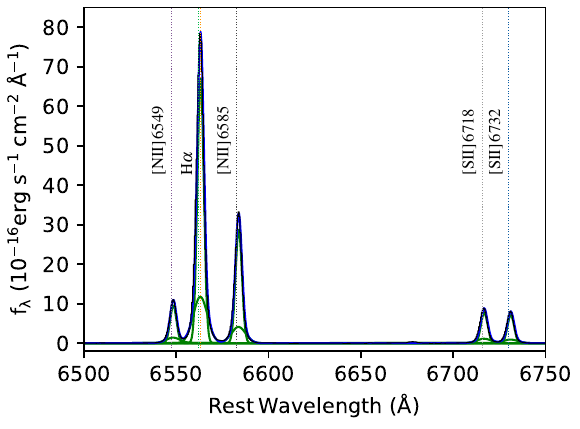}
    \includegraphics[width=0.49\textwidth]{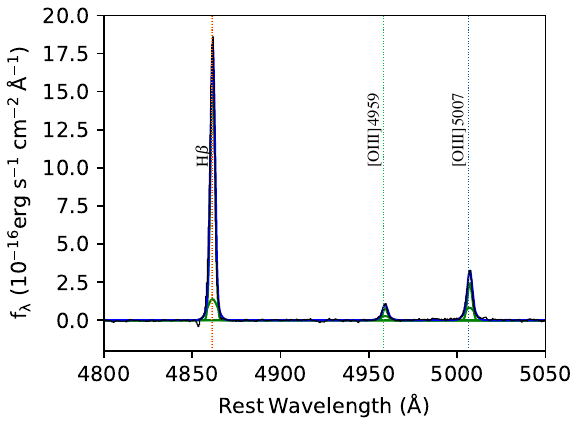}
    \includegraphics[width=0.49\textwidth]{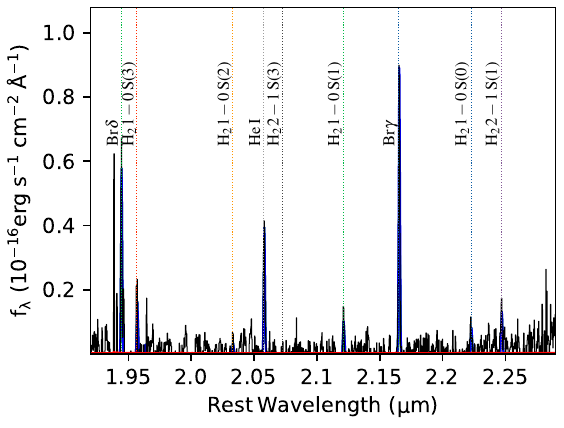}
    \caption{Same as Figure~\ref{fig:obj3_spectra} but showing object~1. %(\textbf{Left}): MUSE \oiii band. White circles indicate the extraction radius for nuclear 1-D spectra; \textbf{Right}: 
    }
    \label{fig:obj1_spectra}
\end{figure*}

\begin{deluxetable*}{lccc}
\tablecaption{Physical Parameters of Cores identified in MUSE-AO Observations \label{tab:core_params_clean}}
\tablehead{
\colhead{Parameter} & 
\colhead{Core 1 (SE)} & 
\colhead{Core 2 (SW)} & 
\colhead{Core 3 (N)} \\
\colhead{(1)} & \colhead{(2)} & \colhead{(3)} & \colhead{(4)}
}
\startdata
Redshift ($z$)                              & 0.0322                         & 0.0320                         & 0.0318 \\
$L_{\mathrm{2-8\,keV,core}}$ (erg s$^{-1}$) & $<3.97\times10^{39}$           & $<3.05\times10^{39}$           & $(1.33\pm0.25)\times10^{40}$ \\
$F_{\mathrm{33\,GHz, core}}$ (mJy)          & $<0.035$                       & $<0.035$                       & $0.68\pm0.01$ \\
$E(B-V)$                              & 0.82                           & 0.55                           & 0.82 \\
SFR ($M_{\odot}$ yr$^{-1}$)                 & $3.55\pm0.54$                  & $2.63\pm0.49$                  & $14.70\pm1.53$ \\
$\sigma_{\star}$ (km s$^{-1}$)              & $170\pm10$                     & $89.4\pm9.7$                   & $220.3\pm9.8$ \\
$M_{\mathrm{BH}}$ ($M_{\odot}$)             & $(1.52\pm0.08)\times10^{8}$    & $(9.08\pm1.01)\times10^{6\dagger}$ & $(4.73\pm0.24)\times10^{8}$ \\
$M_{\star}$ ($M_{\odot}$)                   & $6.5\times10^{9}$              & $4.8\times10^{9}$              & $2.32\times 10^{10}$ \\
\enddata
\tablecomments{
\textbf{Row descriptions:}
(Redshift): Adopted redshift, measured from stellar absorption lines.
($L_{\mathrm{2-8\,keV,core}}$): X-ray luminosity estimated in 2--8\,keV from unresolved cores.
($F_{\mathrm{33\,GHz}}$): 33\,GHz radio flux from high resolution (0\secpoint06), archival VLA A-array data; 3-$\sigma$ limits indicated for objects 1 and 2.
($E(B-V)$): Color excess estimated with optical Balmer lines in a 1\secpoint5 diameter aperture, assuming \citet{Cardelli:1989:245} Galactic attenuation.
(SFR): Extended SFR estimated from archival VLA C-array data within a 1\secpoint5 aperture.
($\sigma_{\star}$): Stellar velocity dispersion measured from Ca~II absorption lines utilizing 1D spectra extracted from 1\secpoint5 apertures.
($M_{\mathrm{BH}}$): Black hole mass (assuming a SMBH does exist) estimated from the $M_{\mathrm{BH}}-\sigma_{\star}$ relation \citep{Kormendy:2013:511}. Note that none of the three cores have clear evidence of BH activity.
($M_{\star}$): Stellar mass estimated from three bands of HST images (1\secpoint5 aperture); typical uncertainty is $\sim30\%$ primarily from $M/L$ calibration \citep{Zibetti2009MNRAS}.\\
$\dagger$ Object~2 may not follow the standard $M_\mathrm{BH}-\sigma$ relation due to tidal disruption.
}
\end{deluxetable*}

\begin{deluxetable*}{llccc}
\tablecaption{Emission Line Measurements \label{tab:line_fluxes_nist}}
\tablehead{
\colhead{Line ID} & 
\colhead{$\lambda_\mathrm{vac}$} & 
\colhead{Object 1} & 
\colhead{Object 2} & 
\colhead{Object 3} \\
\colhead{} & 
\colhead{(\AA\ or $\mu$m)} & 
\colhead{} & 
\colhead{$10^{-17}$ erg s$^{-1}$ cm$^{-2}$} & 
\colhead{}
}
\startdata
\sidehead{\textbf{Optical/UV Lines}}
\oii$^{*}$       & 3727, 3729 \AA    & $367\pm8$ & $472\pm10$ & $546\pm11$ \\
\hb\            & 4862.68 \AA    & $620\pm1$ & $605\pm1$ & $769\pm1$ \\
\oiii\          & 4960.30 \AA    & $41.4\pm0.5$ & $34.4\pm0.5$ & $130\pm1$ \\
\oiii\          & 5008.24 \AA    & $125.7\pm0.6$ & $107.0\pm0.5$ & $402\pm1$ \\
\oi             & 6302.05 \AA    & $35.1\pm0.3$  & $43.8\pm0.3$ & $118\pm1$ \\
\nii\           & 6549.85 \AA    & $535.9\pm0.6$ & $391.1\pm0.5$ & $1341\pm4$ \\
\ha\            & 6564.61 \AA    & $3995.1\pm0.9$ & $2975.4\pm0.8$ & $4966\pm5$ \\
\nii\           & 6585.28 \AA    & $1632\pm1$ & $1150\pm1$ & $3414\pm2$ \\
\sii\           & 6718.29 \AA    & $450.3\pm0.5$ & $405.3\pm0.5$ & $577\pm1$ \\
\sii\           & 6732.67 \AA    & $410.8\pm0.5$ & $328.3\pm0.5$ & $561\pm1$ \\
\oii$^{*}$           & 7322, 7332 \AA    & $28\pm6$ & $21\pm5$ & $163\pm22$ \\
%\oii\           &  \AA    & \dots & \dots & \dots \\
\sidehead{\textbf{Near-Infrared Lines}}
Br$\delta$      & 1.9451 $\mu$m  & $66.3\pm0.2$ & \nodata & $173.0\pm0.1$ \\
\hei\           & 2.0587 $\mu$m  & $52.3\pm0.2$ & \nodata & $125.1\pm0.1$ \\
Br$\gamma$       & 2.1661 $\mu$m  & $125^{+1}_{-5}$ & \nodata & $301.9\pm0.2$ \\
%\sidehead{\textbf{Molecular Hydrogen ($H_2$)}}
H$_2$ 1-0 S(3)  & 1.9576 \micron & $23.2\pm0.1$&\nodata&$211.1\pm0.1$\\
H$_2$ 1-0 S(2)  & 2.0338 $\mu$m  & $7.4\pm0.1$ & \nodata& $80.7\pm0.1$ \\
H$_2$ 2-1 S(3)  & 2.0735 $\mu$m  & $2.3^{+0.9}_{-2.0}$ & \nodata & $26.8\pm0.1$ \\
H$_2$ 1-0 S(1)  & 2.1218 $\mu$m  & $16.8\pm0.1$ & \nodata & $205.9\pm0.1$ \\
H$_2$ 1-0 S(0)  & 2.2233 $\mu$m  & $12.8\pm0.2$ & \nodata & $57.9\pm0.1$ \\
H$_2$ 2-1 S(1)  & 2.2477 $\mu$m  & $20.1\pm0.2$ & \nodata & $40.9\pm0.1$ \\
\enddata
\tablecomments{Observed flux of all major nebular lines extracted from 1\secpoint5 diameter aperture. Vacuum wavelengths and energy levels are sourced from the NIST Atomic Spectra Database and \citet{Dabrowski1984CaJPh}. Doublet components are reported individually to facilitate line ratio diagnostics.}
$^{*}$The doublets of \oii\ are unresolved, thus the flux is estimated by integrating over the entire continuum subtracted profile.
\end{deluxetable*}

\bibliography{bibfinal,bib_add_here}{}

@ARTICLE{Fukushima2024A&A,
       author = {{Fukushima}, K. and {Kobayashi}, S.~B. and {Matsushita}, K.},
        title = "{Revisiting the abundance pattern and charge-exchange emission in the centre of M 82}",
      journal = {\aap},
         year = 2024,
        month = jun,
       volume = {686},
          eid = {A96},
        pages = {A96},
          doi = {10.1051/0004-6361/202349064},
archivePrefix = {arXiv},
       eprint = {2403.12932},
 primaryClass = {astro-ph.GA},
       adsurl = {https://ui.adsabs.harvard.edu/abs/2024A&A...686A..96F}
}

@ARTICLE{pyneb2015A&A...573A..42L,
       author = {{Luridiana}, V. and {Morisset}, C. and {Shaw}, R.~A.},
        title = "{PyNeb: a new tool for analyzing emission lines. I. Code description and validation of results}",
      journal = {\aap},
         year = 2015,
        month = jan,
       volume = {573},
          eid = {A42},
        pages = {A42},
          doi = {10.1051/0004-6361/201323152},
archivePrefix = {arXiv},
       eprint = {1410.6662},
 primaryClass = {astro-ph.IM},
       adsurl = {https://ui.adsabs.harvard.edu/abs/2015A&A...573A..42L}
}

@ARTICLE{pyneb2020Atoms...8...66M,
       author = {{Morisset}, Christophe and {Luridiana}, Valentina and {Garc{\'\i}a-Rojas}, Jorge and {G{\'o}mez-Llanos}, Ver{\'o}nica and {Bautista}, Manuel and {Mendoza}, Claudio},
        title = "{Atomic Data Assessment with PyNeb}",
      journal = {Atoms},
         year = 2020,
        month = oct,
       volume = {8},
       number = {4},
          eid = {66},
        pages = {66},
          doi = {10.3390/atoms8040066},
archivePrefix = {arXiv},
       eprint = {2009.10586},
 primaryClass = {astro-ph.GA},
       adsurl = {https://ui.adsabs.harvard.edu/abs/2020Atoms...8...66M}
}

@ARTICLE{pyneb2023Atoms..11...63M,
       author = {{Mendoza}, Claudio and {M{\'e}ndez-Delgado}, Jos{\'e} E. and {Bautista}, Manuel and {Garc{\'\i}a-Rojas}, Jorge and {Morisset}, Christophe},
        title = "{Atomic Data Assessment with PyNeb: Radiative and Electron Impact Excitation Rates for [Fe II] and [Fe III]}",
      journal = {Atoms},
         year = 2023,
        month = apr,
       volume = {11},
       number = {4},
          eid = {63},
        pages = {63},
          doi = {10.3390/atoms11040063},
archivePrefix = {arXiv},
       eprint = {2304.01298},
 primaryClass = {astro-ph.GA},
       adsurl = {https://ui.adsabs.harvard.edu/abs/2023Atoms..11...63M}
}

@ARTICLE{Mouhcine2002A&A,
       author = {{Mouhcine}, M. and {Contini}, T.},
        title = "{Chemical evolution of starburst galaxies: How does star formation proceed?}",
      journal = {\aap},
         year = 2002,
        month = jul,
       volume = {389},
        pages = {106-114},
          doi = {10.1051/0004-6361:20020592},
archivePrefix = {arXiv},
       eprint = {astro-ph/0204424},
 primaryClass = {astro-ph},
       adsurl = {https://ui.adsabs.harvard.edu/abs/2002A&A...389..106M}
}

@software{ds9_2000ascl.soft03002S,
       author = {{Smithsonian Astrophysical Observatory}},
        title = "{SAOImage DS9: A utility for displaying astronomical images in the X11 window environment}",
 howpublished = {Astrophysics Source Code Library, record ascl:0003.002},
         year = 2000,
        month = mar,
          eid = {ascl:0003.002},
archivePrefix = {ascl},
       eprint = {0003.002},
       adsurl = {https://ui.adsabs.harvard.edu/abs/2000ascl.soft03002S}
}

@software{photutils_larry_bradley_2025_14889440,
  author       = {Larry Bradley and
                  Brigitta Sip{\H o}cz and
                  Thomas Robitaille and
                  Erik Tollerud and
                  Z\`e Vin{\'{\i}}cius and
                  Christoph Deil and
                  Kyle Barbary and
                  Tom J Wilson and
                  Ivo Busko and
                  Axel Donath and
                  Hans Moritz G{\"u}nther and
                  Mihai Cara and
                  P. L. Lim and
                  Sebastian Me{\ss}linger and
                  Zach Burnett and
                  Simon Conseil and
                  Michael Droettboom and
                  Azalee Bostroem and
                  E. M. Bray and
                  Lars Andersen Bratholm and
                  William Jamieson and
                  Adam Ginsburg and
                  Geert Barentsen and
                  Matt Craig and
                  Sergio Pascual and
                  Shivangee Rathi and
                  Marshall Perrin and
                  Brett M. Morris},
  title        = {astropy/photutils: 2.2.0},
  month        = feb,
  year         = 2025,
  publisher    = {Zenodo},
  version      = {2.2.0},
  doi          = {10.5281/zenodo.14889440},
  url          = {https://doi.org/10.5281/zenodo.14889440},
  swhid        = {swh:1:dir:11159107f27a28985192ed1118b1f2055709d093
                   ;origin=https://doi.org/10.5281/zenodo.596036;visi
                   t=swh:1:snp:ae8c4a55d349d43e53cfe9ce92e678fcfe840f
                   3b;anchor=swh:1:rel:0117f67e8888adcdfc85308287dd9c
                   854b466389;path=astropy-photutils-ffb96c5
                  },
}

@ARTICLE{Astropy2013A&A...558A..33A,
       author = {{Astropy Collaboration} and {Robitaille}, Thomas P. and {Tollerud}, Erik J. and {Greenfield}, Perry and {Droettboom}, Michael and {Bray}, Erik and {Aldcroft}, Tom and {Davis}, Matt and {Ginsburg}, Adam and {Price-Whelan}, Adrian M. and {Kerzendorf}, Wolfgang E. and {Conley}, Alexander and {Crighton}, Neil and {Barbary}, Kyle and {Muna}, Demitri and {Ferguson}, Henry and {Grollier}, Fr{\'e}d{\'e}ric and {Parikh}, Madhura M. and {Nair}, Prasanth H. and {Unther}, Hans M. and {Deil}, Christoph and {Woillez}, Julien and {Conseil}, Simon and {Kramer}, Roban and {Turner}, James E.~H. and {Singer}, Leo and {Fox}, Ryan and {Weaver}, Benjamin A. and {Zabalza}, Victor and {Edwards}, Zachary I. and {Azalee Bostroem}, K. and {Burke}, D.~J. and {Casey}, Andrew R. and {Crawford}, Steven M. and {Dencheva}, Nadia and {Ely}, Justin and {Jenness}, Tim and {Labrie}, Kathleen and {Lim}, Pey Lian and {Pierfederici}, Francesco and {Pontzen}, Andrew and {Ptak}, Andy and {Refsdal}, Brian and {Servillat}, Mathieu and {Streicher}, Ole},
        title = "{Astropy: A community Python package for astronomy}",
      journal = {\aap},
         year = 2013,
        month = oct,
       volume = {558},
          eid = {A33},
        pages = {A33},
          doi = {10.1051/0004-6361/201322068},
archivePrefix = {arXiv},
       eprint = {1307.6212},
 primaryClass = {astro-ph.IM},
       adsurl = {https://ui.adsabs.harvard.edu/abs/2013A&A...558A..33A}
}

@ARTICLE{Astropy2018AJ....156..123A,
       author = {{Astropy Collaboration} and {Price-Whelan}, A.~M. and {Sip{\H{o}}cz}, B.~M. and {G{\"u}nther}, H.~M. and {Lim}, P.~L. and {Crawford}, S.~M. and {Conseil}, S. and {Shupe}, D.~L. and {Craig}, M.~W. and {Dencheva}, N. and {Ginsburg}, A. and {VanderPlas}, J.~T. and {Bradley}, L.~D. and {P{\'e}rez-Su{\'a}rez}, D. and {de Val-Borro}, M. and {Aldcroft}, T.~L. and {Cruz}, K.~L. and {Robitaille}, T.~P. and {Tollerud}, E.~J. and {Ardelean}, C. and {Babej}, T. and {Bach}, Y.~P. and {Bachetti}, M. and {Bakanov}, A.~V. and {Bamford}, S.~P. and {Barentsen}, G. and {Barmby}, P. and {Baumbach}, A. and {Berry}, K.~L. and {Biscani}, F. and {Boquien}, M. and {Bostroem}, K.~A. and {Bouma}, L.~G. and {Brammer}, G.~B. and {Bray}, E.~M. and {Breytenbach}, H. and {Buddelmeijer}, H. and {Burke}, D.~J. and {Calderone}, G. and {Cano Rodr{\'\i}guez}, J.~L. and {Cara}, M. and {Cardoso}, J.~V.~M. and {Cheedella}, S. and {Copin}, Y. and {Corrales}, L. and {Crichton}, D. and {D'Avella}, D. and {Deil}, C. and {Depagne}, {\'E}. and {Dietrich}, J.~P. and {Donath}, A. and {Droettboom}, M. and {Earl}, N. and {Erben}, T. and {Fabbro}, S. and {Ferreira}, L.~A. and {Finethy}, T. and {Fox}, R.~T. and {Garrison}, L.~H. and {Gibbons}, S.~L.~J. and {Goldstein}, D.~A. and {Gommers}, R. and {Greco}, J.~P. and {Greenfield}, P. and {Groener}, A.~M. and {Grollier}, F. and {Hagen}, A. and {Hirst}, P. and {Homeier}, D. and {Horton}, A.~J. and {Hosseinzadeh}, G. and {Hu}, L. and {Hunkeler}, J.~S. and {Ivezi{\'c}}, {\v{Z}}. and {Jain}, A. and {Jenness}, T. and {Kanarek}, G. and {Kendrew}, S. and {Kern}, N.~S. and {Kerzendorf}, W.~E. and {Khvalko}, A. and {King}, J. and {Kirkby}, D. and {Kulkarni}, A.~M. and {Kumar}, A. and {Lee}, A. and {Lenz}, D. and {Littlefair}, S.~P. and {Ma}, Z. and {Macleod}, D.~M. and {Mastropietro}, M. and {McCully}, C. and {Montagnac}, S. and {Morris}, B.~M. and {Mueller}, M. and {Mumford}, S.~J. and {Muna}, D. and {Murphy}, N.~A. and {Nelson}, S. and {Nguyen}, G.~H. and {Ninan}, J.~P. and {N{\"o}the}, M. and {Ogaz}, S. and {Oh}, S. and {Parejko}, J.~K. and {Parley}, N. and {Pascual}, S. and {Patil}, R. and {Patil}, A.~A. and {Plunkett}, A.~L. and {Prochaska}, J.~X. and {Rastogi}, T. and {Reddy Janga}, V. and {Sabater}, J. and {Sakurikar}, P. and {Seifert}, M. and {Sherbert}, L.~E. and {Sherwood-Taylor}, H. and {Shih}, A.~Y. and {Sick}, J. and {Silbiger}, M.~T. and {Singanamalla}, S. and {Singer}, L.~P. and {Sladen}, P.~H. and {Sooley}, K.~A. and {Sornarajah}, S. and {Streicher}, O. and {Teuben}, P. and {Thomas}, S.~W. and {Tremblay}, G.~R. and {Turner}, J.~E.~H. and {Terr{\'o}n}, V. and {van Kerkwijk}, M.~H. and {de la Vega}, A. and {Watkins}, L.~L. and {Weaver}, B.~A. and {Whitmore}, J.~B. and {Woillez}, J. and {Zabalza}, V. and {Astropy Contributors}},
        title = "{The Astropy Project: Building an Open-science Project and Status of the v2.0 Core Package}",
      journal = {\aj},
         year = 2018,
        month = sep,
       volume = {156},
       number = {3},
          eid = {123},
        pages = {123},
          doi = {10.3847/1538-3881/aabc4f},
archivePrefix = {arXiv},
       eprint = {1801.02634},
 primaryClass = {astro-ph.IM},
       adsurl = {https://ui.adsabs.harvard.edu/abs/2018AJ....156..123A}
}

@ARTICLE{matplotlib_2007CSE.....9...90H,
       author = {{Hunter}, John D.},
        title = "{Matplotlib: A 2D Graphics Environment}",
      journal = {Computing in Science and Engineering},
         year = 2007,
        month = jan,
       volume = {9},
       number = {3},
        pages = {90-95},
          doi = {10.1109/MCSE.2007.55},
       adsurl = {https://ui.adsabs.harvard.edu/abs/2007CSE.....9...90H}
}

@ARTICLE{numpy_2020Natur.585..357H,
       author = {{Harris}, Charles R. and {Millman}, K. Jarrod and {van der Walt}, St{\'e}fan J. and {Gommers}, Ralf and {Virtanen}, Pauli and {Cournapeau}, David and {Wieser}, Eric and {Taylor}, Julian and {Berg}, Sebastian and {Smith}, Nathaniel J. and {Kern}, Robert and {Picus}, Matti and {Hoyer}, Stephan and {van Kerkwijk}, Marten H. and {Brett}, Matthew and {Haldane}, Allan and {del R{\'\i}o}, Jaime Fern{\'a}ndez and {Wiebe}, Mark and {Peterson}, Pearu and {G{\'e}rard-Marchant}, Pierre and {Sheppard}, Kevin and {Reddy}, Tyler and {Weckesser}, Warren and {Abbasi}, Hameer and {Gohlke}, Christoph and {Oliphant}, Travis E.},
        title = "{Array programming with NumPy}",
      journal = {\nat},
         year = 2020,
        month = sep,
       volume = {585},
       number = {7825},
        pages = {357-362},
          doi = {10.1038/s41586-020-2649-2},
archivePrefix = {arXiv},
       eprint = {2006.10256},
 primaryClass = {cs.MS},
       adsurl = {https://ui.adsabs.harvard.edu/abs/2020Natur.585..357H}
}

@BOOK{Draine2011piim,
       author = {{Draine}, Bruce T.},
        title = "{Physics of the Interstellar and Intergalactic Medium}",
         year = 2011,
       adsurl = {https://ui.adsabs.harvard.edu/abs/2011piim.book.....D}
}

@ARTICLE{Coziol1999A&A,
       author = {{Coziol}, R. and {Reyes}, R.~E. Carlos and {Consid{\`e}re}, S. and {Davoust}, E. and {Contini}, T.},
        title = "{The abundance of nitrogen in starburst nucleus galaxies}",
      journal = {\aap},
         year = 1999,
        month = may,
       volume = {345},
        pages = {733-746},
          doi = {10.48550/arXiv.astro-ph/9903279},
archivePrefix = {arXiv},
       eprint = {astro-ph/9903279},
 primaryClass = {astro-ph},
       adsurl = {https://ui.adsabs.harvard.edu/abs/1999A&A...345..733C}
}

@ARTICLE{Dabrowski1984CaJPh,
       author = {{Dabrowski}, I.},
        title = "{The Lyman and Werner bands of H$_{2}$}",
      journal = {Canadian Journal of Physics},
         year = 1984,
        month = dec,
       volume = {62},
       number = {12},
        pages = {1639-1664},
          doi = {10.1139/p84-210},
       adsurl = {https://ui.adsabs.harvard.edu/abs/1984CaJPh..62.1639D}
}

@ARTICLE{Trainor2012ApJ,
       author = {{Trainor}, Ryan F. and {Steidel}, Charles C.},
        title = "{The Halo Masses and Galaxy Environments of Hyperluminous QSOs at z \raisebox{-0.5ex}\textasciitilde= 2.7 in the Keck Baryonic Structure Survey}",
      journal = {\apj},
         year = 2012,
        month = jun,
       volume = {752},
       number = {1},
          eid = {39},
        pages = {39},
          doi = {10.1088/0004-637X/752/1/39},
archivePrefix = {arXiv},
       eprint = {1204.3636},
 primaryClass = {astro-ph.CO},
       adsurl = {https://ui.adsabs.harvard.edu/abs/2012ApJ...752...39T}
}

@ARTICLE{Smith1998ApJ,
       author = {{Smith}, Harding E. and {Lonsdale}, Colin J. and {Lonsdale}, Carol J.},
        title = "{The Starburst-AGN Connection. II. The Nature of Luminous Infrared Galaxies as Revealed by VLBI, VLA, Infrared, and Optical Observations}",
      journal = {\apj},
         year = 1998,
        month = jan,
       volume = {492},
       number = {1},
        pages = {137-172},
          doi = {10.1086/305020},
       adsurl = {https://ui.adsabs.harvard.edu/abs/1998ApJ...492..137S}
}

@ARTICLE{Momcheva2013AJ,
       author = {{Momcheva}, Ivelina G. and {Lee}, Janice C. and {Ly}, Chun and {Salim}, Samir and {Dale}, Daniel A. and {Ouchi}, Masami and {Finn}, Rose and {Ono}, Yoshiaki},
        title = "{Nebular Attenuation in H{\ensuremath{\alpha}}-selected Star-forming Galaxies at z = 0.8 from the NewH{\ensuremath{\alpha}} Survey}",
      journal = {\aj},
         year = 2013,
        month = feb,
       volume = {145},
       number = {2},
          eid = {47},
        pages = {47},
          doi = {10.1088/0004-6256/145/2/47},
archivePrefix = {arXiv},
       eprint = {1207.5479},
 primaryClass = {astro-ph.CO},
       adsurl = {https://ui.adsabs.harvard.edu/abs/2013AJ....145...47M}
}

@BOOK{Osterbrock2006agna.book,
       author = {{Osterbrock}, Donald E. and {Ferland}, Gary J.},
        title = "{Astrophysics of gaseous nebulae and active galactic nuclei}",
         year = 2006,
       adsurl = {https://ui.adsabs.harvard.edu/abs/2006agna.book.....O},
      publisher = {University Science Books}
}

@ARTICLE{Chandrasekhar1943,
       author = {{Chandrasekhar}, S.},
        title = "{Dynamical Friction. I. General Considerations: the Coefficient of Dynamical Friction.}",
      journal = {\apj},
         year = 1943,
        month = mar,
       volume = {97},
        pages = {255},
          doi = {10.1086/144517},
       adsurl = {https://ui.adsabs.harvard.edu/abs/1943ApJ....97..255C}
}

@book{mo_galaxy_2012,
	address = {Cambridge},
	title = {Galaxy formation and evolution},
	isbn = {978-0-511-80724-4},
	language = {eng},
	publisher = {Cambridge University Press},
	author = {Mo, Houjun and Van den Bosch, Frank and White, Simon D. M.},
	year = {2012},
	doi = {10.1017/CBO9780511807244},
}

@ARTICLE{Rupke2005ApJ,
       author = {{Rupke}, David S. and {Veilleux}, Sylvain and {Sanders}, D.~B.},
        title = "{Outflows in Active Galactic Nucleus/Starburst-Composite Ultraluminous Infrared Galaxies1,}",
      journal = {\apj},
         year = 2005,
        month = oct,
       volume = {632},
       number = {2},
        pages = {751-780},
          doi = {10.1086/444451},
archivePrefix = {arXiv},
       eprint = {astro-ph/0507037},
 primaryClass = {astro-ph},
       adsurl = {https://ui.adsabs.harvard.edu/abs/2005ApJ...632..751R}
}

@ARTICLE{Harrison2014,
       author = {{Harrison}, C.~M. and {Alexander}, D.~M. and {Mullaney}, J.~R. and {Swinbank}, A.~M.},
        title = "{Kiloparsec-scale outflows are prevalent among luminous AGN: outflows and feedback in the context of the overall AGN population}",
      journal = {\mnras},
         year = 2014,
        month = jul,
       volume = {441},
       number = {4},
        pages = {3306-3347},
          doi = {10.1093/mnras/stu515},
archivePrefix = {arXiv},
       eprint = {1403.3086},
 primaryClass = {astro-ph.GA},
       adsurl = {https://ui.adsabs.harvard.edu/abs/2014MNRAS.441.3306H}
}

@ARTICLE{Shih2010ApJ,
       author = {{Shih}, Hsin-Yi and {Rupke}, David S.~N.},
        title = "{The Complex Structure of the Multi-phase Galactic Wind in a Starburst Merger}",
      journal = {\apj},
         year = 2010,
        month = dec,
       volume = {724},
       number = {2},
        pages = {1430-1440},
          doi = {10.1088/0004-637X/724/2/1430},
archivePrefix = {arXiv},
       eprint = {1009.6020},
 primaryClass = {astro-ph.CO},
       adsurl = {https://ui.adsabs.harvard.edu/abs/2010ApJ...724.1430S}
}

@ARTICLE{Leitherer1995,
       author = {{Leitherer}, Claus and {Heckman}, Timothy M.},
        title = "{Synthetic Properties of Starburst Galaxies}",
      journal = {\apjs},
         year = 1995,
        month = jan,
       volume = {96},
        pages = {9},
          doi = {10.1086/192112},
       adsurl = {https://ui.adsabs.harvard.edu/abs/1995ApJS...96....9L}
}

@ARTICLE{Veilleux2005ARA&A,
       author = {{Veilleux}, Sylvain and {Cecil}, Gerald and {Bland-Hawthorn}, Joss},
        title = "{Galactic Winds}",
      journal = {\araa},
         year = 2005,
        month = sep,
       volume = {43},
       number = {1},
        pages = {769-826},
          doi = {10.1146/annurev.astro.43.072103.150610},
archivePrefix = {arXiv},
       eprint = {astro-ph/0504435},
 primaryClass = {astro-ph},
       adsurl = {https://ui.adsabs.harvard.edu/abs/2005ARA&A..43..769V}
}

@BOOK{Binney2008gady.book,
       author = {{Binney}, James and {Tremaine}, Scott},
        title = "{Galactic Dynamics: Second Edition}",
         year = 2008,
       adsurl = {https://ui.adsabs.harvard.edu/abs/2008gady.book.....B},
    publisher = {Princeton University Press}
}

@ARTICLE{Greene2011ApJ,
       author = {{Greene}, Jenny E. and {Zakamska}, Nadia L. and {Ho}, Luis C. and {Barth}, Aaron J.},
        title = "{Feedback in Luminous Obscured Quasars}",
      journal = {\apj},
         year = 2011,
        month = may,
       volume = {732},
       number = {1},
          eid = {9},
        pages = {9},
          doi = {10.1088/0004-637X/732/1/9},
archivePrefix = {arXiv},
       eprint = {1102.2913},
 primaryClass = {astro-ph.CO},
       adsurl = {https://ui.adsabs.harvard.edu/abs/2011ApJ...732....9G}
}

@ARTICLE{Rodriguez2013MNRAS,
       author = {{Rodr{\'\i}guez Zaur{\'\i}n}, J. and {Tadhunter}, C.~N. and {Rose}, M. and {Holt}, J.},
        title = "{The importance of warm, AGN-driven outflows in the nuclear regions of nearby ULIRGs}",
      journal = {\mnras},
         year = 2013,
        month = jun,
       volume = {432},
       number = {1},
        pages = {138-166},
          doi = {10.1093/mnras/stt423},
archivePrefix = {arXiv},
       eprint = {1303.1400},
 primaryClass = {astro-ph.CO},
       adsurl = {https://ui.adsabs.harvard.edu/abs/2013MNRAS.432..138R}
}

@ARTICLE{KCWI,
       author = {{Morrissey}, Patrick and {Matuszewski}, Matuesz and {Martin}, D. Christopher and {Neill}, James D. and {Epps}, Harland and {Fucik}, Jason and {Weber}, Bob and {Darvish}, Behnam and {Adkins}, Sean and {Allen}, Steve and {Bartos}, Randy and {Belicki}, Justin and {Cabak}, Jerry and {Callahan}, Shawn and {Cowley}, Dave and {Crabill}, Marty and {Deich}, Willian and {Delecroix}, Alex and {Doppman}, Greg and {Hilyard}, David and {James}, Ean and {Kaye}, Steve and {Kokorowski}, Michael and {Kwok}, Shui and {Lanclos}, Kyle and {Milner}, Steve and {Moore}, Anna and {O'Sullivan}, Donal and {Parihar}, Prachi and {Park}, Sam and {Phillips}, Andrew and {Rizzi}, Luca and {Rockosi}, Constance and {Rodriguez}, Hector and {Salaun}, Yves and {Seaman}, Kirk and {Sheikh}, David and {Weiss}, Jason and {Zarzaca}, Ray},
        title = "{The Keck Cosmic Web Imager Integral Field Spectrograph}",
      journal = {\apj},
         year = 2018,
        month = sep,
       volume = {864},
       number = {1},
          eid = {93},
        pages = {93},
          doi = {10.3847/1538-4357/aad597},
archivePrefix = {arXiv},
       eprint = {1807.10356},
 primaryClass = {astro-ph.IM},
       adsurl = {https://ui.adsabs.harvard.edu/abs/2018ApJ...864...93M}
}

@ARTICLE{1993ApJ...405L..63G,
       author = {{Guedel}, Manuel and {Benz}, Arnold O.},
        title = "{X-Ray/Microwave Relation of Different Types of Active Stars}",
      journal = {\apjl},
         year = 1993,
        month = mar,
       volume = {405},
        pages = {L63},
          doi = {10.1086/186766},
       adsurl = {https://ui.adsabs.harvard.edu/abs/1993ApJ...405L..63G}
}

@ARTICLE{Cazzoli2016A&A,
       author = {{Cazzoli}, S. and {Arribas}, S. and {Maiolino}, R. and {Colina}, L.},
        title = "{Neutral gas outflows in nearby [U]LIRGs via optical NaD feature}",
      journal = {\aap},
         year = 2016,
        month = may,
       volume = {590},
          eid = {A125},
        pages = {A125},
          doi = {10.1051/0004-6361/201526788},
archivePrefix = {arXiv},
       eprint = {1602.08505},
 primaryClass = {astro-ph.GA},
       adsurl = {https://ui.adsabs.harvard.edu/abs/2016A&A...590A.125C}
}

@misc{KcwiKit,
       author = {{Prusinski}, Nikolaus Z. and {Chen}, Yuguang},
        title = "{KCWIKit: KCWI Post-Processing and Improvements}",
 howpublished = {Astrophysics Source Code Library, record ascl:2404.003},
         year = 2024,
        month = apr,
          eid = {ascl:2404.003},
       adsurl = {https://ui.adsabs.harvard.edu/abs/2024ascl.soft04003P}
}

@ARTICLE{C21,
       author = {{Chen}, Yuguang and {Steidel}, Charles C. and {Erb}, Dawn K. and {Law}, David R. and {Trainor}, Ryan F. and {Reddy}, Naveen A. and {Shapley}, Alice E. and {Pahl}, Anthony J. and {Strom}, Allison L. and {Lamb}, Noah R. and {Li}, Zhihui and {Rudie}, Gwen C.},
        title = "{The KBSS-KCWI survey: the connection between extended Ly {\ensuremath{\alpha}} haloes and galaxy azimuthal angle at z   2-3}",
      journal = {\mnras},
         year = 2021,
        month = nov,
       volume = {508},
       number = {1},
        pages = {19-43},
          doi = {10.1093/mnras/stab2383},
archivePrefix = {arXiv},
       eprint = {2104.10173},
 primaryClass = {astro-ph.GA},
       adsurl = {https://ui.adsabs.harvard.edu/abs/2021MNRAS.508...19C}
}

@article{pypeit:joss_pub,
    doi = {10.21105/joss.02308},
    url = {https://doi.org/10.21105/joss.02308},
    year = {2020},
    publisher = {The Open Journal},
    volume = {5},
    number = {56},
    pages = {2308},
    author = {{Prochaska}, J. Xavier and Joseph F. Hennawi and Kyle B. Westfall and Ryan J. Cooke and Feige Wang and Tiffany Hsyu and Frederick B. Davies and Emanuele Paolo Farina and Debora Pelliccia},
    title = {PypeIt: The Python Spectroscopic Data Reduction Pipeline},
    journal = {Journal of Open Source Software}
}

@MISC{pypeit:zenodo,
       author = {{Prochaska}, J. Xavier and {Hennawi}, Joseph and {Cooke}, Ryan and
         {Westfall}, Kyle and {Wang}, Feige and {EmAstro} and {Tiffanyhsyu} and
         {Wasserman}, Asher and {Villaume}, Alexa and {Marijana777} and
         {Schindler}, JT and {Young}, David and {Simha}, Sunil and
         {Wilde}, Matt and {Tejos}, Nicolas and {Isbell}, Jacob and
         {Fl{\"o}rs}, Andreas and {Sandford}, Nathan and {Vasovi{\'c}}, Zlatan and
         {Betts}, Edward and {Holden}, Brad},
        title = "{pypeit/PypeIt: Release 1.0.0}",
         year = 2020,
        month = apr,
          eid = {10.5281/zenodo.3743493},
          doi = {10.5281/zenodo.3743493},
      version = {v1.0.0},
    publisher = {Zenodo},
       adsurl = {https://ui.adsabs.harvard.edu/abs/2020zndo...3743493P}
}

@ARTICLE{2016MNRAS.458.3210S,
       author = {{Soto}, Kurt T. and {Lilly}, Simon J. and {Bacon}, Roland and {Richard}, Johan and {Conseil}, Simon},
        title = "{ZAP - enhanced PCA sky subtraction for integral field spectroscopy}",
      journal = {\mnras},
         year = 2016,
        month = may,
       volume = {458},
       number = {3},
        pages = {3210-3220},
          doi = {10.1093/mnras/stw474},
archivePrefix = {arXiv},
       eprint = {1602.08037},
 primaryClass = {astro-ph.IM},
       adsurl = {https://ui.adsabs.harvard.edu/abs/2016MNRAS.458.3210S}
}

@ARTICLE{Baron2022MNRAS.509.4457B,
       author = {{Baron}, Dalya and {Netzer}, Hagai and {Lutz}, Dieter and {Prochaska}, J. Xavier and {Davies}, Ric I.},
        title = "{Multiphase outflows in post-starburst E+A galaxies - I. General sample properties and the prevalence of obscured starbursts}",
      journal = {\mnras},
         year = 2022,
        month = jan,
       volume = {509},
       number = {3},
        pages = {4457-4479},
          doi = {10.1093/mnras/stab3232},
archivePrefix = {arXiv},
       eprint = {2105.08071},
 primaryClass = {astro-ph.GA},
       adsurl = {https://ui.adsabs.harvard.edu/abs/2022MNRAS.509.4457B}
}

@ARTICLE{Cappellari2023,
       author = {{Cappellari}, Michele},
        title = "{Full spectrum fitting with photometry in PPXF: stellar population versus dynamical masses, non-parametric star formation history and metallicity for 3200 LEGA-C galaxies at redshift z {\ensuremath{\approx}} 0.8}",
      journal = {\mnras},
         year = 2023,
        month = dec,
       volume = {526},
       number = {3},
        pages = {3273-3300},
          doi = {10.1093/mnras/stad2597},
archivePrefix = {arXiv},
       eprint = {2208.14974},
 primaryClass = {astro-ph.GA},
       adsurl = {https://ui.adsabs.harvard.edu/abs/2023MNRAS.526.3273C}
}

@INPROCEEDINGS{Roper2010AAS,
       author = {{Roper}, Brian W. and {Kim}, D. and {Evans}, A.},
        title = "{GOALS Observations of Star Formation and Nuclear Activity in the Luminous Infrared Galaxy UGC 2369}",
    booktitle = {American Astronomical Society Meeting Abstracts \#215},
         year = 2010,
       series = {American Astronomical Society Meeting Abstracts},
       volume = {215},
        month = jan,
          eid = {602.03},
        pages = {602.03},
       adsurl = {https://ui.adsabs.harvard.edu/abs/2010AAS...21560203R}
}

@ARTICLE{Rupke2005ApJS,
       author = {{Rupke}, David S. and {Veilleux}, Sylvain and {Sanders}, D.~B.},
        title = "{Outflows in Infrared-Luminous Starbursts at z < 0.5. II. Analysis and Discussion}",
      journal = {\apjs},
         year = 2005,
        month = sep,
       volume = {160},
       number = {1},
        pages = {115-148},
          doi = {10.1086/432889},
archivePrefix = {arXiv},
       eprint = {astro-ph/0506611},
 primaryClass = {astro-ph},
       adsurl = {https://ui.adsabs.harvard.edu/abs/2005ApJS..160..115R}
}

@ARTICLE{Soifer1987ApJ,
       author = {{Soifer}, B.~T. and {Sanders}, D.~B. and {Madore}, B.~F. and {Neugebauer}, G. and {Danielson}, G.~E. and {Elias}, J.~H. and {Lonsdale}, Carol J. and {Rice}, W.~L.},
        title = "{The IRAS Bright Galaxy Sample. II. The Sample and Luminosity Function}",
      journal = {\apj},
         year = 1987,
        month = sep,
       volume = {320},
        pages = {238},
          doi = {10.1086/165536},
       adsurl = {https://ui.adsabs.harvard.edu/abs/1987ApJ...320..238S}
}

@ARTICLE{Grajales-Medina2023,
       author = {{Grajales-Medina}, D. and {Argudo-Fern{\'a}ndez}, M. and {V{\'a}squez-Bustos}, P. and {Verley}, S. and {Boquien}, M. and {Salim}, S. and {Duarte Puertas}, S. and {Lisenfeld}, U. and {Espada}, D. and {Salas-Olave}, H.},
        title = "{SIT 45: An interacting, compact, and star-forming isolated galaxy triplet}",
      journal = {\aap},
         year = 2023,
        month = jan,
       volume = {669},
          eid = {A23},
        pages = {A23},
          doi = {10.1051/0004-6361/202244492},
archivePrefix = {arXiv},
       eprint = {2209.12850},
 primaryClass = {astro-ph.GA},
       adsurl = {https://ui.adsabs.harvard.edu/abs/2023A&A...669A..23G}
}

@ARTICLE{Foord2021ApJ,
       author = {{Foord}, Adi and {G{\"u}ltekin}, Kayhan and {Runnoe}, Jessie C. and {Koss}, Michael J.},
        title = "{AGN Triality of Triple Mergers: Detection of Faint X-Ray Point Sources}",
      journal = {\apj},
         year = 2021,
        month = feb,
       volume = {907},
       number = {2},
          eid = {71},
        pages = {71},
          doi = {10.3847/1538-4357/abce5d},
archivePrefix = {arXiv},
       eprint = {2012.00761},
 primaryClass = {astro-ph.HE},
       adsurl = {https://ui.adsabs.harvard.edu/abs/2021ApJ...907...71F}
}

@ARTICLE{Koss2012ApJ,
       author = {{Koss}, Michael and {Mushotzky}, Richard and {Treister}, Ezequiel and {Veilleux}, Sylvain and {Vasudevan}, Ranjan and {Trippe}, Margaret},
        title = "{Understanding Dual Active Galactic Nucleus Activation in the nearby Universe}",
      journal = {\apjl},
         year = 2012,
        month = feb,
       volume = {746},
       number = {2},
          eid = {L22},
        pages = {L22},
          doi = {10.1088/2041-8205/746/2/L22},
archivePrefix = {arXiv},
       eprint = {1201.2944},
 primaryClass = {astro-ph.HE},
       adsurl = {https://ui.adsabs.harvard.edu/abs/2012ApJ...746L..22K}
}

@ARTICLE{King1962,
       author = {{King}, Ivan},
        title = "{The structure of star clusters. I. an empirical density law}",
      journal = {\aj},
         year = 1962,
        month = oct,
       volume = {67},
        pages = {471},
          doi = {10.1086/108756},
       adsurl = {https://ui.adsabs.harvard.edu/abs/1962AJ.....67..471K}
}

@ARTICLE{Koehn2023A&A,
       author = {{Koehn}, H. and {Just}, A. and {Berczik}, P. and {Tremmel}, M.},
        title = "{Dynamics of supermassive black hole triples in the ROMULUS25 cosmological simulation}",
      journal = {\aap},
         year = 2023,
        month = oct,
       volume = {678},
          eid = {A11},
        pages = {A11},
          doi = {10.1051/0004-6361/202347093},
archivePrefix = {arXiv},
       eprint = {2308.10894},
 primaryClass = {astro-ph.GA},
       adsurl = {https://ui.adsabs.harvard.edu/abs/2023A&A...678A..11K}
}

@ARTICLE{Yadav2021,
       author = {{Yadav}, Jyoti and {Das}, Mousumi and {Barway}, Sudhanshu and {Combes}, Francoise},
        title = "{A triple active galactic nucleus in the NGC 7733-7734 merging group}",
      journal = {\aap},
         year = 2021,
        month = jul,
       volume = {651},
          eid = {L9},
        pages = {L9},
          doi = {10.1051/0004-6361/202141210},
       adsurl = {https://ui.adsabs.harvard.edu/abs/2021A&A...651L...9Y}
}

@ARTICLE{Liu2013MNRASb,
       author = {{Liu}, Guilin and {Zakamska}, Nadia L. and {Greene}, Jenny E. and {Nesvadba}, Nicole P.~H. and {Liu}, Xin},
        title = "{Observations of feedback from radio-quiet quasars - II. Kinematics of ionized gas nebulae}",
      journal = {\mnras},
         year = 2013,
        month = dec,
       volume = {436},
       number = {3},
        pages = {2576-2597},
          doi = {10.1093/mnras/stt1755},
archivePrefix = {arXiv},
       eprint = {1305.6922},
 primaryClass = {astro-ph.CO},
       adsurl = {https://ui.adsabs.harvard.edu/abs/2013MNRAS.436.2576L}
}

@ARTICLE{Liu2013MNRASa,
       author = {{Liu}, Guilin and {Zakamska}, Nadia L. and {Greene}, Jenny E. and {Nesvadba}, Nicole P.~H. and {Liu}, Xin},
        title = "{Observations of feedback from radio-quiet quasars - I. Extents and morphologies of ionized gas nebulae}",
      journal = {\mnras},
         year = 2013,
        month = apr,
       volume = {430},
       number = {3},
        pages = {2327-2345},
          doi = {10.1093/mnras/stt051},
archivePrefix = {arXiv},
       eprint = {1301.1677},
 primaryClass = {astro-ph.CO},
       adsurl = {https://ui.adsabs.harvard.edu/abs/2013MNRAS.430.2327L}
}

@article{Hill2019, title={The SCUBA-2 web survey: I. Observations of CO(3–2) in hyper-luminous QSO fields}, volume={485}, ISSN={0035-8711}, DOI={10.1093/mnras/stz429}, number={1}, journal={Monthly Notices of the Royal Astronomical Society}, author={Hill, Ryley and Chapman, Scott C and Scott, Douglas and Smail, Ian and Steidel, Charles C and Krips, Melanie and Babul, Arif and Berg, Trystyn and Bertoldi, Frank and Gao, Yu and Lacaille, Kevin and Matsuda, Yuichi and Ross, Colin and Rudie, Gwen and Trainor, Ryan}, year={2019}, month=feb, pages={753–769} }

@ARTICLE{Prochaska2014ApJ,
       author = {{Prochaska}, J. Xavier and {Lau}, Marie Wingyee and {Hennawi}, Joseph F.},
        title = "{Quasars Probing Quasars. VII. The Pinnacle of the Cool Circumgalactic Medium Surrounds Massive z \raisebox{-0.5ex}\textasciitilde 2 Galaxies}",
      journal = {\apj},
         year = 2014,
        month = dec,
       volume = {796},
       number = {2},
          eid = {140},
        pages = {140},
          doi = {10.1088/0004-637X/796/2/140},
archivePrefix = {arXiv},
       eprint = {1409.6344},
 primaryClass = {astro-ph.GA},
       adsurl = {https://ui.adsabs.harvard.edu/abs/2014ApJ...796..140P}
}

@ARTICLE{Gebhardt2000ApJ,
       author = {{Gebhardt}, Karl and {Bender}, Ralf and {Bower}, Gary and {Dressler}, Alan and {Faber}, S.~M. and {Filippenko}, Alexei V. and {Green}, Richard and {Grillmair}, Carl and {Ho}, Luis C. and {Kormendy}, John and {Lauer}, Tod R. and {Magorrian}, John and {Pinkney}, Jason and {Richstone}, Douglas and {Tremaine}, Scott},
        title = "{A Relationship between Nuclear Black Hole Mass and Galaxy Velocity Dispersion}",
      journal = {\apjl},
         year = 2000,
        month = aug,
       volume = {539},
       number = {1},
        pages = {L13-L16},
          doi = {10.1086/312840},
archivePrefix = {arXiv},
       eprint = {astro-ph/0006289},
 primaryClass = {astro-ph},
       adsurl = {https://ui.adsabs.harvard.edu/abs/2000ApJ...539L..13G}
}

@ARTICLE{Neugebauer1984,
       author = {{Neugebauer}, G. and {Habing}, H.~J. and {van Duinen}, R. and {Aumann}, H.~H. and {Baud}, B. and {Beichman}, C.~A. and {Beintema}, D.~A. and {Boggess}, N. and {Clegg}, P.~E. and {de Jong}, T. and {Emerson}, J.~P. and {Gautier}, T.~N. and {Gillett}, F.~C. and {Harris}, S. and {Hauser}, M.~G. and {Houck}, J.~R. and {Jennings}, R.~E. and {Low}, F.~J. and {Marsden}, P.~L. and {Miley}, G. and {Olnon}, F.~M. and {Pottasch}, S.~R. and {Raimond}, E. and {Rowan-Robinson}, M. and {Soifer}, B.~T. and {Walker}, R.~G. and {Wesselius}, P.~R. and {Young}, E.},
        title = "{The Infrared Astronomical Satellite (IRAS) mission.}",
      journal = {\apjl},
         year = 1984,
        month = mar,
       volume = {278},
        pages = {L1-L6},
          doi = {10.1086/184209},
       adsurl = {https://ui.adsabs.harvard.edu/abs/1984ApJ...278L...1N}
}

@ARTICLE{Tumlinson2017ARA&A,
       author = {{Tumlinson}, Jason and {Peeples}, Molly S. and {Werk}, Jessica K.},
        title = "{The Circumgalactic Medium}",
      journal = {\araa},
         year = 2017,
        month = aug,
       volume = {55},
       number = {1},
        pages = {389-432},
          doi = {10.1146/annurev-astro-091916-055240},
archivePrefix = {arXiv},
       eprint = {1709.09180},
 primaryClass = {astro-ph.GA},
       adsurl = {https://ui.adsabs.harvard.edu/abs/2017ARA&A..55..389T}
}

@ARTICLE{Ni2022ApJ,
       author = {{Ni}, Yueying and {Di Matteo}, Tiziana and {Chen}, Nianyi and {Croft}, Rupert and {Bird}, Simeon},
        title = "{Ultramassive Black Holes Formed by Triple Quasar Mergers at z   2}",
      journal = {\apjl},
         year = 2022,
        month = dec,
       volume = {940},
       number = {2},
          eid = {L49},
        pages = {L49},
          doi = {10.3847/2041-8213/aca160},
archivePrefix = {arXiv},
       eprint = {2209.01249},
 primaryClass = {astro-ph.GA},
       adsurl = {https://ui.adsabs.harvard.edu/abs/2022ApJ...940L..49N}
}

@ARTICLE{Prochaska2011ApJ,
       author = {{Prochaska}, J. Xavier and {Kasen}, Daniel and {Rubin}, Kate},
        title = "{Simple Models of Metal-line Absorption and Emission from Cool Gas Outflows}",
      journal = {\apj},
         year = 2011,
        month = jun,
       volume = {734},
       number = {1},
          eid = {24},
        pages = {24},
          doi = {10.1088/0004-637X/734/1/24},
archivePrefix = {arXiv},
       eprint = {1102.3444},
 primaryClass = {astro-ph.GA},
       adsurl = {https://ui.adsabs.harvard.edu/abs/2011ApJ...734...24P}
}

@ARTICLE{Baron2020MNRAS,
       author = {{Baron}, Dalya and {Netzer}, Hagai and {Davies}, Ric I. and {Xavier Prochaska}, J.},
        title = "{Multiphase outflows in post-starburst E+A galaxies - II. A direct connection between the neutral and ionized outflow phases}",
      journal = {\mnras},
         year = 2020,
        month = jun,
       volume = {494},
       number = {4},
        pages = {5396-5420},
          doi = {10.1093/mnras/staa1018},
archivePrefix = {arXiv},
       eprint = {2004.04749},
 primaryClass = {astro-ph.GA},
       adsurl = {https://ui.adsabs.harvard.edu/abs/2020MNRAS.494.5396B}
}

@article{Boorman2024, title={The NuSTAR Local AGN $N_{\rm H}$ Distribution Survey (NuLANDS) I: Towards a Truly Representative Column Density Distribution in the Local Universe}, url={http://arxiv.org/abs/2410.07339}, DOI={10.48550/arXiv.2410.07339}, note={arXiv:2410.07339}, number={arXiv:2410.07339}, publisher={arXiv}, journal={arXiv preprint}, author={Boorman, Peter G. and Gandhi, Poshak and Buchner, Johannes and Stern, Daniel and Ricci, Claudio and Baloković, Mislav and Asmus, Daniel and Harrison, Fiona A. and Svoboda, Jiří and Greenwell, Claire and Koss, Michael and Alexander, David M. and Annuar, Adlyka and Bauer, Franz and Brandt, William N. and Brightman, Murray and Panessa, Francesca and Chen, Chien-Ting J. and Farrah, Duncan and Forster, Karl and Grefenstette, Brian and Hönig, Sebastian F. and Hill, Adam B. and Kammoun, Elias and Lansbury, George and Lanz, Lauranne and LaMassa, Stephanie and Madsen, Kristin and Marchesi, Stefano and Middleton, Matthew and Mingo, Beatriz and Parker, Michael L. and Treister, Ezequiel and Ueda, Yoshihiro and Urry, C. Megan and Zappacosta, Luca}, year={2024}, month=oct }

@ARTICLE{Hoffman2023,
       author = {{Hoffman}, Calvin and {Chen}, Nianyi and {Di Matteo}, Tiziana and {Ni}, Yueying and {Bird}, Simeon and {Croft}, Rupert and {Loeb}, Abraham},
        title = "{Triple and quadruple black holes in the ASTRID simulation at z=2}",
      journal = {\mnras},
         year = 2023,
        month = sep,
       volume = {524},
       number = {2},
        pages = {1987-1996},
          doi = {10.1093/mnras/stad1929},
archivePrefix = {arXiv},
       eprint = {2303.04825},
 primaryClass = {astro-ph.GA},
       adsurl = {https://ui.adsabs.harvard.edu/abs/2023MNRAS.524.1987H}
}

@ARTICLE{Bittner2019A&A,
       author = {{Bittner}, A. and {Falc{\'o}n-Barroso}, J. and {Nedelchev}, B. and {Dorta}, A. and {Gadotti}, D.~A. and {Sarzi}, M. and {Molaeinezhad}, A. and {Iodice}, E. and {Rosado-Belza}, D. and {de Lorenzo-C{\'a}ceres}, A. and {Fragkoudi}, F. and {Gal{\'a}n-de Anta}, P.~M. and {Husemann}, B. and {M{\'e}ndez-Abreu}, J. and {Neumann}, J. and {Pinna}, F. and {Querejeta}, M. and {S{\'a}nchez-Bl{\'a}zquez}, P. and {Seidel}, M.~K.},
        title = "{The GIST pipeline: A multi-purpose tool for the analysis and visualisation of (integral-field) spectroscopic data}",
      journal = {\aap},
         year = 2019,
        month = aug,
       volume = {628},
          eid = {A117},
        pages = {A117},
          doi = {10.1051/0004-6361/201935829},
archivePrefix = {arXiv},
       eprint = {1906.04746},
 primaryClass = {astro-ph.GA},
       adsurl = {https://ui.adsabs.harvard.edu/abs/2019A&A...628A.117B}
}

@misc{pyqsofit,
       author = {{Guo}, Hengxiao and {Shen}, Yue and {Wang}, Shu},
        title = "{PyQSOFit: Python code to fit the spectrum of quasars}",
 howpublished = {Astrophysics Source Code Library, record ascl:1809.008},
         year = 2018,
        month = sep,
          eid = {ascl:1809.008},
       adsurl = {https://ui.adsabs.harvard.edu/abs/2018ascl.soft09008G}
}

@ARTICLE{emcee,
       author = {{Foreman-Mackey}, Daniel and {Hogg}, David W. and {Lang}, Dustin and {Goodman}, Jonathan},
        title = "{emcee: The MCMC Hammer}",
      journal = {\pasp},
         year = 2013,
        month = mar,
       volume = {125},
       number = {925},
        pages = {306},
          doi = {10.1086/670067},
archivePrefix = {arXiv},
       eprint = {1202.3665},
 primaryClass = {astro-ph.IM},
       adsurl = {https://ui.adsabs.harvard.edu/abs/2013PASP..125..306F}
}

@ARTICLE{Murphy2012ApJ,
       author = {{Murphy}, E.~J. and {Bremseth}, J. and {Mason}, B.~S. and {Condon}, J.~J. and {Schinnerer}, E. and {Aniano}, G. and {Armus}, L. and {Helou}, G. and {Turner}, J.~L. and {Jarrett}, T.~H.},
        title = "{The Star Formation in Radio Survey: GBT 33 GHz Observations of Nearby Galaxy Nuclei and Extranuclear Star-forming Regions}",
      journal = {\apj},
         year = 2012,
        month = dec,
       volume = {761},
       number = {2},
          eid = {97},
        pages = {97},
          doi = {10.1088/0004-637X/761/2/97},
archivePrefix = {arXiv},
       eprint = {1210.3360},
 primaryClass = {astro-ph.CO},
       adsurl = {https://ui.adsabs.harvard.edu/abs/2012ApJ...761...97M}
}

@ARTICLE{Strom2017,
       author = {{Strom}, Allison L. and {Steidel}, Charles C. and {Rudie}, Gwen C. and {Trainor}, Ryan F. and {Pettini}, Max and {Reddy}, Naveen A.},
        title = "{Nebular Emission Line Ratios in z ≃ 2-3 Star-forming Galaxies with KBSS-MOSFIRE: Exploring the Impact of Ionization, Excitation, and Nitrogen-to-Oxygen Ratio}",
      journal = {\apj},
         year = 2017,
        month = feb,
       volume = {836},
       number = {2},
          eid = {164},
        pages = {164},
          doi = {10.3847/1538-4357/836/2/164},
archivePrefix = {arXiv},
       eprint = {1608.02587},
 primaryClass = {astro-ph.GA},
       adsurl = {https://ui.adsabs.harvard.edu/abs/2017ApJ...836..164S}
}

@ARTICLE{Steidel2014,
       author = {{Steidel}, Charles C. and {Rudie}, Gwen C. and {Strom}, Allison L. and {Pettini}, Max and {Reddy}, Naveen A. and {Shapley}, Alice E. and {Trainor}, Ryan F. and {Erb}, Dawn K. and {Turner}, Monica L. and {Konidaris}, Nicholas P. and {Kulas}, Kristin R. and {Mace}, Gregory and {Matthews}, Keith and {McLean}, Ian S.},
        title = "{Strong Nebular Line Ratios in the Spectra of z \raisebox{-0.5ex}\textasciitilde 2-3 Star Forming Galaxies: First Results from KBSS-MOSFIRE}",
      journal = {\apj},
         year = 2014,
        month = nov,
       volume = {795},
       number = {2},
          eid = {165},
        pages = {165},
          doi = {10.1088/0004-637X/795/2/165},
archivePrefix = {arXiv},
       eprint = {1405.5473},
 primaryClass = {astro-ph.GA},
       adsurl = {https://ui.adsabs.harvard.edu/abs/2014ApJ...795..165S}
}

@ARTICLE{BFPR1984,
       author = {{Blumenthal}, G.~R. and {Faber}, S.~M. and {Primack}, J.~R. and {Rees}, M.~J.},
        title = "{Formation of galaxies and large-scale structure with cold dark matter.}",
      journal = {\nat},
         year = 1984,
        month = oct,
       volume = {311},
        pages = {517-525},
          doi = {10.1038/311517a0},
       adsurl = {https://ui.adsabs.harvard.edu/abs/1984Natur.311..517B}
}

@ARTICLE{Garnavich1998ApJ,
       author = {{Garnavich}, Peter M. and {Jha}, Saurabh and {Challis}, Peter and {Clocchiatti}, Alejandro and {Diercks}, Alan and {Filippenko}, Alexei V. and {Gilliland}, Ron L. and {Hogan}, Craig J. and {Kirshner}, Robert P. and {Leibundgut}, Bruno and {Phillips}, M.~M. and {Reiss}, David and {Riess}, Adam G. and {Schmidt}, Brian P. and {Schommer}, Robert A. and {Smith}, R. Chris and {Spyromilio}, Jason and {Stubbs}, Chris and {Suntzeff}, Nicholas B. and {Tonry}, John and {Carroll}, Sean M.},
        title = "{Supernova Limits on the Cosmic Equation of State}",
      journal = {\apj},
         year = 1998,
        month = dec,
       volume = {509},
       number = {1},
        pages = {74-79},
          doi = {10.1086/306495},
archivePrefix = {arXiv},
       eprint = {astro-ph/9806396},
 primaryClass = {astro-ph},
       adsurl = {https://ui.adsabs.harvard.edu/abs/1998ApJ...509...74G}
}

@ARTICLE{Black1987ApJ,
       author = {{Black}, John H. and {van Dishoeck}, Ewine F.},
        title = "{Fluorescent Excitation of Interstellar H 2}",
      journal = {\apj},
         year = 1987,
        month = nov,
       volume = {322},
        pages = {412},
          doi = {10.1086/165740},
       adsurl = {https://ui.adsabs.harvard.edu/abs/1987ApJ...322..412B}
}

@ARTICLE{Steidel2016,
       author = {{Steidel}, Charles C. and {Strom}, Allison L. and {Pettini}, Max and {Rudie}, Gwen C. and {Reddy}, Naveen A. and {Trainor}, Ryan F.},
        title = "{Reconciling the Stellar and Nebular Spectra of High-redshift Galaxies}",
      journal = {\apj},
         year = 2016,
        month = aug,
       volume = {826},
       number = {2},
          eid = {159},
        pages = {159},
          doi = {10.3847/0004-637X/826/2/159},
archivePrefix = {arXiv},
       eprint = {1605.07186},
 primaryClass = {astro-ph.GA},
       adsurl = {https://ui.adsabs.harvard.edu/abs/2016ApJ...826..159S}
}

@ARTICLE{Pettini2004,
       author = {{Pettini}, Max and {Pagel}, Bernard E.~J.},
        title = "{[OIII]/[NII] as an abundance indicator at high redshift}",
      journal = {\mnras},
         year = 2004,
        month = mar,
       volume = {348},
       number = {3},
        pages = {L59-L63},
          doi = {10.1111/j.1365-2966.2004.07591.x},
archivePrefix = {arXiv},
       eprint = {astro-ph/0401128},
 primaryClass = {astro-ph},
       adsurl = {https://ui.adsabs.harvard.edu/abs/2004MNRAS.348L..59P}
}

@BOOK{2016era..book.....C,
       author = {{Condon}, James J. and {Ransom}, Scott M.},
        title = "{Essential Radio Astronomy}",
         year = 2016,
       adsurl = {https://ui.adsabs.harvard.edu/abs/2016era..book.....C}
}

@ARTICLE{2024MNRAS.528.5346A,
       author = {{An}, Fangxia and {Vaccari}, M. and {Best}, P.~N. and {Ocran}, E.~F. and {Ishwara-Chandra}, C.~H. and {Taylor}, A.~R. and {Leslie}, S.~K. and {R{\"o}ttgering}, H.~J.~A. and {Kondapally}, R. and {Haskell}, Paul and {Collier}, J.~D. and {Bonato}, M.},
        title = "{Radio spectral properties of star-forming galaxies between 150 and 5000 MHz in the ELAIS-N1 field}",
      journal = {\mnras},
         year = 2024,
        month = mar,
       volume = {528},
       number = {3},
        pages = {5346-5363},
          doi = {10.1093/mnras/stae364},
archivePrefix = {arXiv},
       eprint = {2303.06941},
 primaryClass = {astro-ph.GA},
       adsurl = {https://ui.adsabs.harvard.edu/abs/2024MNRAS.528.5346A}
}

@ARTICLE{2019A&A...630A..83Z,
       author = {{Zaja{\v{c}}ek}, Michal and {Busch}, Gerold and {Valencia-S.}, M{\'o}nica and {Eckart}, Andreas and {Britzen}, Silke and {Fuhrmann}, Lars and {Schneeloch}, Jana and {Fazeli}, Nastaran and {Harrington}, Kevin C. and {Zensus}, J. Anton},
        title = "{Radio spectral index distribution of SDSS-FIRST sources across optical diagnostic diagrams}",
      journal = {\aap},
         year = 2019,
        month = oct,
       volume = {630},
          eid = {A83},
        pages = {A83},
          doi = {10.1051/0004-6361/201833388},
archivePrefix = {arXiv},
       eprint = {1906.08877},
 primaryClass = {astro-ph.GA},
       adsurl = {https://ui.adsabs.harvard.edu/abs/2019A&A...630A..83Z}
}

@ARTICLE{2025MNRAS.539..808M,
       author = {{Mutie}, Isaac M. and {del Palacio}, Santiago and {Beswick}, Robert J. and {Williams-Baldwin}, David and {Gallimore}, Jack F. and {Gallagher}, John S. and {Aalto}, Susanne E. and {Baki}, Paul O.},
        title = "{A consistent radio to sub-mm pc-scale study of the nucleus of NGC 1068}",
      journal = {\mnras},
         year = 2025,
        month = may,
       volume = {539},
       number = {2},
        pages = {808-819},
          doi = {10.1093/mnras/staf524},
archivePrefix = {arXiv},
       eprint = {2503.20303},
 primaryClass = {astro-ph.GA},
       adsurl = {https://ui.adsabs.harvard.edu/abs/2025MNRAS.539..808M}
}

@ARTICLE{Verro2022A&A,
       author = {{Verro}, K. and {Trager}, S.~C. and {Peletier}, R.~F. and {Lan{\c{c}}on}, A. and {Arentsen}, A. and {Chen}, Y. -P. and {Coelho}, P.~R.~T. and {Dries}, M. and {Falc{\'o}n-Barroso}, J. and {Gonneau}, A. and {Lyubenova}, M. and {Martins}, L. and {Prugniel}, P. and {S{\'a}nchez-Bl{\'a}zquez}, P. and {Vazdekis}, A.},
        title = "{Modelling simple stellar populations in the near-ultraviolet to near-infrared with the X-shooter Spectral Library (XSL)}",
      journal = {\aap},
         year = 2022,
        month = may,
       volume = {661},
          eid = {A50},
        pages = {A50},
          doi = {10.1051/0004-6361/202142387},
archivePrefix = {arXiv},
       eprint = {2110.10190},
 primaryClass = {astro-ph.GA},
       adsurl = {https://ui.adsabs.harvard.edu/abs/2022A&A...661A..50V}
}

@ARTICLE{Ren2022ApJ,
       author = {{Ren}, Wenke and {Wang}, Junxian and {Cai}, Zhenyi and {Guo}, Hengxiao},
        title = "{Extreme Variability Quasars in Their Various States. I. The Sample Selection and Composite SDSS Spectra}",
      journal = {\apj},
         year = 2022,
        month = jan,
       volume = {925},
       number = {1},
          eid = {50},
        pages = {50},
          doi = {10.3847/1538-4357/ac3828},
archivePrefix = {arXiv},
       eprint = {2111.07057},
 primaryClass = {astro-ph.GA},
       adsurl = {https://ui.adsabs.harvard.edu/abs/2022ApJ...925...50R}
}

@ARTICLE{Mannerkoski2021ApJ,
       author = {{Mannerkoski}, Matias and {Johansson}, Peter H. and {Rantala}, Antti and {Naab}, Thorsten and {Liao}, Shihong},
        title = "{Resolving the Complex Evolution of a Supermassive Black Hole Triplet in a Cosmological Simulation}",
      journal = {\apjl},
         year = 2021,
        month = may,
       volume = {912},
       number = {2},
          eid = {L20},
        pages = {L20},
          doi = {10.3847/2041-8213/abf9a5},
archivePrefix = {arXiv},
       eprint = {2103.16254},
 primaryClass = {astro-ph.GA},
       adsurl = {https://ui.adsabs.harvard.edu/abs/2021ApJ...912L..20M}
}

@article{Pfeifle_2019,
doi = {10.3847/1538-4357/ab3a9b},
url = {https://dx.doi.org/10.3847/1538-4357/ab3a9b},
year = {2019},
month = {oct},
publisher = {The American Astronomical Society},
volume = {883},
number = {2},
pages = {167},
author = {Ryan W. Pfeifle and Shobita Satyapal and Christina Manzano-King and Jenna Cann and Remington O. Sexton and Barry Rothberg and Gabriela Canalizo and Claudio Ricci and Laura Blecha and Sara L. Ellison and Mario Gliozzi and Nathan J. Secrest and Anca Constantin and Jenna B. Harvey},
title = {A Triple AGN in a Mid-infrared Selected Late-stage Galaxy Merger},
journal = {The Astrophysical Journal}
}

@ARTICLE{2022MNRAS.512.3703B,
       author = {{Bird}, Simeon and {Ni}, Yueying and {Di Matteo}, Tiziana and {Croft}, Rupert and {Feng}, Yu and {Chen}, Nianyi},
        title = "{The ASTRID simulation: galaxy formation and reionization}",
      journal = {\mnras},
         year = 2022,
        month = may,
       volume = {512},
       number = {3},
        pages = {3703-3716},
          doi = {10.1093/mnras/stac648},
archivePrefix = {arXiv},
       eprint = {2111.01160},
 primaryClass = {astro-ph.GA},
       adsurl = {https://ui.adsabs.harvard.edu/abs/2022MNRAS.512.3703B}
}

@article{Kass1995,
author = {Robert E. Kass and Adrian E. Raftery},
title = {Bayes Factors},
journal = {Journal of the American Statistical Association},
volume = {90},
number = {430},
pages = {773--795},
year = {1995},
publisher = {ASA Website},
doi = {10.1080/01621459.1995.10476572}

}

@BOOK{Spitzer1978ppim.book.....S,
       author = {{Spitzer}, Lyman},
        title = "{Physical processes in the interstellar medium}",
         year = 1978,
          doi = {10.1002/9783527617722},
       adsurl = {https://ui.adsabs.harvard.edu/abs/1978ppim.book.....S},publisher = {Wiley}
}

@ARTICLE{Rupke2002ApJ,
       author = {{Rupke}, David S. and {Veilleux}, Sylvain and {Sanders}, D.~B.},
        title = "{Keck Absorption-Line Spectroscopy of Galactic Winds in Ultraluminous Infrared Galaxies}",
      journal = {\apj},
         year = 2002,
        month = may,
       volume = {570},
       number = {2},
        pages = {588-609},
          doi = {10.1086/339789},
archivePrefix = {arXiv},
       eprint = {astro-ph/0201371},
 primaryClass = {astro-ph},
       adsurl = {https://ui.adsabs.harvard.edu/abs/2002ApJ...570..588R}
}

@ARTICLE{Bellocchi2013A&A,
       author = {{Bellocchi}, Enrica and {Arribas}, Santiago and {Colina}, Luis and {Miralles-Caballero}, Daniel},
        title = "{VLT/VIMOS integral field spectroscopy of luminous and ultraluminous infrared galaxies: 2D kinematic properties}",
      journal = {\aap},
         year = 2013,
        month = sep,
       volume = {557},
          eid = {A59},
        pages = {A59},
          doi = {10.1051/0004-6361/201221019},
archivePrefix = {arXiv},
       eprint = {1307.1659},
 primaryClass = {astro-ph.CO},
       adsurl = {https://ui.adsabs.harvard.edu/abs/2013A&A...557A..59B}
}

@ARTICLE{2010ApJ...724.1373G,
       author = {{Gon{\c{c}}alves}, Thiago S. and {Basu-Zych}, Antara and {Overzier}, Roderik and {Martin}, D. Christopher and {Law}, David R. and {Schiminovich}, David and {Wyder}, Ted K. and {Mallery}, Ryan and {Rich}, R. Michael and {Heckman}, Timothy H.},
        title = "{The Kinematics of Ionized Gas in Lyman-break Analogs at z \raisebox{-0.5ex}\textasciitilde 0.2}",
      journal = {\apj},
         year = 2010,
        month = dec,
       volume = {724},
       number = {2},
        pages = {1373-1388},
          doi = {10.1088/0004-637X/724/2/1373},
archivePrefix = {arXiv},
       eprint = {1009.4934},
 primaryClass = {astro-ph.CO},
       adsurl = {https://ui.adsabs.harvard.edu/abs/2010ApJ...724.1373G}
}

@ARTICLE{Arribas2014,
       author = {{Arribas}, S. and {Colina}, L. and {Bellocchi}, E. and {Maiolino}, R. and {Villar-Mart{\'\i}n}, M.},
        title = "{Ionized gas outflows and global kinematics of low-z luminous star-forming galaxies}",
      journal = {\aap},
         year = 2014,
        month = aug,
       volume = {568},
          eid = {A14},
        pages = {A14},
          doi = {10.1051/0004-6361/201323324},
archivePrefix = {arXiv},
       eprint = {1404.1082},
 primaryClass = {astro-ph.GA},
       adsurl = {https://ui.adsabs.harvard.edu/abs/2014A&A...568A..14A}
}

@ARTICLE{Zanella2019MNRAS,
       author = {{Zanella}, A. and {Le Floc'h}, E. and {Harrison}, C.~M. and {Daddi}, E. and {Bernhard}, E. and {Gobat}, R. and {Strazzullo}, V. and {Valentino}, F. and {Cibinel}, A. and {S{\'a}nchez Almeida}, J. and {Kohandel}, M. and {Fensch}, J. and {Behrendt}, M. and {Burkert}, A. and {Onodera}, M. and {Bournaud}, F. and {Scholtz}, J.},
        title = "{A contribution of star-forming clumps and accreting satellites to the mass assembly of z {\ensuremath{\sim}} 2 galaxies}",
      journal = {\mnras},
         year = 2019,
        month = oct,
       volume = {489},
       number = {2},
        pages = {2792-2818},
          doi = {10.1093/mnras/stz2099},
archivePrefix = {arXiv},
       eprint = {1907.12136},
 primaryClass = {astro-ph.GA},
       adsurl = {https://ui.adsabs.harvard.edu/abs/2019MNRAS.489.2792Z}
}

@ARTICLE{Raginski2016MNRAS,
       author = {{Raginski}, I. and {Laor}, Ari},
        title = "{AGN coronal emission models - I. The predicted radio emission}",
      journal = {\mnras},
         year = 2016,
        month = jun,
       volume = {459},
       number = {2},
        pages = {2082-2096},
          doi = {10.1093/mnras/stw772},
archivePrefix = {arXiv},
       eprint = {1604.00646},
 primaryClass = {astro-ph.GA},
       adsurl = {https://ui.adsabs.harvard.edu/abs/2016MNRAS.459.2082R}
}

@ARTICLE{Lehmer2016ApJ,
       author = {{Lehmer}, B.~D. and {Basu-Zych}, A.~R. and {Mineo}, S. and {Brandt}, W.~N. and {Eufrasio}, R.~T. and {Fragos}, T. and {Hornschemeier}, A.~E. and {Luo}, B. and {Xue}, Y.~Q. and {Bauer}, F.~E. and {Gilfanov}, M. and {Ranalli}, P. and {Schneider}, D.~P. and {Shemmer}, O. and {Tozzi}, P. and {Trump}, J.~R. and {Vignali}, C. and {Wang}, J. -X. and {Yukita}, M. and {Zezas}, A.},
        title = "{The Evolution of Normal Galaxy X-Ray Emission through Cosmic History: Constraints from the 6 MS Chandra Deep Field-South}",
      journal = {\apj},
         year = 2016,
        month = jul,
       volume = {825},
       number = {1},
          eid = {7},
        pages = {7},
          doi = {10.3847/0004-637X/825/1/7},
archivePrefix = {arXiv},
       eprint = {1604.06461},
 primaryClass = {astro-ph.GA},
       adsurl = {https://ui.adsabs.harvard.edu/abs/2016ApJ...825....7L}
}

@ARTICLE{Vardoulaki2015A,
       author = {{Vardoulaki}, E. and {Charmandaris}, V. and {Murphy}, E.~J. and {Diaz-Santos}, T. and {Armus}, L. and {Evans}, A.~S. and {Mazzarella}, J.~M. and {Privon}, G.~C. and {Stierwalt}, S. and {Barcos-Mu{\~n}oz}, L.},
        title = "{Radio continuum properties of luminous infrared galaxies. Identifying the presence of an AGN in the radio}",
      journal = {\aap},
         year = 2015,
        month = feb,
       volume = {574},
          eid = {A4},
        pages = {A4},
          doi = {10.1051/0004-6361/201424125},
archivePrefix = {arXiv},
       eprint = {1408.4177},
 primaryClass = {astro-ph.GA},
       adsurl = {https://ui.adsabs.harvard.edu/abs/2015A&A...574A...4V}
}

@ARTICLE{Linden2019ApJ,
       author = {{Linden}, S.~T. and {Song}, Y. and {Evans}, A.~S. and {Murphy}, E.~J. and {Armus}, L. and {Barcos-Mu{\~n}oz}, L. and {Larson}, K. and {D{\'\i}az-Santos}, T. and {Privon}, G.~C. and {Howell}, J. and {Surace}, J.~A. and {Charmandaris}, V. and {U}, Vivian and {Medling}, A.~M. and {Chu}, J. and {Momjian}, E.},
        title = "{A Very Large Array Survey of Luminous Extranuclear Star-forming Regions in Luminous Infrared Galaxies in GOALS}",
      journal = {\apj},
         year = 2019,
        month = aug,
       volume = {881},
       number = {1},
          eid = {70},
        pages = {70},
          doi = {10.3847/1538-4357/ab2872},
archivePrefix = {arXiv},
       eprint = {1906.05182},
 primaryClass = {astro-ph.GA},
       adsurl = {https://ui.adsabs.harvard.edu/abs/2019ApJ...881...70L}
}

@ARTICLE{Song2022ApJ,
       author = {{Song}, Y. and {Linden}, S.~T. and {Evans}, A.~S. and {Barcos-Mu{\~n}oz}, L. and {Murphy}, E.~J. and {Momjian}, E. and {D{\'\i}az-Santos}, T. and {Larson}, K.~L. and {Privon}, G.~C. and {Huang}, X. and {Armus}, L. and {Mazzarella}, J.~M. and {U}, V. and {Inami}, H. and {Charmandaris}, V. and {Ricci}, C. and {Emig}, K.~L. and {McKinney}, J. and {Yoon}, I. and {Kunneriath}, D. and {Lai}, T.~S. -Y. and {Rodas-Quito}, E.~E. and {Saravia}, A. and {Gao}, T. and {Meynardie}, W. and {Sanders}, D.~B.},
        title = "{Characterizing Compact 15-33 GHz Radio Continuum Sources in Local U/LIRGs}",
      journal = {\apj},
         year = 2022,
        month = nov,
       volume = {940},
       number = {1},
          eid = {52},
        pages = {52},
          doi = {10.3847/1538-4357/ac923b},
archivePrefix = {arXiv},
       eprint = {2209.04002},
 primaryClass = {astro-ph.GA},
       adsurl = {https://ui.adsabs.harvard.edu/abs/2022ApJ...940...52S}
}

@ARTICLE{Murphy2011ApJ,
       author = {{Murphy}, E.~J. and {Condon}, J.~J. and {Schinnerer}, E. and {Kennicutt}, R.~C. and {Calzetti}, D. and {Armus}, L. and {Helou}, G. and {Turner}, J.~L. and {Aniano}, G. and {Beir{\~a}o}, P. and {Bolatto}, A.~D. and {Brandl}, B.~R. and {Croxall}, K.~V. and {Dale}, D.~A. and {Donovan Meyer}, J.~L. and {Draine}, B.~T. and {Engelbracht}, C. and {Hunt}, L.~K. and {Hao}, C. -N. and {Koda}, J. and {Roussel}, H. and {Skibba}, R. and {Smith}, J. -D.~T.},
        title = "{Calibrating Extinction-free Star Formation Rate Diagnostics with 33 GHz Free-free Emission in NGC 6946}",
      journal = {\apj},
         year = 2011,
        month = aug,
       volume = {737},
       number = {2},
          eid = {67},
        pages = {67},
          doi = {10.1088/0004-637X/737/2/67},
archivePrefix = {arXiv},
       eprint = {1105.4877},
 primaryClass = {astro-ph.CO},
       adsurl = {https://ui.adsabs.harvard.edu/abs/2011ApJ...737...67M}
}

@ARTICLE{Perna2020A&A,
       author = {{Perna}, M. and {Arribas}, S. and {Catal{\'a}n-Torrecilla}, C. and {Colina}, L. and {Bellocchi}, E. and {Fluetsch}, A. and {Maiolino}, R. and {Cazzoli}, S. and {Hern{\'a}n Caballero}, A. and {Pereira Santaella}, M. and {Piqueras L{\'o}pez}, J. and {Rodr{\'\i}guez del Pino}, B.},
        title = "{MUSE view of Arp220: Kpc-scale multi-phase outflow and evidence for positive feedback}",
      journal = {\aap},
         year = 2020,
        month = nov,
       volume = {643},
          eid = {A139},
        pages = {A139},
          doi = {10.1051/0004-6361/202038328},
archivePrefix = {arXiv},
       eprint = {2009.03353},
 primaryClass = {astro-ph.GA},
       adsurl = {https://ui.adsabs.harvard.edu/abs/2020A&A...643A.139P}
}

@ARTICLE{Mardling2001MNRAS,
       author = {{Mardling}, Rosemary A. and {Aarseth}, Sverre J.},
        title = "{Tidal interactions in star cluster simulations}",
      journal = {\mnras},
         year = 2001,
        month = mar,
       volume = {321},
       number = {3},
        pages = {398-420},
          doi = {10.1046/j.1365-8711.2001.03974.x},
       adsurl = {https://ui.adsabs.harvard.edu/abs/2001MNRAS.321..398M}
}

@ARTICLE{Zibetti2009MNRAS,
       author = {{Zibetti}, Stefano and {Charlot}, St{\'e}phane and {Rix}, Hans-Walter},
        title = "{Resolved stellar mass maps of galaxies - I. Method and implications for global mass estimates}",
      journal = {\mnras},
         year = 2009,
        month = dec,
       volume = {400},
       number = {3},
        pages = {1181-1198},
          doi = {10.1111/j.1365-2966.2009.15528.x},
archivePrefix = {arXiv},
       eprint = {0904.4252},
 primaryClass = {astro-ph.CO},
       adsurl = {https://ui.adsabs.harvard.edu/abs/2009MNRAS.400.1181Z}
}

@ARTICLE{Perna2024arXiv,
       author = {{Perna}, Michele and {Arribas}, Santiago and {Lamperti}, Isabella and {Pereira-Santaella}, Miguel and {Ulivi}, Lorenzo and {B{\"o}ker}, Torsten and {Maiolino}, Roberto and {Bunker}, Andrew J. and {Charlot}, St{\'e}phane and {Cresci}, Giovanni and {Rodr{\'\i}guez Del Pino}, Bruno and {D'Eugenio}, Francesco and {{\"U}bler}, Hannah and {Fahrion}, Katja and {Ceci}, Matteo},
        title = "{No evidence of AGN features in the nuclei of Arp 220 from JWST/NIRSpec IFS}",
      journal = {arXiv e-prints},
         year = 2024,
        month = mar,
          eid = {arXiv:2403.13948},
        pages = {arXiv:2403.13948},
          doi = {10.48550/arXiv.2403.13948},
archivePrefix = {arXiv},
       eprint = {2403.13948},
 primaryClass = {astro-ph.GA},
       adsurl = {https://ui.adsabs.harvard.edu/abs/2024arXiv240313948P}
}

@ARTICLE{Ricci_2023ApJL,
       author = {{Ricci}, Claudio and {Chang}, Chin-Shin and {Kawamuro}, Taiki and {Privon}, George C. and {Mushotzky}, Richard and {Trakhtenbrot}, Benny and {Laor}, Ari and {Koss}, Michael J. and {Smith}, Krista L. and {Gupta}, Kriti K. and {Dimopoulos}, Georgios and {Aalto}, Susanne and {Ros}, Eduardo},
        title = "{A Tight Correlation between Millimeter and X-Ray Emission in Accreting Massive Black Holes from <100 mas Resolution ALMA Observations}",
      journal = {\apjl},
         year = 2023,
        month = aug,
       volume = {952},
       number = {2},
          eid = {L28},
        pages = {L28},
          doi = {10.3847/2041-8213/acda27},
archivePrefix = {arXiv},
       eprint = {2306.04679},
 primaryClass = {astro-ph.HE},
       adsurl = {https://ui.adsabs.harvard.edu/abs/2023ApJ...952L..28R}
}

@ARTICLE{Planck2020A&A,
       author = {{Planck Collaboration} and {Aghanim}, N. and {Akrami}, Y. and {Ashdown}, M. and {Aumont}, J. and {Baccigalupi}, C. and {Ballardini}, M. and {Banday}, A.~J. and {Barreiro}, R.~B. and {Bartolo}, N. and {Basak}, S. and {Battye}, R. and {Benabed}, K. and {Bernard}, J. -P. and {Bersanelli}, M. and {Bielewicz}, P. and {Bock}, J.~J. and {Bond}, J.~R. and {Borrill}, J. and {Bouchet}, F.~R. and {Boulanger}, F. and {Bucher}, M. and {Burigana}, C. and {Butler}, R.~C. and {Calabrese}, E. and {Cardoso}, J. -F. and {Carron}, J. and {Challinor}, A. and {Chiang}, H.~C. and {Chluba}, J. and {Colombo}, L.~P.~L. and {Combet}, C. and {Contreras}, D. and {Crill}, B.~P. and {Cuttaia}, F. and {de Bernardis}, P. and {de Zotti}, G. and {Delabrouille}, J. and {Delouis}, J. -M. and {Di Valentino}, E. and {Diego}, J.~M. and {Dor{\'e}}, O. and {Douspis}, M. and {Ducout}, A. and {Dupac}, X. and {Dusini}, S. and {Efstathiou}, G. and {Elsner}, F. and {En{\ss}lin}, T.~A. and {Eriksen}, H.~K. and {Fantaye}, Y. and {Farhang}, M. and {Fergusson}, J. and {Fernandez-Cobos}, R. and {Finelli}, F. and {Forastieri}, F. and {Frailis}, M. and {Fraisse}, A.~A. and {Franceschi}, E. and {Frolov}, A. and {Galeotta}, S. and {Galli}, S. and {Ganga}, K. and {G{\'e}nova-Santos}, R.~T. and {Gerbino}, M. and {Ghosh}, T. and {Gonz{\'a}lez-Nuevo}, J. and {G{\'o}rski}, K.~M. and {Gratton}, S. and {Gruppuso}, A. and {Gudmundsson}, J.~E. and {Hamann}, J. and {Handley}, W. and {Hansen}, F.~K. and {Herranz}, D. and {Hildebrandt}, S.~R. and {Hivon}, E. and {Huang}, Z. and {Jaffe}, A.~H. and {Jones}, W.~C. and {Karakci}, A. and {Keih{\"a}nen}, E. and {Keskitalo}, R. and {Kiiveri}, K. and {Kim}, J. and {Kisner}, T.~S. and {Knox}, L. and {Krachmalnicoff}, N. and {Kunz}, M. and {Kurki-Suonio}, H. and {Lagache}, G. and {Lamarre}, J. -M. and {Lasenby}, A. and {Lattanzi}, M. and {Lawrence}, C.~R. and {Le Jeune}, M. and {Lemos}, P. and {Lesgourgues}, J. and {Levrier}, F. and {Lewis}, A. and {Liguori}, M. and {Lilje}, P.~B. and {Lilley}, M. and {Lindholm}, V. and {L{\'o}pez-Caniego}, M. and {Lubin}, P.~M. and {Ma}, Y. -Z. and {Mac{\'\i}as-P{\'e}rez}, J.~F. and {Maggio}, G. and {Maino}, D. and {Mandolesi}, N. and {Mangilli}, A. and {Marcos-Caballero}, A. and {Maris}, M. and {Martin}, P.~G. and {Martinelli}, M. and {Mart{\'\i}nez-Gonz{\'a}lez}, E. and {Matarrese}, S. and {Mauri}, N. and {McEwen}, J.~D. and {Meinhold}, P.~R. and {Melchiorri}, A. and {Mennella}, A. and {Migliaccio}, M. and {Millea}, M. and {Mitra}, S. and {Miville-Desch{\^e}nes}, M. -A. and {Molinari}, D. and {Montier}, L. and {Morgante}, G. and {Moss}, A. and {Natoli}, P. and {N{\o}rgaard-Nielsen}, H.~U. and {Pagano}, L. and {Paoletti}, D. and {Partridge}, B. and {Patanchon}, G. and {Peiris}, H.~V. and {Perrotta}, F. and {Pettorino}, V. and {Piacentini}, F. and {Polastri}, L. and {Polenta}, G. and {Puget}, J. -L. and {Rachen}, J.~P. and {Reinecke}, M. and {Remazeilles}, M. and {Renzi}, A. and {Rocha}, G. and {Rosset}, C. and {Roudier}, G. and {Rubi{\~n}o-Mart{\'\i}n}, J.~A. and {Ruiz-Granados}, B. and {Salvati}, L. and {Sandri}, M. and {Savelainen}, M. and {Scott}, D. and {Shellard}, E.~P.~S. and {Sirignano}, C. and {Sirri}, G. and {Spencer}, L.~D. and {Sunyaev}, R. and {Suur-Uski}, A. -S. and {Tauber}, J.~A. and {Tavagnacco}, D. and {Tenti}, M. and {Toffolatti}, L. and {Tomasi}, M. and {Trombetti}, T. and {Valenziano}, L. and {Valiviita}, J. and {Van Tent}, B. and {Vibert}, L. and {Vielva}, P. and {Villa}, F. and {Vittorio}, N. and {Wandelt}, B.~D. and {Wehus}, I.~K. and {White}, M. and {White}, S.~D.~M. and {Zacchei}, A. and {Zonca}, A.},
        title = "{Planck 2018 results. VI. Cosmological parameters}",
      journal = {\aap},
         year = 2020,
        month = sep,
       volume = {641},
          eid = {A6},
        pages = {A6},
          doi = {10.1051/0004-6361/201833910},
archivePrefix = {arXiv},
       eprint = {1807.06209},
 primaryClass = {astro-ph.CO},
       adsurl = {https://ui.adsabs.harvard.edu/abs/2020A&A...641A...6P}
}

@ARTICLE{Wolniewicz1998,
       author = {{Wolniewicz}, L. and {Simbotin}, I. and {Dalgarno}, A.},
        title = "{Quadrupole Transition Probabilities for the Excited Rovibrational States of H$_{2}$}",
      journal = {\apjs},
         year = 1998,
        month = apr,
       volume = {115},
       number = {2},
        pages = {293-313},
          doi = {10.1086/313091},
       adsurl = {https://ui.adsabs.harvard.edu/abs/1998ApJS..115..293W}
}

@ARTICLE{Mazzalay2013,
       author = {{Mazzalay}, X. and {Saglia}, R.~P. and {Erwin}, Peter and {Fabricius}, M.~H. and {Rusli}, S.~P. and {Thomas}, J. and {Bender}, R. and {Opitsch}, M. and {Nowak}, N. and {Williams}, Michael J.},
        title = "{Molecular gas in the centre of nearby galaxies from VLT/SINFONI integral field spectroscopy - I. Morphology and mass inventory}",
      journal = {\mnras},
         year = 2013,
        month = jan,
       volume = {428},
       number = {3},
        pages = {2389-2406},
          doi = {10.1093/mnras/sts204},
archivePrefix = {arXiv},
       eprint = {1210.4171},
 primaryClass = {astro-ph.CO},
       adsurl = {https://ui.adsabs.harvard.edu/abs/2013MNRAS.428.2389M}
}

@ARTICLE{Scoville1982,
       author = {{Scoville}, N.~Z. and {Hall}, D.~N.~B. and {Ridgway}, S.~T. and {Kleinmann}, S.~G.},
        title = "{Velocity, reddening, and temperature structure of the H2 emission in Orion}",
      journal = {\apj},
         year = 1982,
        month = feb,
       volume = {253},
        pages = {136-148},
          doi = {10.1086/159618},
       adsurl = {https://ui.adsabs.harvard.edu/abs/1982ApJ...253..136S}
}

@ARTICLE{Vega2008A&A,
       author = {{Vega}, O. and {Clemens}, M.~S. and {Bressan}, A. and {Granato}, G.~L. and {Silva}, L. and {Panuzzo}, P.},
        title = "{Modelling the spectral energy distribution of ULIRGs. II. The energetic environment and the dense interstellar medium}",
      journal = {\aap},
         year = 2008,
        month = jun,
       volume = {484},
       number = {3},
        pages = {631-653},
          doi = {10.1051/0004-6361:20078883},
archivePrefix = {arXiv},
       eprint = {0712.1202},
 primaryClass = {astro-ph},
       adsurl = {https://ui.adsabs.harvard.edu/abs/2008A&A...484..631V}
}

@article{CAVANAUGH1997201,
author = {Joseph E. Cavanaugh},
title = {Unifying the derivations for the Akaike and corrected Akaike information criteria},
journal = {Statistics \& Probability Letters},
volume = {33},
number = {2},
pages = {201-208},
year = {1997},
issn = {0167-7152}
}

@ARTICLE{Akaike1974,
  author={Akaike, H.},
  journal={IEEE Transactions on Automatic Control}, 
  title={A new look at the statistical model identification}, 
  year={1974},
  volume={19},
  number={6},
  pages={716-723},
  doi={10.1109/TAC.1974.1100705}}

@ARTICLE{Rodriguez-Ardila2005MNRAS,
       author = {{Rodr{\'\i}guez-Ardila}, A. and {Riffel}, R. and {Pastoriza}, M.~G.},
        title = "{Molecular hydrogen and [FeII] in active galactic nuclei - II. Results for Seyfert 2 galaxies}",
      journal = {\mnras},
         year = 2005,
        month = dec,
       volume = {364},
       number = {3},
        pages = {1041-1053},
          doi = {10.1111/j.1365-2966.2005.09638.x},
       adsurl = {https://ui.adsabs.harvard.edu/abs/2005MNRAS.364.1041R}
}

@INPROCEEDINGS{Kepler:2004,
  author={Altintas, I. and Berkley, C. and Jaeger, E. and Jones, M. and Ludascher, B. and Mock, S.},
  booktitle={Proceedings. 16th International Conference on Scientific and Statistical Database Management, 2004.}, 
  title={Kepler: an extensible system for design and execution of scientific workflows}, 
  year={2004},
  volume={},
  number={},
  pages={423-424},
  doi={10.1109/SSDM.2004.1311241}}

@software{ESO_CPL:2015:04003E,
       author = {{ESO CPL Development Team}},
        title = "{EsoRex: ESO Recipe Execution Tool}",
 howpublished = {Astrophysics Source Code Library, record ascl:1504.003},
         year = 2015,
        month = apr,
          eid = {ascl:1504.003},
       adsurl = {https://ui.adsabs.harvard.edu/abs/2015ascl.soft04003E}
}

@INPROCEEDINGS{Banse:2004:314,
       author = {{Banse}, K. and {Ballester}, P. and {Izzo}, C. and {Jung}, Y. and {Lundin}, L.~K. and {McKay}, D.~J. and {Modigliani}, A. and {Palsa}, R.~M. and {Kiesgen}, M. and {Sabet}, C.},
        title = "{The Common Pipeline Library - a silver bullet for standardising pipelines?}",
    booktitle = {Astronomical Data Analysis Software and Systems (ADASS) XIII},
         year = 2004,
       editor = {{Ochsenbein}, Francois and {Allen}, Mark G. and {Egret}, Daniel},
       series = {Astronomical Society of the Pacific Conference Series},
       volume = {314},
        month = jul,
        pages = {392},
       adsurl = {https://ui.adsabs.harvard.edu/abs/2004ASPC..314..392B}
}

@ARTICLE{Weilbacher:2020:641A,
       author = {{Weilbacher}, Peter M. and {Palsa}, Ralf and {Streicher}, Ole and {Bacon}, Roland and {Urrutia}, Tanya and {Wisotzki}, Lutz and {Conseil}, Simon and {Husemann}, Bernd and {Jarno}, Aur{\'e}lien and {Kelz}, Andreas and {P{\'e}contal-Rousset}, Arlette and {Richard}, Johan and {Roth}, Martin M. and {Selman}, Fernando and {Vernet}, Jo{\"e}l},
        title = "{The data processing pipeline for the MUSE instrument}",
      journal = {\aap},
         year = 2020,
        month = sep,
       volume = {641},
          eid = {A28},
        pages = {A28},
          doi = {10.1051/0004-6361/202037855},
archivePrefix = {arXiv},
       eprint = {2006.08638},
 primaryClass = {astro-ph.IM},
       adsurl = {https://ui.adsabs.harvard.edu/abs/2020A&A...641A..28W}
}

@INPROCEEDINGS{CIAO2006SPIE,
       author = {{Fruscione}, Antonella and {McDowell}, Jonathan C. and {Allen}, Glenn E. and {Brickhouse}, Nancy S. and {Burke}, Douglas J. and {Davis}, John E. and {Durham}, Nick and {Elvis}, Martin and {Galle}, Elizabeth C. and {Harris}, Daniel E. and {Huenemoerder}, David P. and {Houck}, John C. and {Ishibashi}, Bish and {Karovska}, Margarita and {Nicastro}, Fabrizio and {Noble}, Michael S. and {Nowak}, Michael A. and {Primini}, Frank A. and {Siemiginowska}, Aneta and {Smith}, Randall K. and {Wise}, Michael},
        title = "{CIAO: Chandra's data analysis system}",
    booktitle = {Observatory Operations: Strategies, Processes, and Systems},
         year = 2006,
       editor = {{Silva}, David R. and {Doxsey}, Rodger E.},
       series = {Society of Photo-Optical Instrumentation Engineers (SPIE) Conference Series},
       volume = {6270},
        month = jun,
          eid = {62701V},
        pages = {62701V},
          doi = {10.1117/12.671760},
       adsurl = {https://ui.adsabs.harvard.edu/abs/2006SPIE.6270E..1VF}
}

@article{Kalberla:2005:775,
title = {The {Leiden}/{Argentine}/{Bonn} ({LAB}) {Survey} of {Galactic} {HI}. {Final} data release of the combined {LDS} and {IAR} surveys with improved stray-radiation corrections},
volume = {440},
url = {http://adsabs.harvard.edu/cgi-bin/nph-data_query?bibcode=2005A%26A...440..775K&link_type=ABSTRACT},
doi = {10.1051/0004-6361:20041864},
journal = {\aap},
author = {Kalberla, P M W and Burton, W B and Hartmann, Dap and Arnal, E M and Bajaja, E and Morras, R and Pöppel, W G L},
month = sep,
year = {2005},
pages = {775},
}

@article{Haan:2011:100,
title = {The {Nuclear} {Structure} in {Nearby} {Luminous} {Infrared} {Galaxies}: {Hubble} {Space} {Telescope} {NICMOS} {Imaging} of the {GOALS} {Sample}},
volume = {141},
url = {http://adsabs.harvard.edu/cgi-bin/nph-data_query?bibcode=2011AJ....141..100H&link_type=EJOURNAL},
doi = {10.1088/0004-6256/141/3/100},
number = {3},
journal = {\apj},
author = {Haan, S and Surace, J A and Armus, L and Evans, A S and Howell, J H and Mazzarella, J M and Kim, D-C and Vavilkin, T and Inami, H and Sanders, D B and Petric, A and Bridge, C R and Melbourne, J L and Charmandaris, V and Díaz-Santos, T and Murphy, E J and U, V and Stierwalt, S and Marshall, J A},
month = mar,
year = {2011},
note = {Publisher: IOP Publishing},
pages = {100},
}

@article{DiMatteo:2005:604,
title = {Energy input from quasars regulates the growth and activity of black holes and their host galaxies},
volume = {433},
url = {http://adsabs.harvard.edu/cgi-bin/nph-data_query?bibcode=2005Natur.433..604D&link_type=ABSTRACT},
doi = {10.1038/nature03335},
journal = {\nat},
author = {Di Matteo, Tiziana and Springel, Volker and Hernquist, Lars},
month = jan,
year = {2005},
pages = {604},
}

@article{Kennicutt:2012:531,
title = {Star {Formation} in the {Milky} {Way} and {Nearby} {Galaxies}},
volume = {50},
url = {http://adsabs.harvard.edu/cgi-bin/nph-data_query?bibcode=2012ARA%26A..50..531K&link_type=EJOURNAL},
doi = {10.1146/annurev-astro-081811-125610},
number = {1},
journal = {\araa},
author = {Kennicutt, Robert C and Evans, Neal J},
month = sep,
year = {2012},
pages = {531--608},
}

@article{Arnaud:1996:17,
title = {{XSPEC}: {The} {First} {Ten} {Years}},
volume = {101},
url = {http://adsabs.harvard.edu/cgi-bin/nph-data_query?bibcode=1996ASPC..101...17A&link_type=EJOURNAL},
journal = {Astronomical Data Analysis Software and Systems V},
author = {Arnaud, K A},
month = jan,
year = {1996},
pages = {17},
}

@article{Magorrian:1998:2285,
title = {The {Demography} of {Massive} {Dark} {Objects} in {Galaxy} {Centers}},
volume = {115},
url = {http://adsabs.harvard.edu/cgi-bin/nph-data_query?bibcode=1998AJ....115.2285M&link_type=EJOURNAL},
doi = {10.1086/300353},
number = {6},
journal = {\apj},
author = {Magorrian, John and Tremaine, Scott and Richstone, Douglas and Bender, Ralf and Bower, Gary and Dressler, Alan and Faber, S M and Gebhardt, Karl and Green, Richard and Grillmair, Carl and Kormendy, John and Lauer, Tod},
month = jun,
year = {1998},
pages = {2285--2305},
}

@article{Sanders:1988:74,
title = {Ultraluminous infrared galaxies and the origin of quasars},
volume = {325},
url = {http://adsabs.harvard.edu/cgi-bin/nph-data_query?bibcode=1988ApJ...325...74S&link_type=ABSTRACT},
doi = {10.1086/165983},
journal = {\apj},
author = {Sanders, D B and Soifer, B T and Elias, J H and Madore, B F and Matthews, K and Neugebauer, G and Scoville, N Z},
month = feb,
year = {1988},
pages = {74--91},
}

@article{Sanders:1988:L35,
title = {Warm ultraluminous galaxies in the {IRAS} survey - {The} transition from galaxy to quasar?},
volume = {328},
url = {http://adsabs.harvard.edu/cgi-bin/nph-data_query?bibcode=1988ApJ...328L..35S&link_type=ABSTRACT},
doi = {10.1086/185155},
journal = {\apj},
author = {Sanders, D B and Soifer, B T and Elias, J H and Neugebauer, G and Matthews, K},
month = may,
year = {1988},
pages = {L35},
}

@article{Hopkins:2006:1,
title = {Fueling {Low}‐{Level} {AGN} {Activity} through {Stochastic} {Accretion} of {Cold} {Gas}},
volume = {166},
url = {http://stacks.iop.org/0067-0049/166/i=1/a=1},
doi = {10.1086/505753},
number = {1},
journal = {\apjs},
author = {Hopkins, Philip F and Hernquist, Lars},
month = sep,
year = {2006},
note = {Publisher: IOP Publishing},
pages = {1--36},
}

@article{Bertin:1996:393,
title = {{SExtractor}: {Software} for source extraction.},
volume = {117},
url = {http://adsabs.harvard.edu/cgi-bin/nph-data_query?bibcode=1996A%26AS..117..393B&link_type=ABSTRACT},
journal = {\aaps},
author = {Bertin, E and Arnouts, S},
month = may,
year = {1996},
pages = {393},
}

@article{Kewley:2001:37,
title = {Optical {Classification} of {Southern} {Warm} {Infrared} {Galaxies}},
volume = {132},
url = {http://adsabs.harvard.edu/cgi-bin/nph-data_query?bibcode=2001ApJS..132...37K&link_type=ABSTRACT},
doi = {10.1086/318944},
journal = {\apjs},
author = {Kewley, L J and Heisler, C A and Dopita, M A and Lumsden, S},
month = jan,
year = {2001},
note = {Publisher: IOP Publishing},
pages = {37},
}

@article{Baldwin:1981:5,
title = {Classification parameters for the emission-line spectra of extragalactic objects},
volume = {93},
url = {http://adsabs.harvard.edu/cgi-bin/nph-data_query?bibcode=1981PASP...93....5B&link_type=ABSTRACT},
doi = {10.1086/130766},
journal = {Astronomical Society of the Pacific},
author = {Baldwin, J A and Phillips, M M and Terlevich, R},
month = feb,
year = {1981},
pages = {5},
}

@article{Ferrarese:2000:L9a,
title = {A {Fundamental} {Relation} between {Supermassive} {Black} {Holes} and {Their} {Host} {Galaxies}},
volume = {539},
url = {http://adsabs.harvard.edu/cgi-bin/nph-data_query?bibcode=2000ApJ...539L...9F&link_type=ABSTRACT},
doi = {10.1086/312838},
number = {1},
journal = {\apj},
author = {Ferrarese, Laura and Merritt, David},
month = aug,
year = {2000},
pages = {L9--L12},
}

@article{Kauffmann:2003:1055,
title = {The host galaxies of active galactic nuclei},
volume = {346},
url = {http://adsabs.harvard.edu/cgi-bin/nph-data_query?bibcode=2003MNRAS.346.1055K&link_type=ABSTRACT},
doi = {10.1111/j.1365-2966.2003.07154.x},
journal = {\mnras},
author = {Kauffmann, Guinevere and Heckman, Timothy M and Tremonti, Christy and Brinchmann, Jarle and Charlot, Stéphane and White, Simon D M and Ridgway, Susan E and Brinkmann, Jon and Fukugita, Masataka and Hall, Patrick B and Ivezić, Željko and Richards, Gordon T and Schneider, Donald P},
month = nov,
year = {2003},
pages = {1055},
}

@article{Cardelli:1989:245,
title = {The relationship between infrared, optical, and ultraviolet extinction},
volume = {345},
url = {http://adsabs.harvard.edu/cgi-bin/nph-data_query?bibcode=1989ApJ...345..245C&link_type=ABSTRACT},
doi = {10.1086/167900},
journal = {\apj},
author = {Cardelli, Jason A and Clayton, Geoffrey C and Mathis, John S},
month = sep,
year = {1989},
pages = {245},
}

@article{Lamperti:2017:540,
title = {{BAT} {AGN} {Spectroscopic} {Survey} - {IV}: {Near}-{Infrared} {Coronal} {Lines}, {Hidden} {Broad} {Lines}, and {Correlation} with {Hard} {X}-ray {Emission}},
volume = {467},
url = {http://adsabs.harvard.edu/cgi-bin/nph-data_query?bibcode=2017MNRAS.467..540L&link_type=EJOURNAL},
doi = {10.1093/mnras/stx055},
number = {1},
journal = {\mnras},
author = {Lamperti, Isabella and Koss, Michael and Trakhtenbrot, Benny and Schawinski, Kevin and Ricci, Claudio and Oh, Kyuseok and Landt, Hermine and Riffel, Rogério and Rodríguez-Ardila, Alberto and Gehrels, Neil and Harrison, Fiona and Masetti, Nicola and Mushotzky, Richard and Treister, Ezequiel and Ueda, Yoshihiro and Veilleux, Sylvain},
month = may,
year = {2017},
pages = {540--572},
}

@article{Calzetti:2000:682,
title = {The {Dust} {Content} and {Opacity} of {Actively} {Star}-forming {Galaxies}},
volume = {533},
url = {http://iopscience.iop.org/0004-637X/533/2/682},
doi = {10.1086/308692},
journal = {\apj},
author = {Calzetti, Daniela and Armus, Lee and Bohlin, Ralph and Kinney, Anne and Koornneef, Jan and Storchi-Bergmann, Thaisa},
month = apr,
year = {2000},
pages = {682},
}

@article{Cappellari:2017:798,
title = {Improving the full spectrum fitting method: accurate convolution with {Gauss}-{Hermite} functions},
volume = {466},
url = {http://adsabs.harvard.edu/cgi-bin/nph-data_query?bibcode=2017MNRAS.466..798C&link_type=EJOURNAL},
doi = {10.1093/mnras/stw3020},
number = {1},
journal = {\mnras},
author = {Cappellari, Michele},
month = apr,
year = {2017},
pages = {798--811},
}

@article{Schawinski:2007:512,
title = {The {Effect} of {Environment} on the {Ultraviolet} {Color}-{Magnitude} {Relation} of {Early}-{Type} {Galaxies}},
volume = {173},
url = {http://adsabs.harvard.edu/cgi-bin/nph-data_query?bibcode=2007ApJS..173..512S&link_type=ABSTRACT},
doi = {10.1086/516631},
number = {2},
journal = {\apjs},
author = {Schawinski, K and Kaviraj, S and Khochfar, S and Yoon, S J and Yi, S K and Deharveng, J M and Boselli, A and Barlow, T and Conrow, T and Forster, K and Friedman, P G and Martin, D C and Morrissey, P and Neff, S and Schiminovich, D and Seibert, M and Small, T and Wyder, T and Bianchi, L and Donas, J and Heckman, T and Lee, Y W and Madore, B and Milliard, B and Rich, R M and Szalay, A},
month = dec,
year = {2007},
note = {Publisher: IOP Publishing},
pages = {512--523},
}

@article{Kewley:2006:961,
title = {The host galaxies and classification of active galactic nuclei},
volume = {372},
url = {http://mnras.oxfordjournals.org/cgi/doi/10.1111/j.1365-2966.2006.10859.x},
doi = {10.1111/j.1365-2966.2006.10859.x},
number = {3},
journal = {\mnras},
author = {Kewley, L J and Groves, B and Kauffmann, G and Heckman, T},
month = nov,
year = {2006},
pages = {961--976},
}

@article{Lynden-Bell:1969:690,
title = {Galactic {Nuclei} as {Collapsed} {Old} {Quasars}},
volume = {223},
url = {http://adsabs.harvard.edu/cgi-bin/nph-data_query?bibcode=1969Natur.223..690L&link_type=ABSTRACT},
doi = {10.1038/223690a0},
journal = {\nat},
author = {Lynden-Bell, D},
month = aug,
year = {1969},
pages = {690},
}

@article{Freudling:2013:A96,
title = {Automated data reduction workflows for astronomy. {The} {ESO} {Reflex} environment},
volume = {559},
url = {http://adsabs.harvard.edu/cgi-bin/nph-data_query?bibcode=2013A%26A...559A..96F&link_type=EJOURNAL},
doi = {10.1051/0004-6361/201322494},
journal = {\aap},
author = {Freudling, W and Romaniello, M and Bramich, D M and Ballester, P and Forchi, V and García-Dabló, C E and Moehler, S and Neeser, M J},
month = nov,
year = {2013},
note = {Publisher: EDP Sciences},
pages = {A96},
}

@article{Armus:2009:559,
title = {{GOALS}: {The} {Great} {Observatories} {All}-{Sky} {LIRG} {Survey}},
volume = {121},
url = {http://adsabs.harvard.edu/cgi-bin/nph-data_query?bibcode=2009PASP..121..559A&link_type=ABSTRACT},
doi = {10.1086/600092},
journal = {PASP},
author = {Armus, L and Mazzarella, J M and Evans, A S and Surace, J A and Sanders, D B and Iwasawa, K and Frayer, D T and Howell, J H and Chan, B and Petric, A and Vavilkin, T and Kim, D-C and Haan, S and Inami, H and Murphy, E J and Appleton, P N and Barnes, J E and Bothun, G and Bridge, C R and Charmandaris, V and Jensen, J B and Kewley, L J and Lord, S and Madore, B F and Marshall, J A and Melbourne, J E and Rich, J and Satyapal, S and Schulz, B and Spoon, H W W and Sturm, E and U, V and Veilleux, S and Xu, K},
month = jan,
year = {2009},
pages = {559},
}

@article{Laor:2008:847,
title = {On the origin of radio emission in radio-quiet quasars},
volume = {390},
url = {http://adsabs.harvard.edu/cgi-bin/nph-data_query?bibcode=2008MNRAS.390..847L&link_type=ABSTRACT},
doi = {10.1111/j.1365-2966.2008.13806.x},
journal = {\mnras},
author = {Laor, Ari and Behar, Ehud},
month = sep,
year = {2008},
pages = {847},
}

@article{Hopkins:2008:356,
title = {A {Cosmological} {Framework} for the {Co}‐evolution of {Quasars}, {Supermassive} {Black} {Holes}, and {Elliptical} {Galaxies}. {I}. {Galaxy} {Mergers} and {Quasar} {Activity}},
volume = {175},
url = {http://stacks.iop.org/0067-0049/175/i=2/a=356},
doi = {10.1086/524362},
number = {2},
journal = {\apjs},
author = {Hopkins, Philip F and Hernquist, Lars and Cox, Thomas J and Kereš, Dušan},
month = apr,
year = {2008},
note = {Publisher: IOP Publishing},
pages = {356--389},
}

@article{Kormendy:2013:511,
title = {Coevolution ({Or} {Not}) of {Supermassive} {Black} {Holes} and {Host} {Galaxies}},
volume = {51},
url = {http://www.annualreviews.org/doi/abs/10.1146/annurev-astro-082708-101811},
doi = {10.1146/annurev-astro-082708-101811},
number = {1},
journal = {\araa},
author = {Kormendy, John and Ho, Luis C},
month = aug,
year = {2013},
note = {Publisher: Annual Reviews},
pages = {511--653},
}

@article{Assef:2018:23,
title = {The {WISE} {AGN} {Catalog}},
volume = {234},
issn = {0067-0049},
url = {https://ui.adsabs.harvard.edu/abs/2018ApJS..234...23A},
doi = {10.3847/1538-4365/aaa00a},
urldate = {2021-08-04},
journal = {\apjs},
author = {Assef, R. J. and Stern, D. and Noirot, G. and Jun, H. D. and Cutri, R. M. and Eisenhardt, P. R. M.},
month = feb,
year = {2018},
pages = {23},
}

@article{Liu:2019:21,
title = {A comprehensive and uniform sample of broad-line active galactic nuclei from the {SDSS} {DR7}},
volume = {243},
issn = {1538-4365},
url = {http://arxiv.org/abs/1906.05597},
doi = {10.3847/1538-4365/ab298b},
number = {2},
urldate = {2021-08-11},
journal = {\apjs},
author = {Liu, He-Yang and Liu, Wen-Juan and Dong, Xiao-Bo and Zhou, Hongyan and Wang, Tinggui and Lu, Honglin and Yuan, Weimin},
month = jul,
year = {2019},
note = {arXiv: 1906.05597},
pages = {21},
}

@article{Cushing:2004:362,
title = {Spextool: {A} {Spectral} {Extraction} {Package} for {SpeX}, a 0.8-5.5 {Micron} {Cross}-{Dispersed} {Spectrograph}},
volume = {116},
issn = {0004-6280},
shorttitle = {Spextool},
url = {https://ui.adsabs.harvard.edu/abs/2004PASP..116..362C},
doi = {10.1086/382907},
urldate = {2021-08-27},
journal = {PASP},
author = {Cushing, Michael C. and Vacca, William D. and Rayner, John T.},
month = apr,
year = {2004},
note = {ADS Bibcode: 2004PASP..116..362C},
pages = {362--376},
}

@article{Hoffman:2007:957,
title = {Dynamics of triple black hole systems in hierarchically merging massive galaxies},
volume = {377},
issn = {0035-8711},
url = {https://doi.org/10.1111/j.1365-2966.2007.11694.x},
doi = {10.1111/j.1365-2966.2007.11694.x},
number = {3},
urldate = {2021-09-09},
journal = {\mnras},
author = {Hoffman, Loren and Loeb, Abraham},
month = may,
year = {2007},
pages = {957--976},
}

@article{Koss:2022:1,
title = {{BASS}. {XXI}. {The} {Data} {Release} 2 {Overview}},
volume = {261},
issn = {0067-0049},
url = {https://doi.org/10.3847/1538-4365/ac6c8f},
doi = {10.3847/1538-4365/ac6c8f},
language = {en},
number = {1},
urldate = {2022-07-22},
journal = {\apjs},
author = {Koss, Michael J. and Trakhtenbrot, Benny and Ricci, Claudio and Bauer, Franz E. and Treister, Ezequiel and Mushotzky, Richard and Urry, C. Megan and Ananna, Tonima T. and Baloković, Mislav and Brok, Jakob S. den and Cenko, S. Bradley and Harrison, Fiona and Ichikawa, Kohei and Lamperti, Isabella and Lein, Amy and Mejía-Restrepo, Julian E. and Oh, Kyuseok and Pacucci, Fabio and Pfeifle, Ryan W. and Powell, Meredith C. and Privon, George C. and Ricci, Federica and Salvato, Mara and Schawinski, Kevin and Shimizu, Taro and Smith, Krista L. and Stern, Daniel},
month = jul,
year = {2022},
note = {Publisher: American Astronomical Society},
pages = {1},
}

@article{McMullin:2007:127,
title = {{CASA} {Architecture} and {Applications}},
journal = {Astronomical Data Analysis Software and Systems XVI},
volume = {376},
url = {https://ui.adsabs.harvard.edu/abs/2007ASPC..376..127M},
urldate = {2022-08-06},
author = {McMullin, J. P. and Waters, B. and Schiebel, D. and Young, W. and Golap, K.},
month = oct,
year = {2007},
note = {Conference Name: Astronomical Data Analysis Software and Systems XVI 
ADS Bibcode: 2007ASPC..376..127M},
pages = {127},
}

@inproceedings{Larkin:2006:441,
title = {{OSIRIS}: a diffraction limited integral field spectrograph for {Keck}},
volume = {6269},
shorttitle = {{OSIRIS}},
url = {https://www.spiedigitallibrary.org/conference-proceedings-of-spie/6269/62691A/OSIRIS-a-diffraction-limited-integral-field-spectrograph-for-Keck/10.1117/12.672061.full},
doi = {10.1117/12.672061},
urldate = {2022-08-06},
booktitle = {Ground-based and {Airborne} {Instrumentation} for {Astronomy}},
publisher = {SPIE},
author = {Larkin, James and Barczys, Matthew and Krabbe, Alfred and Adkins, Sean and Aliado, Ted and Amico, Paola and Brims, George and Campbell, Randy and Canfield, John and Gasaway, Thomas and Honey, Allan and Iserlohe, Christof and Johnson, Christopher A. and Kress, Evan and LaFreniere, David and Lyke, James and Magnone, Ken and Magnone, Nick and McElwain, Michael and Moon, Juleen and Quirrenbach, Andreas and Skulason, Gunnar and Song, Inseok and Spencer, Michael and Weiss, Jason and Wright, Shelley},
month = jun,
year = {2006},
pages = {441--445},
}

@article{Bacon:2010:773508a,
title = {The {MUSE} second-generation {VLT} instrument},
volume = {7735},
url = {https://ui.adsabs.harvard.edu/abs/2010SPIE.7735E..08B},
doi = {10.1117/12.856027},
urldate = {2022-08-14},
author = {Bacon, R. and Accardo, M. and Adjali, L. and Anwand, H. and Bauer, S. and Biswas, I. and Blaizot, J. and Boudon, D. and Brau-Nogue, S. and Brinchmann, J. and Caillier, P. and Capoani, L. and Carollo, C. M. and Contini, T. and Couderc, P. and Daguisé, E. and Deiries, S. and Delabre, B. and Dreizler, S. and Dubois, J. and Dupieux, M. and Dupuy, C. and Emsellem, E. and Fechner, T. and Fleischmann, A. and François, M. and Gallou, G. and Gharsa, T. and Glindemann, A. and Gojak, D. and Guiderdoni, B. and Hansali, G. and Hahn, T. and Jarno, A. and Kelz, A. and Koehler, C. and Kosmalski, J. and Laurent, F. and Le Floch, M. and Lilly, S. J. and Lizon, J. -L. and Loupias, M. and Manescau, A. and Monstein, C. and Nicklas, H. and Olaya, J. -C. and Pares, L. and Pasquini, L. and Pécontal-Rousset, A. and Pelló, R. and Petit, C. and Popow, E. and Reiss, R. and Remillieux, A. and Renault, E. and Roth, M. and Rupprecht, G. and Serre, D. and Schaye, J. and Soucail, G. and Steinmetz, M. and Streicher, O. and Stuik, R. and Valentin, , H. and Vernet, J. and Weilbacher, P. and Wisotzki, L. and Yerle, N.},
month = jul,
year = {2010},journal = {Proc. SPIE},
note = {Conference Name: Ground-based and Airborne Instrumentation for Astronomy III 
ADS Bibcode: 2010SPIE.7735E..08B},
pages = {773508},
}

@article{Kollatschny:2020:A79,
title = {{NGC} 6240: {A} triple nucleus system in the advanced or final state of merging},
volume = {633},
issn = {0004-6361},
shorttitle = {{NGC} 6240},
url = {https://ui.adsabs.harvard.edu/abs/2020A&A...633A..79K/abstract},
doi = {10.1051/0004-6361/201936540},
language = {en},
urldate = {2023-03-12},
journal = {\aap},
author = {Kollatschny, W. and Weilbacher, P. M. and Ochmann, M. W. and Chelouche, D. and Monreal-Ibero, A. and Bacon, R. and Contini, T.},
month = jan,
year = {2020},
pages = {A79},
}
\bibliographystyle{myapj}

\end{document}